\documentclass[fleqn,usenatbib]{rasti}

\usepackage{newtxtext,newtxmath}
\usepackage[T1]{fontenc}

\DeclareRobustCommand{\VAN}[3]{#2}
\let\VANthebibliography\thebibliography
\def\thebibliography{\DeclareRobustCommand{\VAN}[3]{##3}\VANthebibliography}

\newcommand{\FITNAperiod}{\mbox{$0.9689951$}}
\newcommand{\FITNArb}{\mbox{$0.1070$}}
\newcommand{\FITNArsuma}{\mbox{$0.342724$}}
\newcommand{\FITNAtc}{\mbox{$2458392.587691$}}
\newcommand{\FITNAqone}{\mbox{$0.323$}}
\newcommand{\FITNAqtwo}{\mbox{$0.376$}}
\newcommand{\FITNAinc}{\mbox{$0.097583$}}

\newcommand{\CMOSNArb}{\mbox{$0.10905_{-0.0013}^{+0.00067}$}}
\newcommand{\CMOSNArsuma}{\mbox{$0.3475_{-0.0021}^{+0.0017}$}}%
\newcommand{\CMOSNAinc}{\mbox{$0.09744_{-0.00028}^{+0.00036}$}}
\newcommand{\CMOSNAtc}{\mbox{$2460892.5977_{-0.0039}^{+0.0055}$}}
\newcommand{\CMOSNAperiod}{\mbox{$0.9679_{-0.0018}^{+0.0013}$}}
\newcommand{\CMOSNAqone}{\mbox{$0.3234_{-0.0029}^{+0.0033}$}}
\newcommand{\CMOSNAqtwo}{\mbox{$0.353_{-0.037}^{+0.040}$}}

\newcommand{\FITNBperiod}{\mbox{$1.21749415$}}
\newcommand{\FITNBrb}{\mbox{$0.0771$}}
\newcommand{\FITNBrsuma}{\mbox{$0.291897$}}
\newcommand{\FITNBtc}{\mbox{$2457727.01162$}}
\newcommand{\FITNBqone}{\mbox{$0.318$}}
\newcommand{\FITNBqtwo}{\mbox{$0.369$}}
\newcommand{\FITNBinc}{\mbox{$0.055822$}}

\newcommand{\CMOSNBrb}{\mbox{$0.07732_{-0.0024}^{+0.00077}$}}
\newcommand{\CMOSNBrsuma}{\mbox{$0.2732_{-0.0011}^{+0.0033}$}}%
\newcommand{\CMOSNBinc}{\mbox{$0.0496_{-0.0081}^{+0.016}$}}
\newcommand{\CMOSNBtc}{\mbox{$2460894.9301_{-0.0021}^{+1.2}$}}
\newcommand{\CMOSNBperiod}{\mbox{$1.21845_{-0.0024}^{+0.00051}$}}
\newcommand{\CMOSNBqone}{\mbox{$0.3263_{-0.018}^{+0.0012}$}}
\newcommand{\CMOSNBqtwo}{\mbox{$0.410_{-0.027}^{+0.022}$}}

\newcommand{\FITNCperiod}{\mbox{$1.855557555$}}
\newcommand{\FITNCrb}{\mbox{$0.10253$}}
\newcommand{\FITNCrsuma}{\mbox{$0.218193$}}
\newcommand{\FITNCtc}{\mbox{$2455664.159537$}}
\newcommand{\FITNCqone}{\mbox{$0.369$}}
\newcommand{\FITNCqtwo}{\mbox{$0.396$}}
\newcommand{\FITNCinc}{\mbox{$0.135473$}}

\newcommand{\CMOSNCrb}{\mbox{$0.10160_{-0.0034}^{+0.00026}$}}
\newcommand{\CMOSNCrsuma}{\mbox{$0.2151_{-0.0016}^{+0.0048}$}}%
\newcommand{\CMOSNCinc}{\mbox{$0.13789_{-0.0069}^{+0.00041}$}}
\newcommand{\CMOSNCtc}{\mbox{$2460893.1_{-3.7}^{+5.6}$}}
\newcommand{\CMOSNCperiod}{\mbox{$1.85368_{-0.00020}^{+0.0059}$}}
\newcommand{\CMOSNCqone}{\mbox{$0.3557_{-0.0035}^{+0.015}$}}
\newcommand{\CMOSNCqtwo}{\mbox{$0.4155_{-0.058}^{+0.0067}$}}

\newcommand{\FITNEperiod}{\mbox{$2.816504852$}}
\newcommand{\FITNErb}{\mbox{$0.121703$}}
\newcommand{\FITNErsuma}{\mbox{$0.117198$}}
\newcommand{\FITNEtc}{\mbox{$2455003.96270000$}}
\newcommand{\FITNEqone}{\mbox{$0.382$}}
\newcommand{\FITNEqtwo}{\mbox{$0.393$}}
\newcommand{\FITNEinc}{\mbox{$0.010995$}}

\newcommand{\CMOSNErb}{\mbox{$0.1150_{-0.0054}^{+0.0028}$}}
\newcommand{\CMOSNErsuma}{\mbox{$0.1146_{-0.0050}^{+0.0027}$}}%
\newcommand{\CMOSNEinc}{\mbox{$0.0142_{-0.0060}^{+0.0043}$}}
\newcommand{\CMOSNEtc}{\mbox{$2460904.5375_{-0.0021}^{+0.0016}$}}
\newcommand{\CMOSNEperiod}{\mbox{$2.8150_{-0.0023}^{+0.0013}$}}
\newcommand{\CMOSNEqone}{\mbox{$0.3809_{-0.0083}^{+0.012}$}}
\newcommand{\CMOSNEqtwo}{\mbox{$0.321_{-0.065}^{+0.11}$}}

\newcommand{\FITNFperiod}{\mbox{$1.544928884$}}
\newcommand{\FITNFrb}{\mbox{$0.1033$}}
\newcommand{\FITNFrsuma}{\mbox{$0.247377$}}
\newcommand{\FITNFtc}{\mbox{$2455654.042121$}}
\newcommand{\FITNFqone}{\mbox{$0.345$}}
\newcommand{\FITNFqtwo}{\mbox{$0.384$}}
\newcommand{\FITNFinc}{\mbox{$0.215588$}}

\newcommand{\CMOSNFrb}{\mbox{$0.10656_{-0.0026}^{+0.00025}$}}
\newcommand{\CMOSNFrsuma}{\mbox{$0.24629_{-0.0025}^{+0.00015}$}}%
\newcommand{\CMOSNFinc}{\mbox{$0.21193_{-0.00050}^{+0.00064}$}}
\newcommand{\CMOSNFtc}{\mbox{$2460934.6_{-9.3}^{+1.5}$}}
\newcommand{\CMOSNFperiod}{\mbox{$1.54583_{-0.0014}^{+0.00075}$}}
\newcommand{\CMOSNFqone}{\mbox{$0.35485_{-0.0050}^{+0.00060}$}}
\newcommand{\CMOSNFqtwo}{\mbox{$0.3297_{-0.0070}^{+0.058}$}}

\usepackage{graphicx}	% Including figure files
\usepackage{amsmath}	% Advanced maths commands
\usepackage{orcidlink}
\usepackage{anyfontsize}
\usepackage{soul}
\usepackage{amsmath}	% Advanced maths commands
\usepackage{siunitx} % For \unit and other SI unit formatting
\newcommand{\eps}{{e$^-$\,pix$^{-1}$\,s$^{-1}$}}
\newcommand{\tess}{\textit{TESS}}

\newcommand{\kepler}{\textit{Kepler}}
\newcommand{\plato}{\textit{PLATO}}

\title[Andor Toran-26B CMOS camera]{High-Precision Photometry with a scientific CMOS Camera: III Lab and on-Sky testing of Andor Toran-26B}

\author[Ioannis Apergis et al.]{
Ioannis~Apergis$^{\orcidlink{0009-0004-7473-4573}}$,$^{1,2,3}$\thanks{E-mail: Ioannis.Apergis@warwick.ac.uk (IA)}
Leonidas~Asimakoulas$^{\orcidlink{0000-0003-3676-5606}}$,$^{4,5}$
Daniel~Bayliss$^{\orcidlink{0000-0001-6023-1335}}$,$^{1,2}$
Paul~Chote,$^{1,3}$
\newauthor
Morgan~A.~Mitchell$^{\orcidlink{0009-0004-6130-7775}}$,$^{1,2}$
James~McCormac$^{\orcidlink{0000-0003-1631-4170}}$,$^{1,2,3}$
Ben~Godson$^{\orcidlink{0000-0003-3766-7266}}$,$^{1}$
Philip~G.~Steen$^{6}$
\\
$^{1}$Department of Physics, University of Warwick, Gibbet Hill Road, Coventry CV4 7AL, UK\\
$^{2}$Centre for Exoplanets and Habitability, University of Warwick, Gibbet Hill Road, Coventry CV4 7AL, UK\\
$^{3}$Centre for Space Domain Awareness, University of Warwick, Gibbet Hill Road, Coventry CV4 7AL, UK\\
$^{4}$School of Computing, Engineering and Intelligent Systems, Ulster University, Magee Campus, Northland Road, Derry\textasciitilde{}Londonderry BT48 7JL, UK\\
$^{5}$Semiconductor Photonics Education and Research Centre (SPEAR), Ulster University, Magee Campus, Northland Road, Derry\textasciitilde{}Londonderry BT48 7JL, UK\\
$^{6}$Andor Technology Ltd, Springvale Business Park, 7 Millennium Way, Belfast BT12 7AL, UK
}

\date{Accepted XXX. Received YYY; in original form ZZZ}

\pubyear{2025}

\begin{document}
\label{firstpage}
\pagerange{\pageref{firstpage}--\pageref{lastpage}}
\maketitle

\begin{abstract}
Scientific CMOS cameras are increasingly used in modern astronomy, offering CCD-like performance with lower power use, faster readout, and reduced cost. We evaluate the photometric precision of the Toran-26B camera, featuring the IMX571 sensor from Sony, a new CMOS camera from Andor Technology, through laboratory and on-sky tests in its two modes: high well capacity (HWC) and low noise (LN). Key metrics include read noise, dark current, photo and dark signal non-uniformity, quantum efficiency, and window transmittance. The median read noise is 1.07\,e$^-$ (LN) and 2.79\,e$^-$ (HWC). Bias shows minor dark signal non-uniformity: 0.076\,e$^-$ (LN), 0.173\,e$^-$ (HWC). Pixel well depth reaches 15700\,e$^-$ (LN) and 50000\,e$^-$ (HWC), with dynamic ranges of 83.39\,dB and 85.05\,dB, respectively. Linearities are 99.93\,\% (LN) and 99.82\,\% (HWC) within the 5-95\,\% range of the saturation signal. Photo response non-uniformity is 0.076\,\% (LN) and 0.05\,\% (HWC). Dark current at 0\,$^\circ$C is 0.001\,\eps, (LN and HWC). On-sky tests show that, for our test telescope configuration, the HWC mode can be camera noise limited for specific magnitude ranges, while the LN mode remains free of such limitations, making it the preferred choice.
\end{abstract}

% Include between one and six keywords.

%%%%%%%%%%%%%%%%%%%%%%%%%%%%%%%%%%%%%%%%%%%%%%%%%%

%%%%%%%%%%%%%%%%% BODY OF PAPER %%%%%%%%%%%%%%%%%%

\section{Introduction}
\label{sec:introduction}
Charge-Coupled Devices (CCDs) have long been the standard for optical astronomical imaging, particularly in applications requiring high-precision photometry. Notable examples include instruments on \kepler\ \citep{2010Sci...327..977B}, Transiting Exoplanet Survey Satellite \citep[\tess]{Ricker2015TESS} and PLAnetary Transits and Oscillations \citep[\plato]{PLATO_MISSION_2025}. CCDs are prized for their high sensitivity, low read noise, minimal dark current, excellent stability, and uniform pixel response. Architectural enhancements, such as deep depletion technology \citep{1480428, peckerar1981x, 1985ITED...32.1525T, gimeno2016sky} and backside illuminated designs, increase their quantum efficiency across a broad spectral range. 

Developed in parallel with CCD sensor technology, Complementary Metal-Oxide Semiconductor (CMOS) sensors are based on a fundamentally different architecture. Unlike the passive pixel array of CCD sensors, CMOS sensors use active pixels, with each pixel containing field-effect transistors for performing charge-to-voltage conversion on a per-pixel basis. CMOS sensors have long been favoured in consumer electronics due to their fast readout speeds, low power consumption, and lower cost. Today CMOS image sensors are typically more affordable and accessible than CCD image sensors, and recent lab and on-sky evaluations demonstrate their increasing suitability for scientific imaging \citep{jorden2014e2v, 2019AN....340..638K, 2023sndd.confE...3A, 2025arXiv250312449X, Layden_2025, Ioannis24, Ioannis26}. CMOS cameras have found significant usage in astronomy \citep{Sako_2018, 2019AN....340..638K, 2020SPIE11525E..2IW, ardilanov2021high, 2021RAA....21..268Q, stefanov2022cmos, townsend2023electro, 2023sndd.confE...3A, Wang_2024, 2024SPIE13103E..0RK, Ioannis26, Roth_2026, Layden_2026}. Their high frame rates and short exposure times enable observations of rapidly varying astronomical phenomena \citep{Richmond_2020}, lucky and speckle imaging \citep{Strakhov_2023}, and wavefront sensing \citep{Guyon_2010, Downing_2016, Hopgood_2023}, while their low readout overheads allow high observing duty cycles.

In this work, we aim to contribute to the growing body of literature evaluating whether modern CMOS detectors can deliver the level of precision required to rival that of traditional CCD detectors. To this end, we conduct a comprehensive characterisation and performance assessment of the Toran 26B CMOS camera, developed by Oxford Instruments Andor\footnote{\url{https://andor.oxinst.com/}}, through both controlled laboratory experiments and on-sky observations using the TWenty Inch Survey Telescope (TWIST) telescope, a University of Warwick facility located at the Observatorio del Roque de los Muchachos in La Palma, Spain \citep{mitchell_search_2025}.

This paper is organised as follows: in Section~\ref{sec:new_camera} we describe the CMOS camera and its capabilities. In Section~\ref{sec:total-equipment} we describe the laboratory set up, the telescope configuration and the observations. Our methodologies of laboratory and on-sky tests are described in Section~\ref{sec:methods}. In Section~\ref{sec:results} we present our results and discussion and in Section~\ref{sec:conclusion} we set out our conclusions.

\section{The Toran-26B CMOS Camera}
\label{sec:new_camera}
The Toran 26B CMOS camera is equipped with the IMX571 sensor. This sensor is the smaller version of the Sony models IMX455 and IMX411 \citep{alarcon2023scientific}. The camera has two imaging modes at different bit depths, referred to the low noise (LN) in 10-bit, 11-bit, 12-bit and 16-bit and the high well capacity (HWC) in 16-bit. For our lab-test and on-sky test applications, which use FITS images format, we primarily operated in 16-bit for both HWC and LN modes. The camera can be connected via either a USB 3.2 or a CoaXPress interface\footnote{More info can be found in \citet{Secull_2026}}.

\begin{table}
\centering
\caption[Specifications of the sensor for the Toran 26B camera]{Specifications of the sensor for the Toran 26B camera taken from the manufacturer sheet, and configuration on TWIST telescope.}
\label{tab:specs_toran}
\begin{tabular}{lllcc}
\hline 
\hline
\textbf{Parameter}             &  &  &     \textbf{Value/Description}         \\ \hline  
Camera                         &  &  &     Toran 26B                               \\
Sensor Type                    &  &  &     CMOS                               \\
Model                          &  &  &     IMX571 Sony                             \\
Min Temperature*        &  &  &     -20\,$^{\circ}$C air-cooled                             \\
Active pixels                  &  &  &     4176 $\times$ 6252                 \\
Pixel size                     &  &  &     3.76 $\times$ 3.76\,$\unit{\um}$   \\
Sensor size                    &  &  &     15.7 $\times$ 23.5\,mm             \\
Peak Response                  &  &  &     $>$90\%  @500\,nm                    \\
Bit depths                     &  &  &     10-bit, 11-bit, 12-bit, 16-bit     \\
Shutter type                   &  &  &     Rolling shutter                    \\
TWIST Pixel angular scale      &  &  &     0.225\,arcsec\,pixel$^{-1}$ \\
TWIST Field Of View            &  &  &     36.1\,$\times$\,24.1 arcmin$^2$ 
\\ \hline         
 \multicolumn{4}{l}{*The current version is available only with air cooling.}\\
\end{tabular}
\end{table}

\section{Optical bench, Telescope and Observations}
\label{sec:total-equipment}
\subsection{Laboratory Setup}
\label{sec:lab-equipment}
\begin{figure*}
\includegraphics[width=\textwidth]{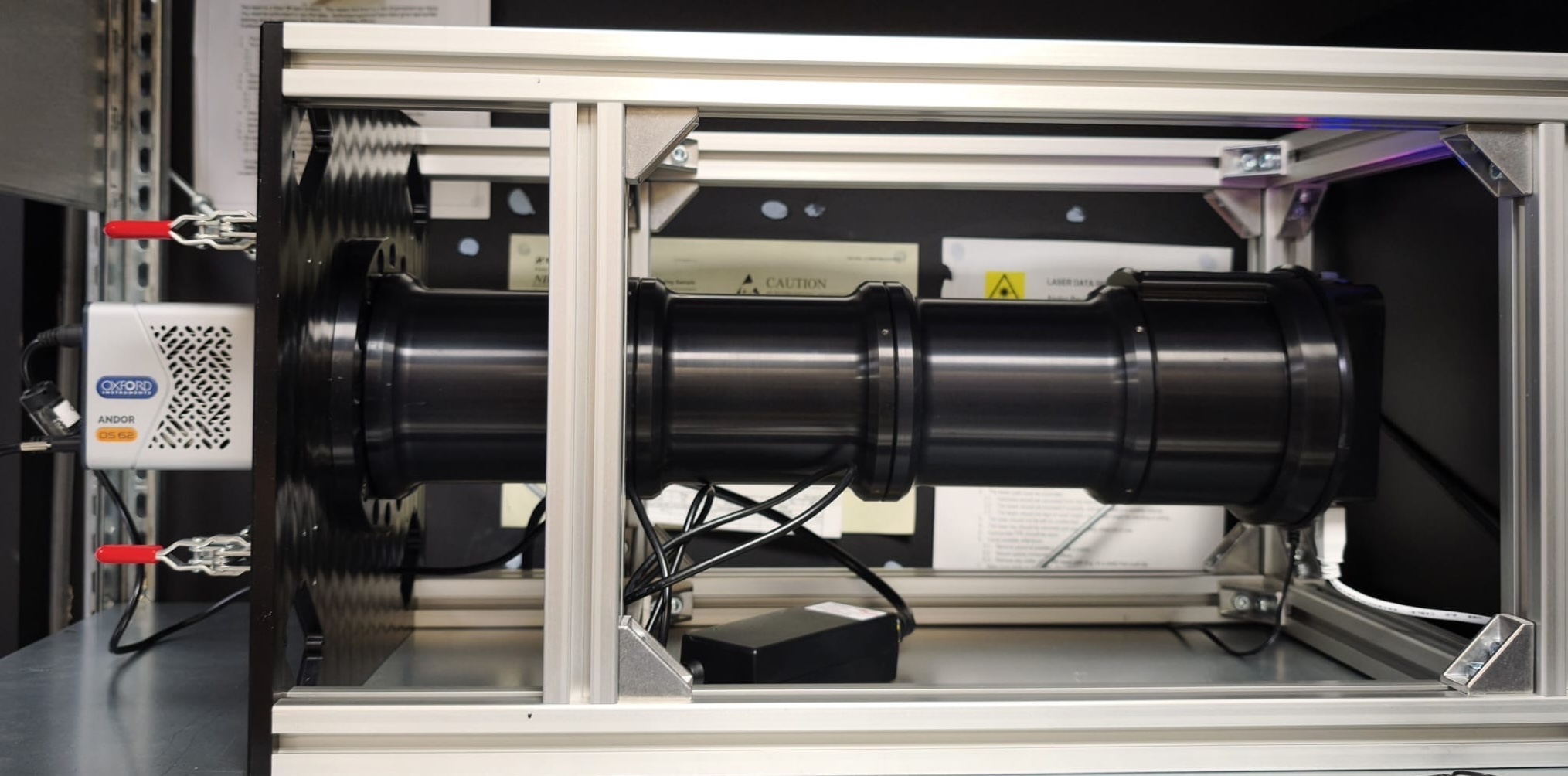}
\caption[Optical bench setup for the Toran 26B camera]{Optical bench setup for the Toran 26B camera. The black cylindrical component is the light tube, which accommodates the light source and the diffuser on the right side. The camera is mounted on the left side of the light tube.}
\label{fig:lab_setup}
\end{figure*}

A DC-powered light rig containing a 4-V blue four-LED array was used
as the light source. The configuration was similar to that described
by \citet{Ioannis24}. The LED array was positioned at one end of a
cylindrical enclosure, with the Toran 26B camera mounted at the
opposite end. This geometry defined the illumination path, while the
black, anti-reflective lining suppressed stray light and internal
reflections. A diffuser placed in front of the LED array improved the
spatial uniformity of the illumination across the detector. The
resulting light output was stable and quasi-uniform, with a peak
wavelength of approximately 465\,nm. This configuration was used to measure the characteristics of the sensor, such as photon transfer curve, photo response and dark signal non-uniformity, linearity read noise and dark current. The full light rig setup, including the Toran 26B camera mounted at the left, is shown in \autoref{fig:lab_setup}. 

To estimate the quantum efficiency (QE) of the camera and the transmittance of the front window, we used a methodology and optical bench setup similar to that described in \citet{Ioannis24}, with the exception of the light source. We used a silicon photodiode (model S1337-1010BQ by Hamamatsu), with sensitivity across the optical range and a peak responsivity at 960\,nm. The photodiode current was measured using a Keithley picoammeter connected via an RS232 interface. For wavelength selection, we employed the Andor Shamrock SR-303i spectrograph as a monochromator. Illumination was provided by the CoolLED pe-4000 light source, which is optimised for fluorescence microscopy and delivers broadband light using 16 LEDs spanning wavelengths from 365\,nm to 770\,nm. The use of the spectrograph allowed us to interpolate between the discrete LED wavelengths, resulting in a finer wavelength sampling and a more accurate QE curve. The total process was controlled and operated via Python using the Andor SDK Python wrapper. The light source was operated manually to select the desired wavelengths during the measurement process.

\subsection{Telescope configuration and Observations}
\label{sec:sky-equipment}

\begin{figure}\centering
\includegraphics[width=\columnwidth]{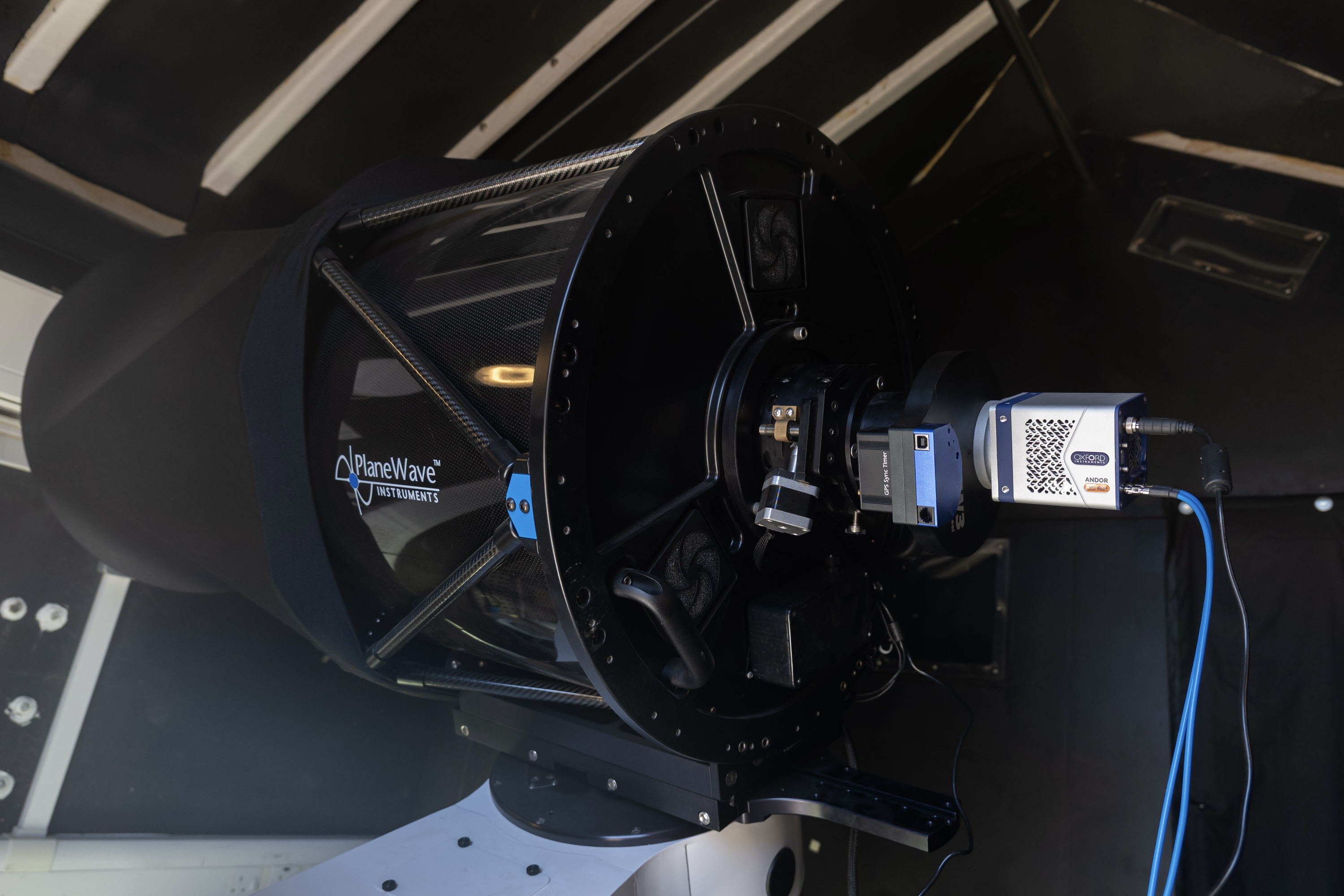}
\caption[The TWenty-Inch Survey Telescope]{The TWenty-Inch Survey Telescope equipped with the Toran 26B camera, shown in its enclosure at La Palma, Canary Islands.}
\label{fig:telescope}
\end{figure}

After completing the laboratory tests, we installed the Toran 26B camera on the 50\,cm aperture TWIST telescope located at the Observatorio del Roque de los Muchachos on La Palma, Canary Islands. The camera was connected to the control computer via a CoaXPress connection. TWIST consists of a PlaneWave Instruments $f$/6.8 CDK20 optical tube assembly (OTA) equatorially mounted on a L500 direct drive mount. The telescope configuration with the camera is shown in \autoref{fig:telescope}.

% Filters (hardware) Rockit to communicate (software).
% Configurations, 10 seconds? observations

Observations were conducted between August and September 2025. We made observations of well-known transiting exoplanets around stars including KELT \citep{KELT2018}, Kepler host stars \citep{2010Sci...327..977B}, as well as confirmed \tess\ Objects of Interest (TOI) systems \citep{Ricker2015TESS}. The target list is shown in \autoref{tab:targets}. The telescope and camera were controlled using the \texttt{Rockit} observatory control system\footnote{\url{https://rockit-astro.github.io/}}. For consistency between nightly measurements, the CMOS camera was configured with a 10 second exposure time and a total cadence of 10.147 seconds, with the remaining time accounting for the readout. In our operation we did not employ filters. The images from the CMOS camera were saved as FITS files onto the CMOS control computer. Instrument parameters were saved into the FITS file headers, including the JD timestamps, telescope coordinates, and camera settings. In this paper we analyse data from 5 nights where a known transiting exoplanet system was observed, as show in \autoref{tab:targets}. We use the field observed for KELT-16 (R.A. = 20$^{\mathrm{h}}$57$^{\mathrm{m}}$04\fs4, Dec = +31$^{\circ}$39\arcmin39\farcs) to analyse the noise performance of the camera. To this end, observations were obtained in both HWC and LN modes during bright nights (2025 August 06 and 08) and dark nights (2025 August 19 and September 21).

\begin{table*}
\centering
\caption{Transit observations of the stars.}
\begin{tabular}{ccccccc}
\hline
\hline
Target   & TIC         & TESS            & Teff             & Observation        & Observation Duration & Transit fraction \\ 
  Star &  ID        &  Magnitude *          &  (K)            &   Date      & (hours) & \\ 
\hline
KELT-16 \citep{KELT-16_2017}    & TIC 236445129 & 11.43  & 5760 $\pm$ 130    & 2025 Aug 6, 8, 19, Sep 21 & 8.8 \& 8.4 \& 6.9   & Full    \\
KELT-1 \citep{kelt1}        & TIC 432549364 & 10.22  & 5540 $\pm$ 140   & 2025 Aug 10 & 5.5   & Full     \\
Kepler-41 \citep{kepler41}  & TIC 271042217 & 13.9   & 5660 $\pm$ 113   & 2025 Aug 12 & 7     & Full     \\ 
% TOI-5688 A \citep{toi5688}  & TIC 193634953 & 14.2   & 3710 $\pm$ 60    & 2025 Aug 15 & 4.8   & Egress   \\ 
Kepler-670 \citep{kepler670}& TIC 27990610  & 14.48  & 5530 $\pm$ 110   & 2025 Aug 16 & 6     & Full     \\
Kepler-76 \citep{kepler76}  & TIC 138644215 & 12.89  & 6300 $\pm$ 126   & 2025 Sep 9  & 6.3   & Full     \\ \hline
 \multicolumn{3}{l}{* source: TIC v8.2 \citep{2018AJ....156..102S}}\\
\label{tab:targets}
\end{tabular}
\end{table*}

\section{Methods}
\label{sec:methods}

\subsection{Bias and Dark}
\label{sec:bias_dark-method}
The bias is measured in analog-to-digital units (ADU) and contains contributions from several sources within the readout chain, including frame-level and column level offsets. A frame-level analog offset prevents the analog to digital converters from clipping negative voltage fluctuations, which would truncate the read noise distribution at zero \citep{2000hccd.book.....H}. We define the bias signal as the camera output under dark conditions and with the minimum possible exposure time. For our analysis, we acquired 75 bias frames. These frames were utilized both individually and as averaged stacks for temporal and spatial noise analysis, particularly for evaluating read noise and Dark Signal Non-Uniformity (DSNU). The resulting bias frames were also used to subtract the bias level from dark and illuminated images.

To measure the temporal noise, we adopted a similar approach to that described in \citet{Ioannis24}. For each pixel, we
calculated the standard deviation of its recorded value across a
sequence of bias frames. This compares the same pixel location between
successive images and therefore provides a per-pixel estimate of the read noise \citep{fowler2011single}. In addition, we examined individual rows and columns across the bias frames to investigate the behaviour of correlated noise arising from electronic interference within the camera circuitry or power supply fluctuations \citep{geurts201798db}. This structured noise, commonly referred to as row banding, is a known artefact in imaging systems \citep{2014SPIE.9154E..2IW, 2019AN....340..638K, Shao_2024}.

To assess the spatial noise, an average bias frame was created by median-stacking the individual bias images. We then assessed the pixel-to-pixel, column-to-column, and row-to-row variations within this averaged frame to quantify non-uniformities \cite{jahne2010emva, Ioannis24}. This is know as DSNU. For the DSNU and readout noise measurements, the exposure time was set to its minimum value at $8.35\times 10^{-5}$\,s for both LN and HWC mode. 

We analyse dark frames which are acquired with a certain exposure time under dark conditions. At non-zero detector temperatures, statistical thermal excitation promotes electrons in the silicon into the conduction band, producing dark current that accumulates during the exposure \citep{PhysRev.87.835, Aberle1992ImpactOI}. We investigate the dark current at 0\,$^{\circ}$C that was also used in our on-sky tests, to quantify its level and to identify any additional artefacts or signals, such as sensor glow \citep{1985SPIE..570....7J} as has been observed in other CMOS camera models \citep{2019AN....340..638K, 2020SPIE11525E..2IW, 2021RAA....21..268Q, Layden_2025, Ioannis24}. 

\subsection{Photo Response and Linearity}
\label{sec:ptc-linearity-prnu-method}
To measure the electron to ADU conversion factor (gain) we produce a photon transfer curve, which is a plot of the flux variance versus the mean flux level, as described by \cite{2007SPIE.6690E..03J}. We maintained stable illumination and varied the exposure time incrementally, capturing two consecutive frames at each step, ranging from low signal levels up to sensor saturation. Each pair of frames was subtracted to eliminate the fixed pattern (or spatial) noise, isolating the temporal noise component. 

% A linear fit to the resulting variance versus signal data yields the inverse of the gain. The point at which the variance reaches its maximum indicates the pixel’s full well capacity.

To evaluate the pixel response uniformity, we estimated the Photo Response Non-Uniformity (PRNU) by acquiring images at approximately half of the sensor’s saturation level. Averaging a large number ($\sim$100) of half-saturation signal level frames helps suppress temporal noise, allowing the fixed-pattern component to be isolated more effectively. Our method follows the procedures outlined in \citet{jahne2010emva} and implemented in \citet{Ioannis24}. PRNU is defined as the dimensionless ratio of fixed-pattern noise to the mean signal, and it quantifies the spatial variation in pixel responsivity across the sensor.

To assess the linearity of the sensor, we evaluated how the recorded signal increases with exposure time under constant illumination. Ideally, in a linear system, the signal should scale proportionally with exposure time. Any deviation from this linear relationship indicates non-linearity in the sensor response. 

% We fit a straight line to the measured signal as a function of exposure time in order to extract the residuals and estimate the system's non-linearity.

\subsection{Quantum Efficiency and Transmission Window}
\label{sec:qe_tra-method}

To estimate the transmission of the camera window, we first measured the current of a photodiode across various wavelengths using different filter configurations, including a single spectrograph grating mirror. The entrance and exit slit widths were each set to 400\,$\mu$m. With the 1200-lines\,mm$^{-1}$ grating used in these measurements, the nominal reciprocal dispersion of the Shamrock SR 303i is approximately 2.6\,nm\,mm$^{-1}$ at 500\,nm\footnote{found in Andor Shamrock 303i manual version 4.2.}, corresponding to an approximate spectral bandwidth of 1.0\,nm. For each selected wavelength, we recorded 25 photodiode current measurements using a Keithley picoammeter and calculated the mean and standard deviation. This measurement procedure was then repeated with the camera window placed in front of the light path. By comparing the photodiode current values with and without the window, we estimated the transmission percentage of the window as a function of wavelength.

To determine the camera’s QE, we acquired 10 images at each selected wavelength using an exposure time of 0.01\,s, which provided a signal level close to half of the sensor’s full-well capacity. The average pixel value across these frames was taken as the camera signal. Additionally, 10 bias frames were acquired and averaged for bias level subtraction. Dark current subtraction was omitted due to the sensor’s negligible dark current at short exposure times. We extract the QE of the photodiode using the responsitivity curve from Hamamatsu website.
We compared the signal values of the camera with the current values and the QE of photodiode and extract the QE for the camera. Our method is follows the steps and equations described in \citet{Ioannis24} and \cite{Layden_2025}.

\subsection{Image reduction and photometry}
\label{sec:image-reduction}
For each observation,  we execute a custom pipeline to extract photometry for each observed field. We solve each image astrometrically using the World Coordinate System (WCS), accounting for the field centre and pixel scale, with the help of \textit{astrometry.net} \citep{astrometry}.  While the software applies a simple polynomial model, we find that the solution can be less reliable near the image corners due to optical distortions. To mitigate this, we cross-match sources with the TESS Input Catalog version 8 \citep[TIC8]{2018AJ....156..102S}, focusing on stars brighter than $\mathrm{T}<16$\,mag. The distortion parameters in the initial WCS solution are iteratively refined until sub-pixel accuracy is reached \citep{2025AdSpR..76..764C}. Images for which the WCS solution fails are excluded, typically representing less than 3\,\% of the night’s observations. For our photometric analysis we selected isolated stars from the TIC8 catalogue. Stars with neighbours within 100 pixels and 1 delta magnitude of brightness were flagged as blended and rejected from our analysis. We also exclude stars located near the edges of the sensor.

For each observing run, a set of calibration images was obtained, including bias and flat-field frames. Dark frames were not acquired since the sensor exhibits a negligible dark current (see Section~\ref{sec:dark_noise_results}). Bias frames were obtained using the minimum exposure time. We used 21 bias frames and average them to construct the master bias frame. We followed the same procedure for the flat images, however, we also apply a normalisation to the master flat image and subtract the master bias signal. We subtract the master bias and divide the master flat from the science images. 

To ensure that autoguiding was operating with sub-pixel accuracy, we used the \texttt{DONUTS} algorithm \citep{DONUTS} to assess the image alignment throughout the dataset. Given the pixel scale of 0.22\,arcsec\,pixel$^{-1}$, few pixel shifts were expected under ideal conditions. However, since large photometric apertures were used, we adopted an upper limit of 4 pixels for allowable image shifts. Frames exceeding this threshold were excluded from the analysis, typically accounting for less than 0.6\,\% of the night’s observations.

\subsection{Photometry}
\label{sec:photometry-noise}
We tested a range of aperture radii from 10 to 50 pixels, in steps of 10. Upon examining the resulting lightcurves, we found that an aperture radius of 30 pixels (6.6\,arcsec) produced the lowest RMS scatter. Although this aperture size may not be optimal for faint stars, we adopted it for constructing the noise models for stars with $10<\mathrm{T}<16$ (see Section~\ref{sec:noise_model}). For specific targets, such as exoplanet transits, the aperture size was adjusted to achieve the best RMS scatter in the light curves. 

We performed aperture photometry by extracting the stellar positions from the catalog and using the \texttt{SEP} Python package \citep{sex, sex2016}. Circular apertures with radii ranging from 10 to 50 pixels in steps of 10 were applied. For each star, an annulus was defined to estimate the local sky background, with the inner and outer radii set to 60 and 65 pixels, respectively, to ensure the background region remained well separated from the largest photometric aperture.

We used comparison stars to remove atmospheric systematics, including airmass-dependent variations, from the target light curves. For each target, candidate comparison stars were selected according to their \textit{Gaia} DR3 ($G_{\mathrm{BP}}-G_{\mathrm{RP}}$) colour indices \citep{Castro-Ginard_2024}, magnitude differences, and angular separations relative to the target. The initial configuration imposed no restrictions on colour or angular separation, but required candidates to be within $\pm3$\,mag of the target. Stars brighter than $T=9$\,mag were excluded because they were likely to be saturated.

The candidates were then filtered iteratively according to their photometric stability, quantified by the root-mean-square (RMS) scatter \citep{Ioannis26}. Each candidate's light curve was corrected for airmass-dependent trends and normalised before the RMS of its residuals was calculated. The resulting RMS values were grouped into magnitude bins, and a spline was used to model the median RMS trend with brightness. Stars with RMS values more than $2\sigma$ above this trend were rejected. This procedure was repeated ten times to progressively exclude variable or noisy stars, as illustrated in \autoref{fig:rms_vs_mags_material}. The fluxes of the remaining comparison stars were summed to construct a master comparison light curve, which was used to detrend the target light curve. The RMS of the detrended target light curve was then measured.

We repeated this procedure using different constraints on colour, angular separation, and magnitude difference. These configurations included progressively tighter selection criteria, as well as combinations of strict colour constraints and relaxed angular-separation limits. For each configuration, the RMS of the detrended target light curve was recorded alongside the corresponding selection parameters. The results were ranked from highest to lowest RMS, and the final configuration was selected after inspecting both the master comparison light curve and the detrended target light curve. This procedure was applied to each target star in the field to identify the comparison-star ensemble that best captured the systematics affecting its light curve.

\subsection{System Noise}
The total photometric noise comprises several components that can influence the quality of the data. As stellar fluxes propagate through the Earth’s atmosphere, they experience fluctuations due to atmospheric turbulence \citep{2015MNRASOsborn}, introducing scintillation noise into the photometric measurements. We estimated the scintillation contribution following the approach described in \citep{mitchell_search_2025}, taking into account the scintillation coefficient as well as the telescope and observatory site parameters. The received flux from the source and the sky background follows the Poisson distribution. Instrumental noise components include the read noise and the dark current. The total noise for a given photometric aperture is summarised as \citep{1989PASP..101..616H}:

\begin{equation}
\label{eq.6}
    \mathrm{\sigma_{total}} = \sqrt{\mathrm{\sigma_{flux}}^2 + \mathrm{\sigma_{sky}}^2 + \mathrm{\sigma_{DC}}^2 + \mathrm{\sigma_{RN}}^2 + f^2\mathrm{\sigma_{sc}}^2},
\end{equation}

where $f$ is the flux from the target star, $\mathrm{\sigma_{flux}}$ is the shot noise from the target star, $\mathrm{\sigma_{sky}}$ is the sky background noise, $\mathrm{\sigma_{DC}}$ is the dark current noise, $\mathrm{\sigma_{RN}}$ is the read noise and $\mathrm{\sigma_{sc}}$ is the scintillation noise.

% Timescale of Noise
We examined the photometry of our target stars to assess the presence of correlated noise in the light curves (see similar work in \citep{Ioannis26} and for correlated noise in \citep{2006MNRAS.373..231P}). This was achieved by binning the data as a function of time and analysing how the RMS value of the residuals changes with increasing bin size. For light curves dominated by uncorrelated (white) noise, the RMS is expected to decrease proportionally to the inverse square root of the number of points per bin, following the relation RMS $\propto 1 / \sqrt{n}$. In contrast, if the RMS deviates from this white noise trend, by decreasing more slowly with binning, it indicates the presence of time-correlated noise. We apply this method by averaging the RMS values of a sample of stars, randomly selected when possible, at different magnitudes.

\section{Results, Analysis and Discussion}
\label{sec:results}
\subsection{PTC, PRNU and Linearity measurements}
\label{sec:ptc_prnu_linearity_results}
\begin{figure*}
\includegraphics[width=\textwidth]{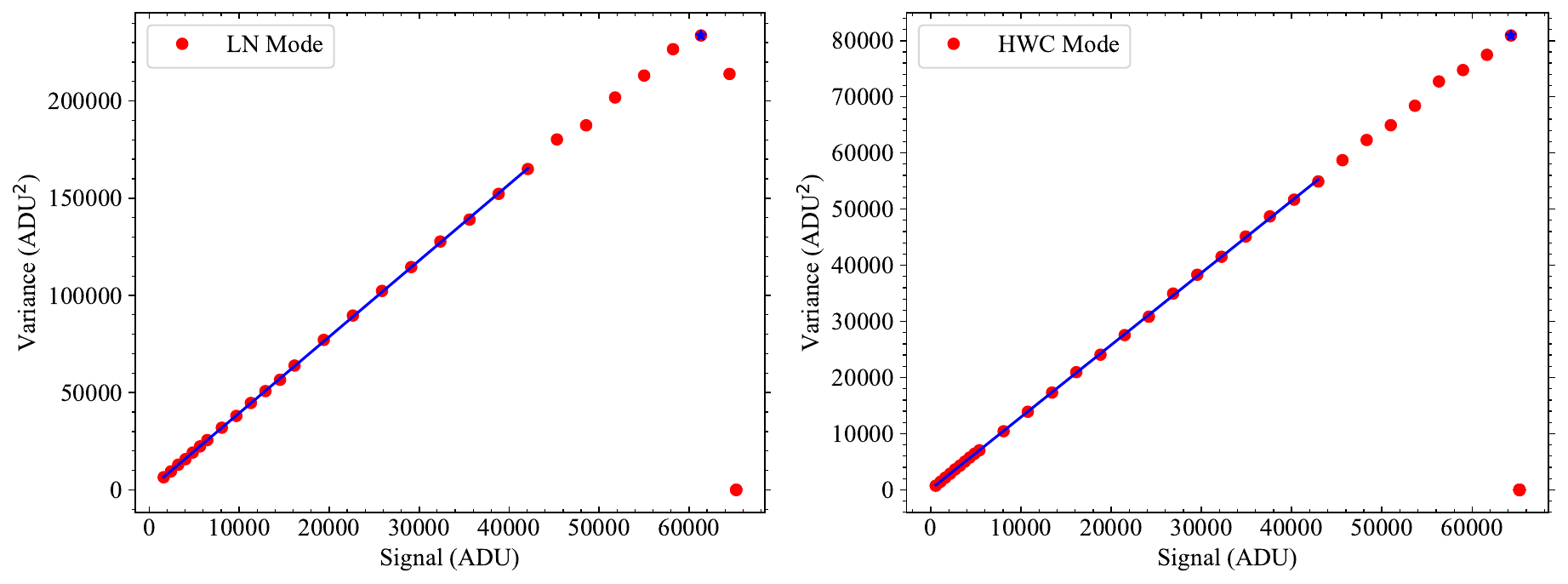}
\caption[Photon transfer curve for Toran 26B]{Photon transfer curves for the Toran 26B camera at 0\,$^{\circ}$C. \textbf{Left:} The LN image mode. \textbf{Right:} The HWC image mode. The solid blue line is a best fit linear model. The data point with the highest variance is marked with a blue star, and indicates the FWC value.}
\label{fig_ptc_toran}
\end{figure*}

To measure the photon transfer curve (PTC) in HWC mode, we began with an exposure of 0.01 seconds, incrementing by 0.01 seconds up to 0.1 seconds. Beyond this point, the step size was increased to 0.05 seconds until saturation was reached at 1.6 seconds. In LN mode, the sequence also started at 0.01 seconds, but with an initial step size of 0.005 seconds up to 0.1 seconds. From there, the exposure time was incremented by 0.02 seconds until saturation occurred at 0.46 seconds. The mean signal of the consecutive frames and the variance of the subtracted same consecutive frames was plotted for each exposure step as show in \autoref{fig_ptc_toran}. The PTC appears fairly linear without significant deviations that have been seen in other CMOS camera models \citep{2021RAA....21..268Q, 2024SPIE13103E..0RK, Ioannis24} that utilise two channels to produce a high dynamic range image. 

A linear fit was applied to the photon-transfer curve over the range 0–70\,\% of the saturation level. The inverse slope of this fit gives the system gain, which was found to be 0.779\,e$^-$/ADU for the HWC mode and 0.255\,e$^-$/ADU for the LN mode. The full well capacity was then estimated from the signal level at the maximum variance point, converted from ADU to electrons using the measured gains. This gave full well capacities of 15,700\,e$^-$ for the LN mode and 50,000\,e$^-$ for the HWC mode.

\begin{figure*}
\includegraphics[width=\textwidth]{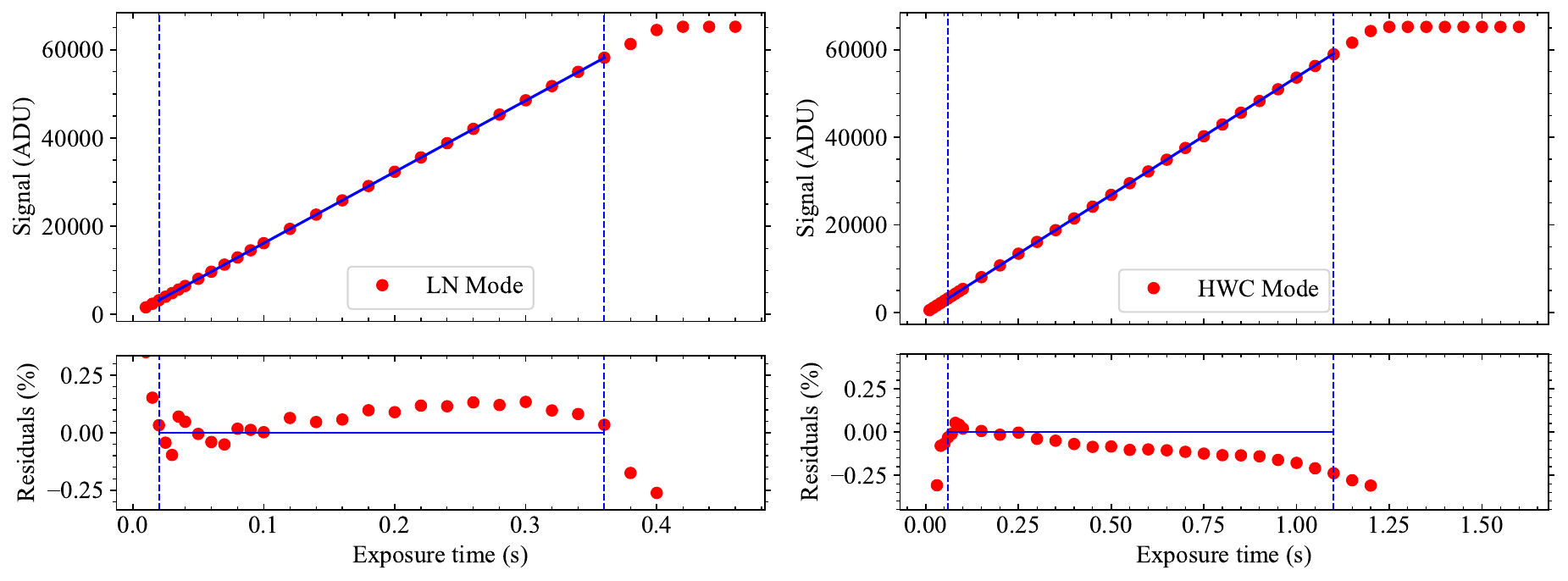}
\caption[Linearity for the Toran 26B]{Linearity measurements for the Toran 26B camera at 0\,$^{\circ}$C. \textbf{Left:} The LN image mode. \textbf{Right:} The HWC image mode. The best fit linear model between 5 and 95 \% of saturation is plotted with a blue solid line. Residuals to the linear model are shown in the bottom panels. The dashed horizontal blue lines indicate the 5 to 95\% boundaries we set to measure fit the linearity.}
\label{fig:linearity_toran}
\end{figure*}

The mean signal was analysed as a function of exposure time to assess the system’s linearity. A line was fitted to the measured signal versus exposure, and the residuals were used to quantify the system’s non-linearity. The fit employed inverse weighting following the method described in \cite{jahne2010emva} and implemented in \autoref{sec:methods} in \citet{Ioannis24}. The non-linearity was measured within 5\,\% to 95\,\% from the saturation as the average absolute deviation of the residuals as show in \autoref{fig:linearity_toran}. The non-linearity was found to be 0.067\,\% for the LN mode and 0.09\,\% for the HWC mode.

We further investigated the detector response below the conventional 5-95\,\% FWC interval using an additional, more finely sampled linearity data set obtained with another Toran 26B camera, as shown in a log-log scale in  \autoref{fig:linearity_toran_0-95}. Characterising this regime is important because low signal non-linearity in CMOS detectors may indicate incomplete charge transfer \citep{janesick2006fundamental} and a corresponding reduction in the effective internal quantum efficiency \citep{Layden_2025}. For example, \citet{Layden_2026} identified two distinct features in the response of the ORCA Quest 2. This is the discontinuity caused by switching between its high- and low-gain signal paths, and a gradual low signal non-linearity attributed to incomplete charge transfer. A gain transition feature has also been reported for the Marana camera by \citet{karpov2020} and later in \citet{Ioannis24}. In contrast, each operating mode of the Toran 26B uses a single gain path, and no comparable discontinuity is observed in our measurements.

The additional measurements extend from 5.12\,ADU (1.3\,e$^-$) in LN mode and 1.7\,ADU (1.3\,e$^-$) in HWC mode to saturation. When the fitted interval was extended from the lowest sampled positive signal to 95\,\% of the FWC, the linearity error increased to 0.341\,\% in LN mode to 0.316\,\% in HWC mode. The corresponding fitted intercepts were small and positive, particularly 0.008\,e$^-$ in LN mode and 0.03\,e$^-$ in HWC mode. We therefore find no evidence for the substantial negative offset associated with the fixed charge-trapping capacity reported by \citet{Layden_2026}, although these measurements cannot exclude smaller or more complex signal dependent charge collection losses. The inclusion of the low signal measurements reveals a modest increase in the overall non-linearity; however, the maximum linearity error remains below 0.35\,\% across the investigated dynamic range.

\begin{figure}\centering
\includegraphics[width=\columnwidth]{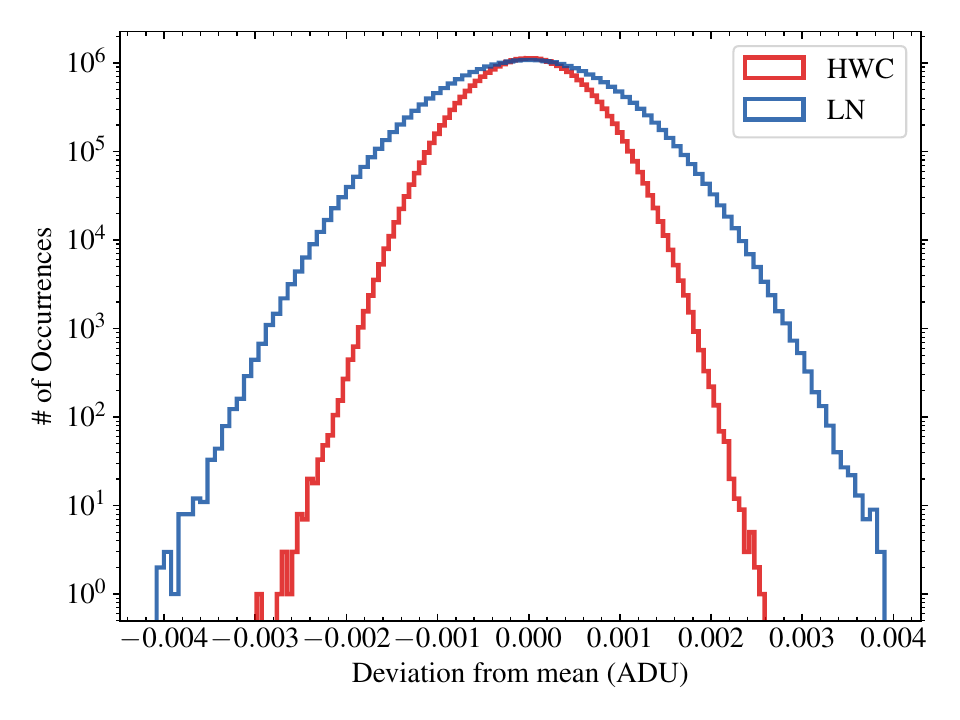}
\caption[PRNU for Toran 26B]{Histogram distributions of the photo response non-uniformities for both modes. The red bins indicate the HWC while the blue the LN mode.}
\label{fig:prnu}
\end{figure}

To ensure sufficient sampling for estimating the PRNU while avoiding system memory issues, we stacked 500 images in a buffer, following the approach described in \citet{jahne2010emva}. The measured PRNU was 0.076\,\% for the LN mode and 0.050\,\% for the HWC mode. The slightly broader distribution in the LN mode is consistent with its higher PRNU value. The PRNU distribution is shown in \autoref{fig:prnu}.

\subsection{Bias level Spatial and Temporal Noise}
\label{sec:temporal_spatial_noise_results}
\begin{figure}\centering
\includegraphics[width=\columnwidth]{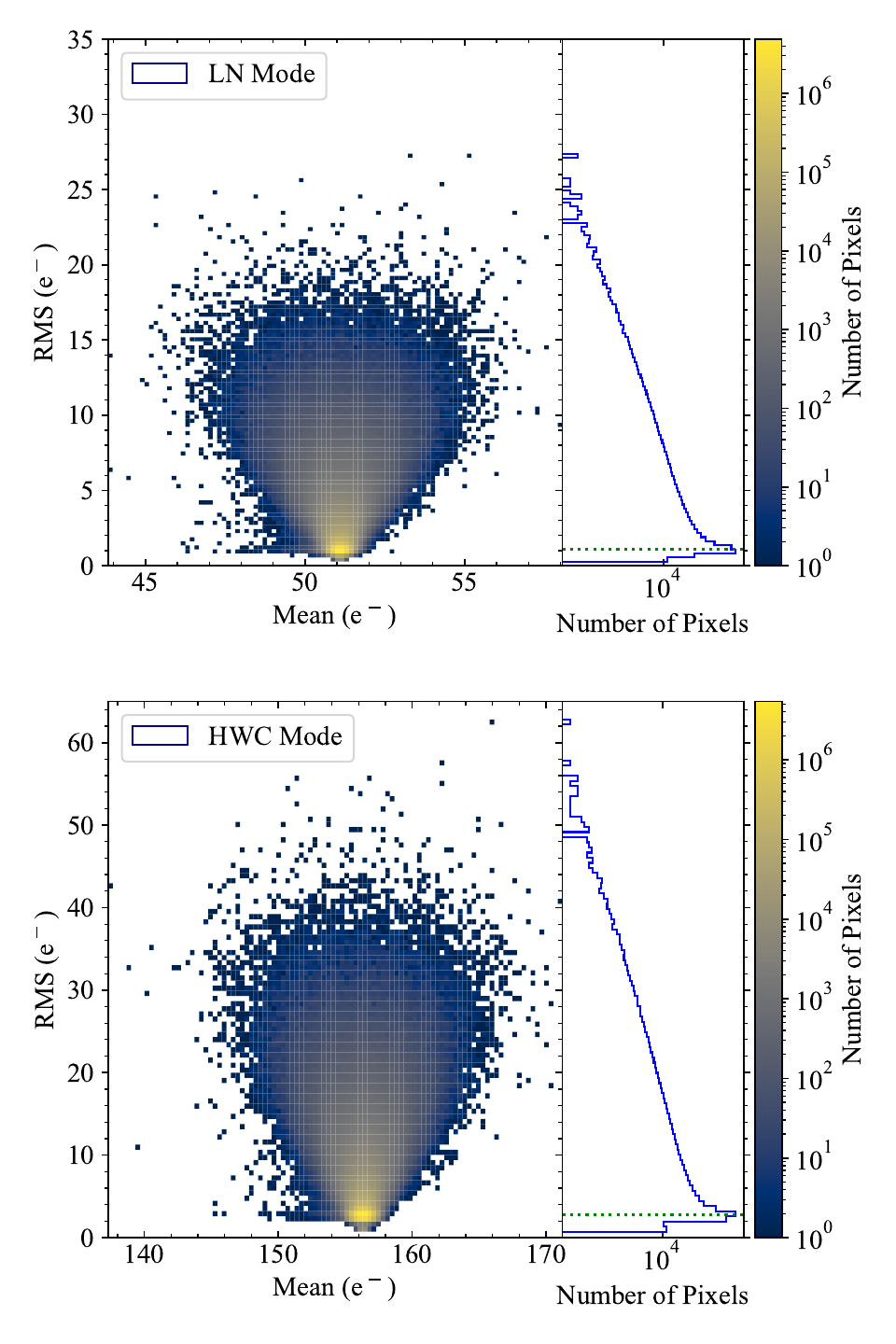}
\caption[Read noise for Toran 26B]{Read noise distribution derived from 75 consecutive bias frames. The 2-D histogram represents the temporal noise calculated from the mean and standard deviation of each pixel across all frames, with the colour bar indicating the number of pixels. The blue histograms show the logarithmic distribution of the read noise, with the median value marked by a green dotted line. \textbf{Top:} Median read noise distribution for the LN mode. \textbf{Bottom:} Median read noise distribution for the HWC mode.}
\label{fig:rn_toran}
\end{figure}

For each of the 75 bias images, we study pixels individually by reporting the standard deviation and mean values. Our results are illustrated in \autoref{fig:rn_toran}. The distribution of the read noise is described with long tail in the histogram, which implies that the pixels are not represented by the same read noise. This feature is a common feature in CMOS sensor since each pixel has its own circuitry. Noisy amplifiers, and random telegraph signal (RTS) behaviours cause noisy pixels \citep{338418, 4154408}. Nevertheless, the RMS, mean and median read noise for the Toran 26B camera was found to be 1.53\,e$^-$, 1.28\,e$^-$ and 1.07\,e$^-$ for the LN mode and 3.61\,e$^-$, 3.19\,e$^-$ and 2.79\,e$^-$ for the HWC respectively.

For each of the 75 bias images, we analysed the temporal noise along rows and columns by computing the standard deviation and mean values. The distributions of the column and row temporal noise are shown in \autoref{fig:col_col} and \autoref{fig:row_row_rn}, respectively.
The mean column temporal noise was measured to be 0.023\,e$^-$ in LN mode and 0.056\,e$^-$ in HWC mode. The corresponding mean row temporal noise was 0.064\,e$^-$ for LN mode and 0.173\,e$^-$ for HWC mode. These values are significantly lower than the overall system read noise, indicating minimal fluctuation in the column-level amplifiers and minimal contribution from the row banding. In contrast, the row temporal noise shows a more pronounced difference between the two modes, suggesting increased row-level variability in HWC mode.

% RESULTS FOR THE DSNU here
To evaluate the spatial noise distribution across the image, we averaged 75 bias frames by computing the mean value at each pixel position, thereby constructing a representative averaged bias image. Given the use of a CMOS sensor, it is important to investigate the different contributions to the DSNU specifically, the total, pixel, row, and column components, to determine whether the dominant spatial noise arises from pixel-level variation or from fixed patterns from row or column. The results are presented in \autoref{fig:dsnu_rn}, where the mean bias level has been subtracted, and the residuals corresponding to the total, row, and column DSNU are shown. The residual distributions are highly uniform and follow a Gaussian profile. For the LN mode, the distribution is narrower compared to the HWC mode. 
% The measured total, pixel, row, and column DSNU values are 0.076\,e$^-$, 0.075\,e$^-$, 0.002\,e$^-$, 0.008\,e$^-$ for the LN mode, and 0.173\,e$^-$, 0.172\,e$^-$, 0.007\,e$^-$, 0.021\,e$^-$ for the HWC mode. 

Additionally, we investigated spatial correlations in the bias images in each mode. For the spatial autocorrelation maps, we subtracted the per-pixel temporal mean and estimated the covariance at spatial separations of up to five pixels in each direction (see \autoref{fig:cross_corr_dark}).

To quantify the nearest neighbour correlations and their uncertainties, we also formed ten disjoint bias-frame pairs in each mode. Each pair was differenced and divided by $\sqrt{2}$, removing the stable bias pattern while retaining the single-frame noise scale. We calculated the normalised spatial autocovariance separately for each difference image and averaged the resulting horizontal, vertical and diagonal coefficients across pairs. The mean diagonal coefficient averages the two diagonal orientations.

Both LN and HWC modes exhibit weak but statistically significant positive nearest neighbour spatial autocorrelations, with the strongest correlation occurring in the vertical direction. The mean diagonal correlation also exceeds the horizontal nearest neighbour correlation in both modes. For the datasets analysed here, the horizontal, vertical and mean diagonal coefficients are lower in HWC mode than in LN mode (see \autoref{tab:specs_results}).

\subsection{Dark frame structure and Dark Current}
\label{sec:dark_noise_results}
The IMX571 sensor exhibits remarkably low dark current even at relatively high operating temperatures operating at 0\,$^{\circ}$C \citep{2023sndd.confE...3A, 2022arXiv220606693G}. The Toran 26B CMOS camera features a low dark current (LDC) mode, which further reduces the dark signal. In CMOS detectors, such suppression can be achieved by placing selected peripheral readout amplifiers in a low-power state, or deactivating the amplifiers during long integration times \citep{greffe2022characterization, Ioannis24}. This may involve reducing or switching selected supply voltages during the exposure, thereby decreasing power dissipation, local heating, and optical or infrared emission from the readout electronics \citep{1985SPIE..570....7J, wang20154m}. The Toran 26B CMOS camera's LDC mode is based on a SONY proprietary readout sequence that Andor has adopted and it is to the authors understanding that it is based on the above method. Additionally, this mode does not alter the readout speed as it happens in parallel to the sensor's integration time. The only limitation is that it cannot be applied at very short exposure times (few milliseconds scale at the start of integration) due to the clocking requirements of the LDC mode.

\begin{figure*}
\includegraphics[width=\textwidth]{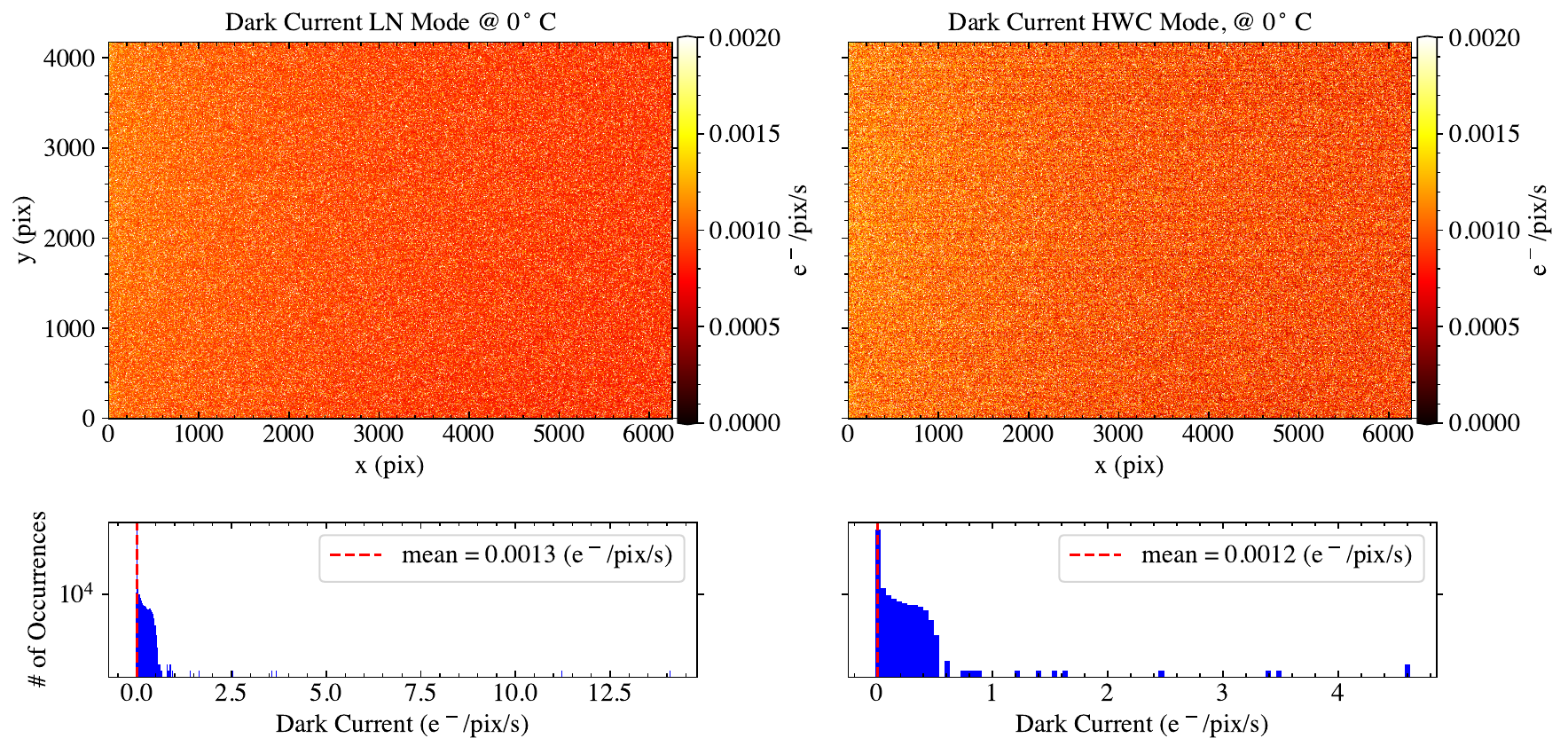}
\caption[Dark images for the Toran 26B]{Dark current maps taken for Toran 26B with LDC at temperature of 0\,$^\circ$\,C. \textbf{Left} is the LN mode. \textbf{Right} is the HWC mode. Below each we show the histograms of pixel value distributions from each frame, along with the mean (red dashed line).}
\label{fig:Dark_LN_HWC_LDC}
\end{figure*}

An example of a one-hour exposure after bias subtraction and normalisation by exposure time, is shown in \autoref{fig:Dark_LN_HWC_LDC}. The mean values are 0.001\,\eps for the LN mode and 0.001\,\eps for the HWC mode. For comparison, dark current frames acquired without the LDC mode are presented in \autoref{fig:Dark_LN_HWC} with mean values of 0.009\,\eps for both modes demonstrating the effectiveness of the LDC mode. In these images, hot pixels are clearly visible, although the mean signal level remains low. The extended tail observed in the pixel histograms likely originates from noisy pixels, hot pixels, or pixels exhibiting RTS behaviour. Additionally, the images reveal a brighter region on the left side of the sensor array, which may indicate either light leakage accumulated over the one-hour exposure or glow emission from the sensor.

\subsection{QE and Window Transmittance}
\label{sec:qe_results}
\begin{figure}\centering
\centering
\includegraphics[width=\columnwidth]{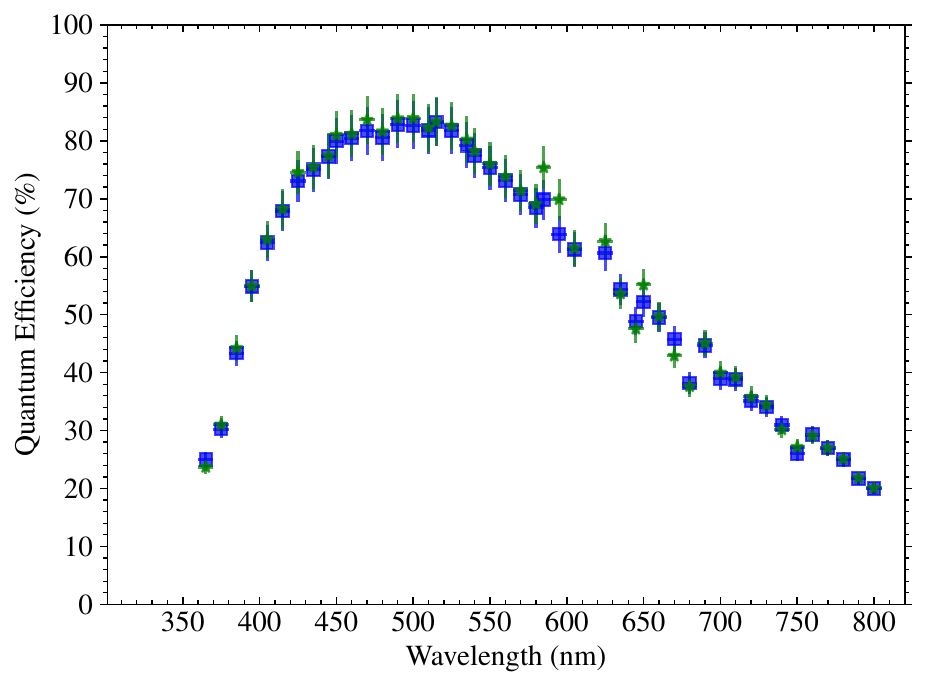}
\caption[QE plot for Toran 26B]{QE for the Toran 26B camera at constant temperature of 0\,$^\circ$C. With squares is the first measurement taken and with stars the second measurement taken.}
\label{fig:QE}
\end{figure}

QE measurements were performed on two separate days, both conducted at a stable temperature of 0$^\circ$C, as shown in \autoref{fig:QE}. The QE uncertainties were calculated by propagating in quadrature the uncertainties in the camera signal and the measured photodiode current. The photodiode QE was estimated from the manufacturer's datasheet (see also appendix in \citet{Ioannis24}). We adopted a relative uncertainty of 5\,\% in this estimate and included this contribution in the camera QE uncertainty by combining it in quadrature with the camera signal and photodiode current uncertainties \footnote{Hamamatsu informed the authors that the QE of individual S1337-1010BQ photodiodes may differ from the advertised QE curve by up to 15\,\%}. To account for the window transmittance, we measured the current of the photodiode with the window in front of the light source. The measured QE curve exhibits a sharp rise toward the blue end of the spectrum, reaching a prominent peak of 84\,\% at 500\,nm. This behaviour is characteristic of a thinned, back-illuminated sensor architecture. Beyond 550\,nm, the QE begins to decline steadily across the green and red portions of the spectrum.

Our results are consistent with previous studies conducted on Sony’s IMX455 and IMX411 sensors. For instance, \citet{2023sndd.confE...3A} and \citet{2023A&A...670A.119B} reported similar QE behaviour, with the IMX411 reaching a peak of 83\,\% at 490\,nm. In the near-infrared region (around 800\,nm), our data align with these findings, showing a QE of approximately 20\,\%. However, both \citet{Gill_2022} and QHYCCD report higher QE values for the IMX455 sensor, peaking at approximately 92\,\% near 490\,nm.

\begin{figure}\centering
\centering
\includegraphics[width=\columnwidth]{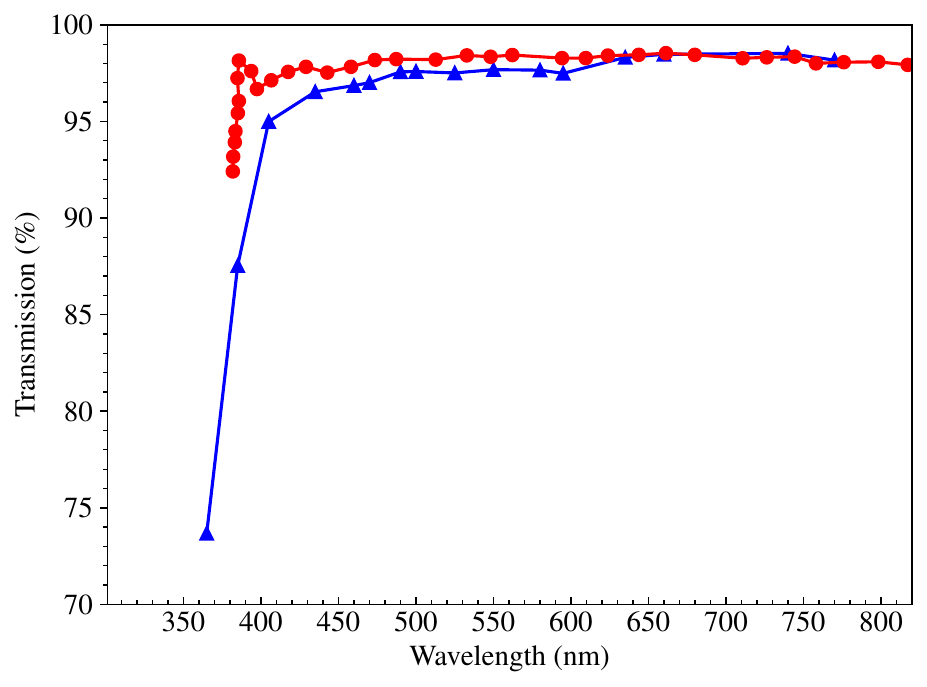}
\caption[Transmission plot for Toran 26B.]{Transmission as a function of wavelength for the window which is mounted on the Toran 26B camera measured as the fractional percentage ratio of the photodiode current measurements with and without the window. The measured data is shown with triangles for comparison is the transmission curve from the window manufacturer performance sheet with circles. The y-axis is zoomed-in for detailed visualisation.}
\label{fig:Transmission_toran}
\end{figure}

The fractional ratio of the photodiode current measured with and without the camera window is plotted in \autoref{fig:Transmission_toran}, illustrating the transmission curve of the camera’s front window. Our measurements generally align with the manufacturer’s model, though we observe slightly higher transmission at bluer wavelengths than reported. Due to the lack of data in the ultraviolet range, we cannot confirm whether a significant transmission drop occurs at shorter wavelengths. Nonetheless, the data suggest that the window is coated to suppress UV and shorter-wavelength light, likely to reduce unwanted signal contributions.

\subsection{Noise models from on-sky tests}
\label{sec:noise_model}
\begin{figure*}
\includegraphics[width=\textwidth]{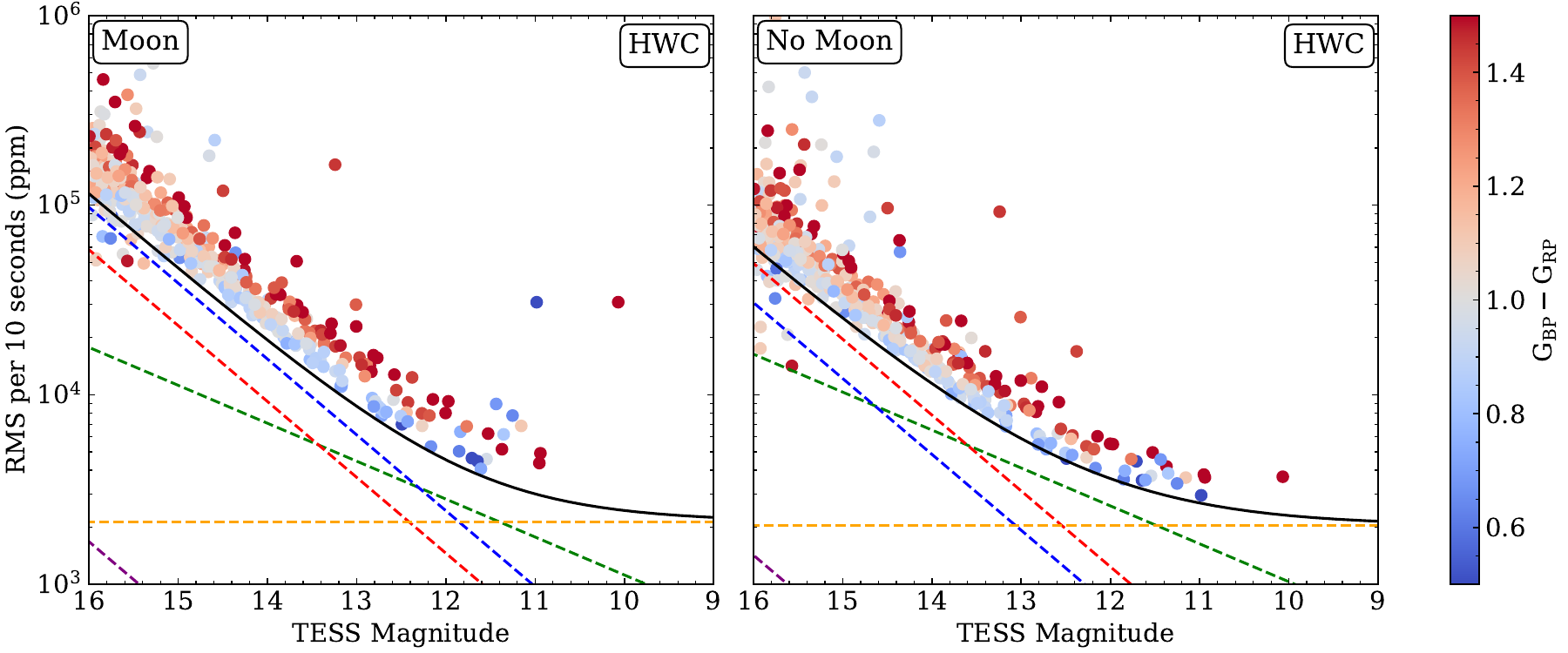}
\caption[Noise models for Toran 26B with HWC mode.]{Photometric precision as a function of stellar magnitude for 10-second exposures, comparing predictions from the theoretical noise model with observations obtained on a bright night (6 August 2025, left) with zero point of 22.25\,mag and a dark night (19 August 2025, right) with zero point of 22.42\,mag. The data was recorded with the HWC mode of the camera. The dark line indicates the total modelled noise, including contributions from scintillation (yellow), photon shot noise (green), dark current noise (purple), sky background noise (blue), and read noise (red). Measured noise values, derived from detrended relative photometry light curves of 457 stars with $9<\mathrm{T}<16$ in a field centered on TIC-236445129 (KELT-16), are plotted as coloured points, where the colour scale represents the \textit{Gaia} DR3 $\mathrm{G_{BP}}-\mathrm{G_{RP}}$ index.}
\label{fig:noise_dark_bright_HWC}
\end{figure*}

\begin{figure*}
\includegraphics[width=\textwidth]{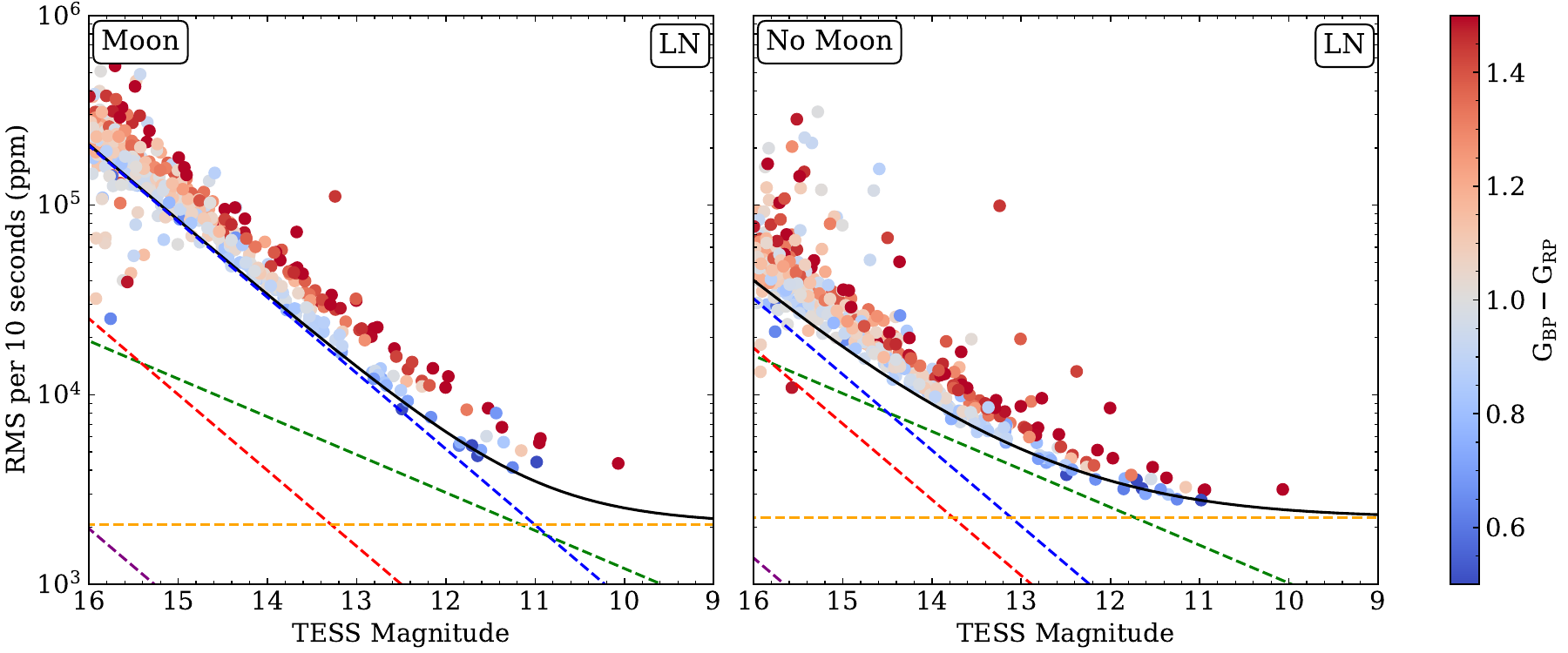}
\caption[Noise models for Toran 26B with LN mode.]{Photometric precision as a function of stellar magnitude for 10-second exposures, comparing predictions from the theoretical noise model with observations obtained on a bright night (8 August 2025, left) with zero point of 22.07\,mag and a dark night (21 September 2025, right)  with zero point of 22.46\,mag. The data was recorded with the LN mode of the camera. The dark line indicates the total modelled noise, including contributions from scintillation (yellow), photon shot noise (green), dark current noise (purple), sky background noise (blue), and read noise (red). Measured noise values, derived from detrended relative photometry light curves of 457 stars with $9<\mathrm{T}<16$ in a field centered on TIC-236445129 (KELT-16), are plotted as coloured points, where the colour scale represents the \textit{Gaia} DR3 $\mathrm{G_{BP}}-\mathrm{G_{RP}}$ index.}
\label{fig:noise_dark_bright_LN}
\end{figure*}

We perform on-sky measurements by comparing the observed stellar noise with the theoretical noise model, which accounts for instrumental contributions such as read noise and dark current, as well as sky background noise, target shot noise, and atmospheric scintillation. We used the read and dark current noise values derived in Sections~\ref{sec:temporal_spatial_noise_results} and~\ref{sec:dark_noise_results}. To model the sky background noise, we extracted the median value of the night’s sky level. Parameters specific to the TWIST telescope were incorporated, including the telescope diameter, elevation above sea level, atmospheric scale height, airmass and exposure used. The median empirical coefficient quantifying the scintillation amplitude for each night was adopted from \citep{2015MNRASOsborn} and was set to 1.30. These quantities were applied to a modified version of Young’s formula \citep{1967AJ.....72..747Y, 2015MNRASOsborn}. Finally, all noise components were scaled using a synthetic flux array representing shot noise. The extracted magnitudes were calibrated against TESS magnitudes using the measured zeropoint, and all noise terms were combined to produce the total noise model. Results were obtained for a field centered at KELT-16 star for four observing nights: two using the HWC mode under both bright (moon) and dark (no-moon) night conditions as shown in \autoref{fig:noise_dark_bright_HWC}, and two using the LN mode under bright and dark night conditions as shown in \autoref{fig:noise_dark_bright_LN}.

For the HWC and at the bright night case, the sky background is the dominant noise source for stars fainter than $\mathrm{T}>12.5$\,mag, while scintillation noise becomes the primary contributor for brighter stars. Under dark night conditions, the dominant noise source for faint stars ($\mathrm{T}>13.5$\,mag) is the read noise. In the intermediate magnitude range ($11.5<\mathrm{T}<13.5$\,mag), shot noise dominates, and scintillation noise prevails for the brightest stars. Consequently, under certain sky brightness levels and with 10 second exposures, the HWC mode in this telescope configuration can become camera-noise limited.

For the LN mode and at the bright night case, the sky background is the dominant noise source for stars fainter than $\mathrm{T}>10.5$\,mag. In dark night conditions, the sky background dominates for stars fainter than $\mathrm{T}>14.5$\,mag, while shot noise is the main contributor for stars in the range $11.5<\mathrm{T}<14.5$ mag. In both LN cases, scintillation noise is dominant for the brightest stars. Under these conditions and with 10 second exposures, the LN mode is never camera-noise limited in this telescope configuration.

\begin{figure}\centering
\includegraphics[width=\columnwidth]{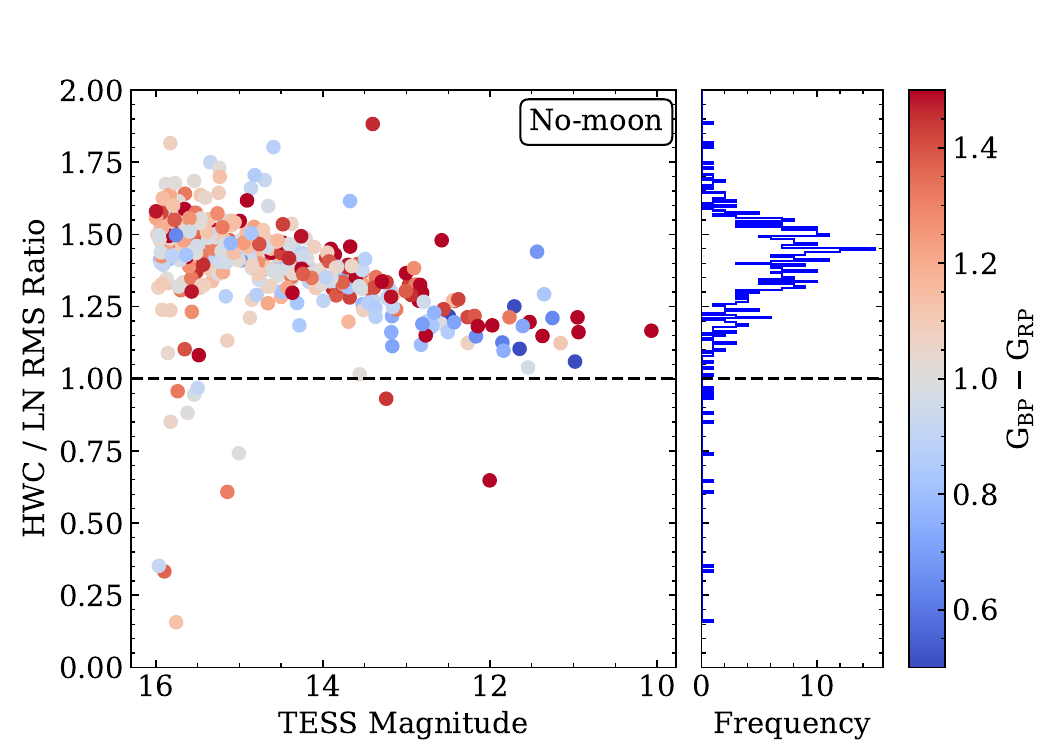}
\caption[Noise ratio for Toran 26B for HWC and LN mode.]{Ratio for HWC and LN camera mode of the noise contribution for different magnitude stars from \autoref{fig:noise_dark_bright_HWC} and \autoref{fig:noise_dark_bright_LN}. The data points are coloured according to the catalogued \textit{Gaia} DR3 $\mathrm{G_{BP}}-\mathrm{G_{RP}}$ colour. The black dashed line represents the ratio 1 which is the case where the noise is the same for both cameras. The noise ratio for a dark night (no-moon). The right panel shows the histogram distribution.}
\label{fig:RMS_ratio}
\end{figure}

We directly compare the noise models by calculating the ratio of the RMS values of the stars between the two modes for the dark night dataset, as shown in \autoref{fig:RMS_ratio}. Our results indicate that the HWC mode consistently produces higher noise compared to the LN mode. However, this comparison should not be considered definitive, as the datasets were acquired approximately one month apart. It is worth noting that, in the fainter regime ($13.5 < \mathrm{T} < 16$), the LN mode exhibits lower limited noise levels because it is dominated by sky background noise rather than read noise, which is prominent in the HWC mode. Under dark-night conditions, this conclusion remains valid, as the one-month separation between the observations has a negligible effect on this regime.

\subsection{Timescale of noise from on-sky tests}
\label{sec:timescale_noise}
\begin{figure*}
\includegraphics[width=\textwidth]{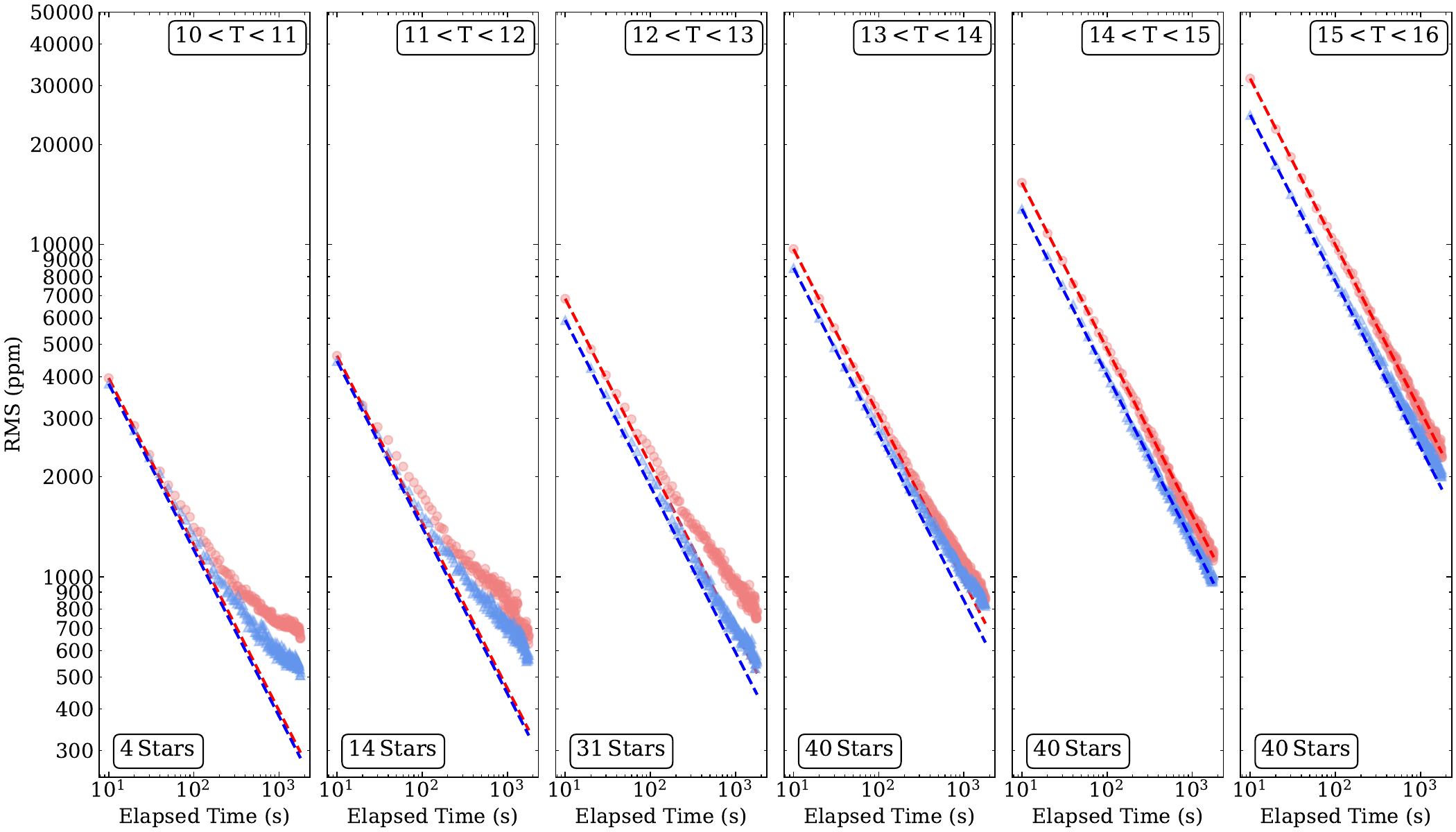}
\caption[Timescale of noise for Toran 26B]{Noise levels for Toran 26B camera with the HWC mode (light red circles) and LN mode (light blue triangles) at varying timescales during observations conducted on the dark night of 2025 August 19 and 2025 September 21 respectively. The data is binned as a function of exposure and readout times. The blue and red dashed lines indicate the white noise model, proportional to $1 / \sqrt{n}$ for the HWC and LN mode respectively. The RMS noise was measured from stars that are non-variable and non-saturated. The plots also show the magnitude regimes, and the numbers of stars used.}
\label{fig:rms_timescale}
\end{figure*}

To evaluate the presence of correlated noise in our light curves, we separate our sample stars into six magnitude regimes. Within each regime, we computed the average RMS of the residuals as a function of binning time, with time bins ranging up to 1800\,seconds. This analysis was performed for both the HWC and LN observing modes under dark night conditions, as illustrated in \autoref{fig:rms_timescale}.

Due to the limited number of bright stars in the range $10<\mathrm{T}<13$, the sample size in these regimes was smaller than 40 and not randomly selected, since all available stars were used. In these cases, we observed noticeable correlated noise, particularly starting at the 200\,seconds binning scale. A few stars in these bright regimes exhibited strong correlated noise contributions, which likely influenced the average noise timescale. Overall, the LN mode showed slightly lower correlated noise levels compared to the HWC mode. Specifically, the minimum RMS at 30-minute binning reached 500 ppm for LN and 700 ppm for HWC.

For the fainter regimes ($13<\mathrm{T}<16$), we randomly selected 40 stars. In these cases, the RMS as a function of binning time follows the expected white-noise behaviour. The binned RMS values closely follow the theoretical uncorrelated noise trend, with the LN mode again exhibiting slightly lower overall RMS than the HWC mode.

We note that the differences observed between the two modes should not be taken as a rule of thumb, since the observations were conducted approximately a month apart and the sky conditions may have differed between the two epochs.

\subsection{Transiting exoplanets from on-sky tests}
\label{sec:transiting_exoplanet}

A total of five successful nights of observations were carried out, targeting five well-known transiting exoplanet host stars, as summarised in \autoref{tab:targets}. All targets are short period gas giants, predominantly classified as hot Jupiters. Relative photometry, described in Section~\ref{sec:photometry-noise}, was applied to correct the light curves for systematic effects. As each target star differs in brightness, the photometric aperture was optimised individually to achieve the lowest RMS scatter in the resulting transit lightcurves. 

For the transit modelling a similar approach was used as in \citet{Ioannis26}. We used \texttt{allesfitter} package for modelling the transits \citep{allesfitter-code, allesfitter-paper}. To sample the model parameters, we applied additional detrending using a Gaussian Process with a Matérn 3/2 covariance kernel. This kernel is described by the logarithmic amplitude, $\ln(\sigma)$, and the logarithmic length scale, $\ln(\rho)$, allowing us to account for time-correlated systematics in the light curves. The transit models were computed with the \texttt{ellc} code \citep{maxted2016ellc}, which accurately represents the geometry and limb-darkening of transiting systems. Parameter estimation was performed using nested sampling \citep{Skilling2004nestedsampling, Skilling2006nestedsampling}, as implemented in \texttt{allesfitter} via the \texttt{dynesty} sampler \citep{speagle2020dynesty}. 

For the initial parameters of our five exoplanet systems, we adopted previously published solutions from the literature along with their associated uncertainties (see \autoref{tab:targets}). These parameters include the orbital period $P$, planet-to-star radius ratio ($R_\text{B} / R_\text{A}$), semi-major axis ($(R_\text{A} + R_\text{B}) / a$), epoch of mid-transit ($T_{0,\text{B}}$), quadratic limb-darkening coefficients, and orbital inclination ($\cos{i_B}$). Given that these exoplanets are short-period systems, we assumed circular orbits and fixed both the eccentricity and argument of periastron to zero.

\begin{figure}
\centering
\includegraphics[width=\columnwidth]{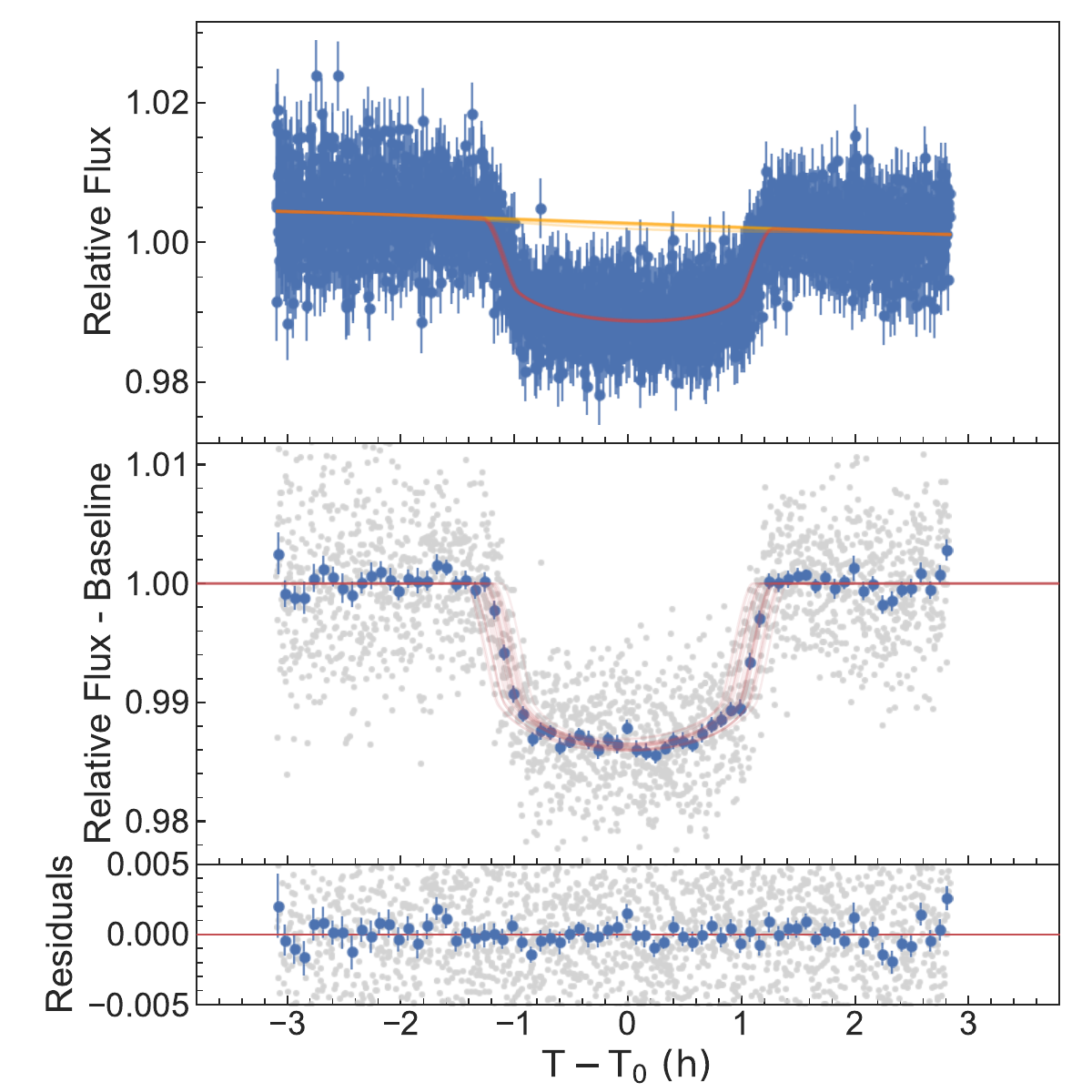}
\caption{Normalized lightcurves for the
transiting system of KELT-16 (TIC 236445129) \citep{KELT-16_2017} from TWIST on 2025 August 6.
\textbf{Top}: Undetrended light curves with the GP detrending function shown as yellow solid line
and transit model shown with red solid line. \textbf{Middle}: Detrended lightcurve with data binned
to 5 minutes (solid blue circles) for clarity. Again the best fit model is shown in red.
\textbf{Bottom}: Residuals from the best fit.}
\label{fig:kelt-16}
\end{figure}

An example of a complete transit light curve is shown in \autoref{fig:kelt-16}. The upper panel presents the unbinned lightcurve obtained from the relative photometry. Owing to the high quality and low level of systematics in the data, the Gaussian Process contribution was minimal, indicating that only limited detrending was required. The middle panel displays the light curve binned to five minute intervals, while the lower panel shows the residuals of the best fit model. The rest of our sample transiting light curves are provided in \autoref{sec:extra_material}. For each transiting exoplanet we provide the initial guesses that we used from existed literature to fit for the transit, as shown in \autoref{tab:planetary-KELT-16_0806}.

\begin{table*}
    \centering
    \caption{\texttt{allesfitter} model priors. Planetary system properties
    for the transiting system KELT-16 (TIC 236445129), as illustrated in
    \autoref{fig:kelt-16}, observed with NGTS on 2025 August 6.}

    {\renewcommand{\arraystretch}{1.25}
    \begin{tabular}{lccc}
        \hline
        \textbf{Parameter} & \textbf{Initial Guess} & \textbf{Prior}
        & \textbf{Fitted value Toran-26B} \\
        \hline
        $R_p/R_{\star}$        & \FITNArb     & $\mathcal{U}(0.1,0.11)$                 & \CMOSNArb \\
        $(R_\star+R_b)/a_b$    & \FITNArsuma  & $\mathcal{U}(0.34,0.35)$                & \CMOSNArsuma \\
        $\cos{i_b}$            & \FITNAinc    & $\mathcal{U}(0.097,0.098)$              & \CMOSNAinc \\
        $T_{\mathrm{c}}$ (BJD) & \FITNAtc     & $\mathcal{U}(2458392.586,2458392.589)$  & \CMOSNAtc \\
        Orbital period (days)  & \FITNAperiod & $\mathcal{U}(0.965,0.97)$               & \CMOSNAperiod \\
        $q_1$                  & \FITNAqone   & $\mathcal{N}(0.323,0.003)$              & \CMOSNAqone \\
        $q_2$                  & \FITNAqtwo   & $\mathcal{N}(0.376,0.047)$              & \CMOSNAqtwo \\
        \hline
    \end{tabular}}
    \label{tab:planetary-KELT-16_0806}
\end{table*}

\begin{figure}
\includegraphics[width=\columnwidth]{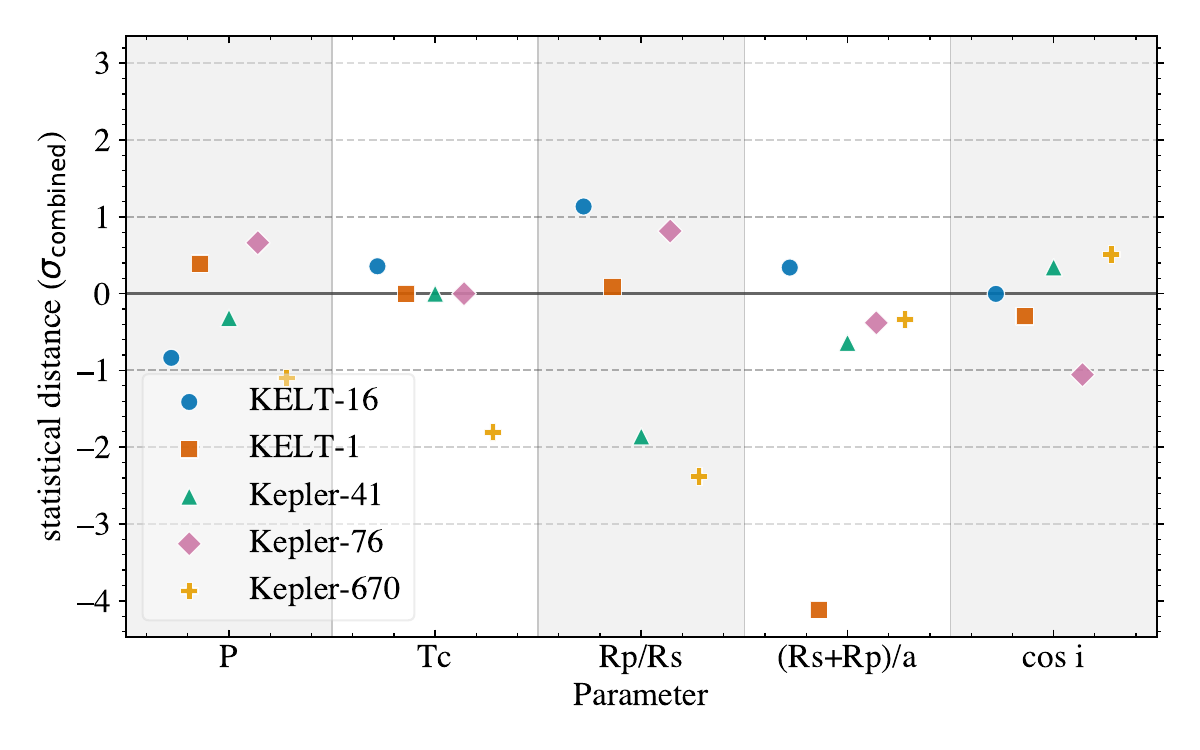}
\caption[Statistical difference of fitted transit parameters]{Statistical distance between the fitted posterior median values and the corresponding literature values for the transiting systems observed with TWIST. For each parameter, the plotted value is defined as $(\theta_{\mathrm{post}} - \theta_{\mathrm{lit}})/\sqrt{\sigma_{\mathrm{post}}^2 + \sigma_{\mathrm{lit}}^2}$. Vertical columns separate the fitted parameters, and horizontal dashed lines mark integer sigma offsets from the literature solution.}
\label{fig:statistical_difference}
\end{figure}

To quantify the consistency of the recovered transit parameters with the published literature values, we computed a statistical distance for each fitted parameter. For each system, the posterior median value returned by \texttt{allesfitter} was compared with the corresponding literature value used as the initial reference solution. The statistical distance was defined as:

\begin{equation}
    \Delta_\sigma =
    \frac{\theta_{\mathrm{post}} - \theta_{\mathrm{lit}}}
    {\sqrt{\sigma_{\mathrm{post}}^2 + \sigma_{\mathrm{lit}}^2}},
\end{equation}
  
where $\theta_{\mathrm{post}}$ is the posterior median of the fitted parameter, $\theta_{\mathrm{lit}}$ is the literature value, $\sigma_{\mathrm{post}}$ is the posterior uncertainty from the fit, and $\sigma_{\mathrm{lit}}$ is the published literature uncertainty. For asymmetric posterior uncertainties, we used the posterior uncertainty in the direction of the literature value. For example, if the literature value was larger than the posterior median, the upper posterior uncertainty was used. For the mid-transit time, the literature ephemeris and its uncertainty were propagated to the observed transit epoch before computing $\Delta_\sigma$, ensuring that the comparison was made at the same orbital cycle. This provides a direct measure of how far the fitted solution lies from the published solution in units of the combined uncertainty, as shown in \autoref{fig:statistical_difference}. 

Most of the parameters recovered from the TWIST observations agree with the literature values within $3\sigma$ combined standard deviations. The principal exception is $(R_\star+R_{\rm p})/a$, which differs by approximately $4\sigma$. This parameter is constrained by the transit duration and shape and is correlated with the impact parameter and orbital inclination. Its recovery from a single transit is consequently sensitive to the available out-of-transit baseline, detrending, time-correlated noise, and modelling assumptions. The discrepancy should therefore not be interpreted as a revision of the published system parameters, but as an indication of the limitations and potentially underestimated systematic uncertainties associated with fitting a single transit. Overall, the comparison provides a consistency check on the parameters recovered from the TWIST data, rather than an independent refinement of the system properties.

\section{Conclusions}
\label{sec:conclusion}
We evaluated the performance of the Toran 26B CMOS camera through both laboratory and on-sky tests. The camera provides two 16-bit operating modes, the LN and HWC, each optimised for different performance characteristics. The HWC produces large well capacity, while the LN produces low read noise. On these combination the dynamic range is similar. In both modes, the read noise exhibited a skewed distribution, indicating that certain pixels are noisier than average, a behaviour commonly observed in many CMOS sensors. The dark current was found to be low and could be considered negligible for short time-series photometry, with no evidence of glow spots in the dark frames. Temporal row and column noise components showed minimal variation. The camera demonstrated good linearity across most of its dynamic range, with only minor non-uniformities detected. The QE measurements of the IMX571 sensor were consistent with previously published results \citep{2023sndd.confE...3A}, though slightly lower than the peak sensitivity reported by some manufacturers. A summary of these results is provided in \autoref{tab:specs_results}.

\begin{table*}
\centering
\caption{Specifications of the New Andor CMOS camera measured in this work in for the LN and HWC modes.}
\renewcommand{\arraystretch}{1.25}
\label{tab:specs_results}
\begin{tabular}{lcc}
\hline
\hline
\textbf{Camera mode}               & \textbf{LN Mode}       & \textbf{HWC Mode}      \\ \hline
Digital output                     & 16-bit                         & 16-bit         \\
Readout speeds                     & 6.8 fps                         & 6.8 fps        \\
Quantum Efficiency Peak, \%        & 84$\pm$4.2 @500\,nm             & 84$\pm$4.2 @500\,nm      \\
Gain, e$^-$ADU$^{-1}$              & 0.255$\pm$0.003                 & 0.779$\pm$0.02          \\
Full Well Capacity, e$^-$          & 15700                           & 50000          \\
Non-Linearity, \%                  & 0.067                           & 0.090          \\
Read Noise median*, e$^-$           & $1.07_{-0.2}^{+0.3}$           & $2.79_{-0.55}^{+0.83}$     \\
Read Noise RMS*, e$^-$        &  $1.53_{-0.37}^{+0.13}$   & $3.61_{-0.83}^{+0.5}$    \\
Dark Current LDC OFF*, \eps         & $0.009_{-0.003}^{+0.002}$ @0$^{\circ}$C   & $0.009_{-0.004}^{+0.002}$ @0$^{\circ}$C   \\
Dark Current LDC ON*,\eps           & $0.001_{-0.0008}^{+0.0003}$ @0$^{\circ}$C   & $0.001_{-0.001}^{+0.0009}$ @0$^{\circ}$C   \\
PRNU, \%                           & 0.076                           & 0.050                \\
Row temporal Noise \,e$^-$         & 0.064                           & 0.173                \\
Column temporal Noise \,e$^-$      & 0.023                           & 0.056                \\
DSNU Row, \,e$^-$                  & 0.008                           & 0.021                \\
DSNU Column, \,e$^-$               & 0.002                           & 0.007                \\
DSNU Pixel, \,e$^-$                & 0.075                           & 0.172                \\
DSNU Total, \,e$^-$                & 0.076                           & 0.173                \\
Dynamic Range, \,dB                & 83.39                           & 85.05  \\
Autocorrelation, horizontal
                                  & $0.00594 \pm 0.00006$ & $0.00383 \pm 0.00006$ \\
Autocorrelation, vertical
                                  & $0.01612 \pm 0.00008$ & $0.01439 \pm 0.00007$ \\
Autocorrelation, mean diagonal
                                  & $0.00639 \pm 0.00004$ & $0.00480 \pm 0.00004$ \\ \hline    
\multicolumn{3}{l}{* Uncertainties are reported as the 16th-84th percentile range.} \\
% \multicolumn{3}{l}{\red{$\dagger$ their uncertainties are frame-jackknife standard errors.}} \\
\end{tabular}
\end{table*}

We successfully installed the Toran 26B CMOS camera on the TWIST telescope at the Roque de los Muchachos Observatory in La Palma, where we conducted photometric observations of well-known transiting exoplanets under a range of sky conditions. We estimated the expected photometric noise and compared it with theoretical predictions, finding an agreement between the two. The results indicate that the HWC mode is expected to be instrument-noise-limited for faint targets, whereas the LN mode does not reach the instrument-noise limit for this telescope configuration.

An analysis of noise as a function of timescale, performed through data binning, revealed that certain bright stars exhibit persistent correlated noise, while fainter stars are predominantly limited by uncorrelated noise. We successfully observed and modelled the transits of well-known exoplanets, confirming their detections and obtaining results consistent with those reported in the literature.

Overall, the Toran 26B CMOS camera demonstrates performance comparable to that of other Sony sensors based on similar architectures. It is capable of delivering high-precision photometry with minimal influence from instrumental systematics, depending on the selected operating mode. These findings support the growing evidence that modern CMOS detectors can achieve photometric performance on par with high-grade CCD cameras.

\section*{Acknowledgements}
This is a project based on the collaboration between the University of Warwick and Andor Technology Ltd under the STFC Industrial CASE (Cooperative Awards in Science and Technology) studentship. This work makes use of data from the TWIST instrument operated on the island of La Palma by the University of Warwick at the Observatorio del Roque de los Muchachos of the Instituto de Astrofísica de Canarias. IA acknowledges the support of the UK Science and Technology Facilities Council (STFC) under the CASE Industry scheme (ST/W005077/1 Project title: Precision Photometry with the new generation of fast readout Scientific CMOS Camera). MAM gratefully acknowledges the support of STFC through the training grant ST/X508871/1. We acknowledge capital funding from the STFC Early Technology Development Capital Funding (ST/W005719/1 Project title: Discovering New Worlds with sCMOS Cameras). While the project benefited from a collaborative funding arrangement between the University of Warwick and Andor Technology Ltd, all data analysis, interpretation, and conclusions presented in this work were carried out independently by the author. The QE experiment was conducted at Andor’s laboratories, using optical equipment provided by the company. The data presented are based on genuine experimental results, with no input from Andor in the selection, alteration, or promotion of specific outcomes. IA affirms that no results were omitted or adjusted to favour the company’s products.
%%%%%%%%%%%%%%%%%%%%%%%%%%%%%%%%%%%%%%%%%%%%%%%%%%
\section*{Conflict of Interest}
The authors declare no conflict of interest.

%%%%%%%%%%%%%%%%%%%%%%%%%%%%%%%%%%%%%%%%%%%%%%%%%%
\section*{Data Availability}
The data the software used is publicly available on the GitHub repository: \url{https://github.com/AperpieGG/Paper_III}. Data will be available upon request to the author.
 


\bibliographystyle{rasti}
\bibliography{marana} % if your bibtex file is called example.bib

@inproceedings{Secull_2026,
        author = {Tom Seccull and Ioannis Apergis and Leonidas Asimakoulas and Colin Coates and Darren Frew and Jouni Jussila},
        title = {{Andor Toran: a versatile sCMOS camera platform for astronomy and physical science}},
        volume = {14149},
        booktitle = {Ground-based and Airborne Instrumentation for Astronomy XI},
        editor = {Jo{\"e}l D. R. Vernet and Julia J. Bryant and Kentaro Motohara},
        organization = {International Society for Optics and Photonics},
        publisher = {SPIE},
        pages = {14149A3},
        year = {2026},
        doi = {10.1117/12.3102852},
        URL = {https://doi.org/10.1117/12.3102852}
}

@ARTICLE{Richmond_2020,
       author = {{Richmond}, Michael W. and {Tanaka}, Masaomi and {Morokuma}, Tomoki and {Sako}, Shigeyuki and {Ohsawa}, Ryou and {Arima}, Noriaki and {Tominaga}, Nozomu and {Doi}, Mamoru and {Aoki}, Tsutomu and {Arimatsu}, Ko and {Ichiki}, Makoto and {Ikeda}, Shiro and {Ita}, Yoshifusa and {Kasuga}, Toshihiro and {Kawabata}, Koji S. and {Kawakita}, Hideyo and {Kobayashi}, Naoto and {Kokubo}, Mitsuru and {Konishi}, Masahiro and {Maehara}, Hiroyuki and {Mito}, Hiroyuki and {Miyata}, Takashi and {Mori}, Yuki and {Morii}, Mikio and {Motohara}, Kentaro and {Nakada}, Yoshikazu and {Okumura}, Shin-Ichiro and {Onozato}, Hiroki and {Sarugaku}, Yuki and {Sato}, Mikiya and {Shigeyama}, Toshikazu and {Soyano}, Takao and {Takahashi}, Hidenori and {Tanikawa}, Ataru and {Tarusawa}, Ken'ichi and {Urakawa}, Seitaro and {Usui}, Fumihiko and {Watanabe}, Junichi and {Yamashita}, Takuya and {Yoshikawa}, Makoto},
        title = "{An optical search for transients lasting a few seconds}",
      journal = {\pasj},
         year = 2020,
        month = feb,
       volume = {72},
       number = {1},
          eid = {3},
        pages = {3},
          doi = {10.1093/pasj/psz120},
archivePrefix = {arXiv},
       eprint = {1910.11343},
 primaryClass = {astro-ph.IM},
       adsurl = {https://ui.adsabs.harvard.edu/abs/2020PASJ...72....3R}
}

@article{Hopgood_2023,
author = {Hopgood, Joshua},
title = {Electro-optical performance of the large visible adaptive optics sensor module},
journal = {Astronomische Nachrichten},
volume = {344},
number = {8-9},
pages = {e20230068},
doi = {https://doi.org/10.1002/asna.20230068},
url = {https://onlinelibrary.wiley.com/doi/abs/10.1002/asna.20230068},
eprint = {https://onlinelibrary.wiley.com/doi/pdf/10.1002/asna.20230068},
year = {2023}
}

@ARTICLE{Strakhov_2023,
       author = {{Strakhov}, I.~A. and {Safonov}, B.~S. and {Cheryasov}, D.~V.},
        title = "{Speckle Interferometry with CMOS Detector}",
      journal = {Astrophysical Bulletin},
         year = 2023,
        month = jun,
       volume = {78},
       number = {2},
        pages = {234-258},
          doi = {10.1134/S1990341323020104},
archivePrefix = {arXiv},
       eprint = {2305.00451},
 primaryClass = {astro-ph.IM},
       adsurl = {https://ui.adsabs.harvard.edu/abs/2023AstBu..78..234S}
}

@article{Roth_2026,
author = {Roth, Martin M. and Roje, Paško and Vješnica, Stella and Cikota, Stefan and Brown, Alex J. and Kretlow, Mike and Azzaro, Marco and Reinhart, Santiago and Aceituno, Jesús and Dawson, Harry and Denker, Carsten and Enßlin, Torsten and Geier, Stephan and Granzer, Thomas and Kupfer, Thomas and Causi, Gianluca Li and Nissanke, Samaya and Ortiz, José Luis and Pedichini, Fernando and Poppenhäger, Katja and Schwope, Axel},
title = {The ORCA-TWIN qCMOS Experiment I. Science Case and Commissioning at Calar Alto Observatory},
journal = {Astronomische Nachrichten},
volume = {347},
number = {7},
pages = {e70110},
doi = {https://doi.org/10.1002/asna.70110},
url = {https://onlinelibrary.wiley.com/doi/abs/10.1002/asna.70110},
eprint = {https://onlinelibrary.wiley.com/doi/pdf/10.1002/asna.70110},
year = {2026}
}

@ARTICLE{Layden_2026,
       author = {{Layden}, Christopher and {Burdge}, Kevin and {Furesz}, Gabor and {Garc{\'\i}a-Mej{\'\i}a}, Juliana and {Dinsmore}, Jack and {Mo}, Geoffrey and {Osip}, David and {Piotrowski}, John J. and {Romani}, Roger W. and {Berne}, August and {Chakrabarty}, Deepto and {Chickles}, Emma},
        title = "{proto-Lightspeed: A High-speed, Ultra-low Read Noise Imager on the Magellan Clay Telescope}",
      journal = {\pasp},
         year = 2026,
        month = jun,
       volume = {138},
       number = {6},
          eid = {065001},
        pages = {065001},
          doi = {10.1088/1538-3873/ae6dfc},
archivePrefix = {arXiv},
       eprint = {2601.16268},
 primaryClass = {astro-ph.IM},
       adsurl = {https://ui.adsabs.harvard.edu/abs/2026PASP..138f5001L}
}

@INPROCEEDINGS{Wang_2024,
       author = {{Wang}, Shiang-Yu and {Wang}, Bo-Jhou and {Huang}, Chung-Kai and {Ling}, Hung-Hsu and {Chang}, Yin-Chang and {Hsu}, Shu-Fu and {Chen}, Hsin-Yo and {Zhang}, Zhi-Wei and {Lehner}, Matthew and {Geary}, John C. and {Amato}, Stephen M. and {Szentgyorgyi}, Andrew and {Norton}, Timothy and {Alcock}, Charles and {Castro-Chac{\'o}n}, Joel and {Reyes-Ru{\'\i}z}, Mauricio},
        title = "{The performance of the Mosaic CMOS Wide Field Camera for Transneptunian Automatic Occultation Survey (TAOS II)}",
    booktitle = {Ground-based and Airborne Instrumentation for Astronomy X},
         year = 2024,
       editor = {{Bryant}, Julia J. and {Motohara}, Kentaro and {Vernet}, Jo{\"e}l. R.~D.},
       series = {Society of Photo-Optical Instrumentation Engineers (SPIE) Conference Series},
       volume = {13096},
        month = jul,
          eid = {130962Y},
        pages = {130962Y},
          doi = {10.1117/12.3018461},
       adsurl = {https://ui.adsabs.harvard.edu/abs/2024SPIE13096E..2YW}
}

@INPROCEEDINGS{Sako_2018,
       author = {{Sako}, Shigeyuki and {Ohsawa}, Ryou and {Takahashi}, Hidenori and {Kojima}, Yuto and {Doi}, Mamoru and {Kobayashi}, Naoto and {Aoki}, Tsutomu and {Arima}, Noriaki and {Arimatsu}, Ko and {Ichiki}, Makoto and {Ikeda}, Shiro and {Inooka}, Kota and {Ita}, Yoshifusa and {Kasuga}, Toshihiro and {Kokubo}, Mitsuru and {Konishi}, Masahiro and {Maehara}, Hiroyuki and {Matsunaga}, Noriyuki and {Mitsuda}, Kazuma and {Miyata}, Takashi and {Mori}, Yuki and {Morii}, Mikio and {Morokuma}, Tomoki and {Motohara}, Kentaro and {Nakada}, Yoshikazu and {Okumura}, Shin-Ichiro and {Sarugaku}, Yuki and {Sato}, Mikiya and {Shigeyama}, Toshikazu and {Soyano}, Takao and {Tanaka}, Masaomi and {Tarusawa}, Ken'ichi and {Tominaga}, Nozomu and {Totani}, Tomonori and {Urakawa}, Seitaro and {Usui}, Fumihiko and {Watanabe}, Junichi and {Yamashita}, Takuya and {Yoshikawa}, Makoto},
        title = "{The Tomo-e Gozen wide field CMOS camera for the Kiso Schmidt telescope}",
    booktitle = {Ground-based and Airborne Instrumentation for Astronomy VII},
         year = 2018,
       editor = {{Evans}, Christopher J. and {Simard}, Luc and {Takami}, Hideki},
       series = {Society of Photo-Optical Instrumentation Engineers (SPIE) Conference Series},
       volume = {10702},
        month = jul,
          eid = {107020J},
        pages = {107020J},
          doi = {10.1117/12.2310049},
       adsurl = {https://ui.adsabs.harvard.edu/abs/2018SPIE10702E..0JS}
}

@book{stefanov2022cmos,
  title={CMOS image sensors},
  author={Stefanov, Konstantin D},
  year={2022},
  publisher={IOP Publishing}
}

@article{Aberle1992ImpactOI,
  title={Impact of illumination level and oxide parameters on Shockley–Read–Hall recombination at the Si‐SiO2 interface},
  author={Armin Gerhard Aberle and Stefan W. Glunz and Wilhelm Warta},
  journal={Journal of Applied Physics},
  year={1992},
  volume={71},
  pages={4422-4431}
}

@article{PhysRev.87.835,
  title = {Statistics of the Recombinations of Holes and Electrons},
  author = {Shockley, W. and Read, W. T.},
  journal = {Phys. Rev.},
  volume = {87},
  issue = {5},
  pages = {835--842},
  numpages = {0},
  year = {1952},
  month = {Sep},
  publisher = {American Physical Society},
  doi = {10.1103/PhysRev.87.835},
  url = {https://link.aps.org/doi/10.1103/PhysRev.87.835}
}

@article{Layden_2025,
    author = {Christopher Layden and Jill Juneau and Gustav Pettersson and Nathan Lourie and Benjamin Schneider and Beverly LaMarr and Francesco Elio Angile and Fadi Farag and Michelle Luo and Zhi Zheng Ong and G{\'a}bor Fur{\'e}sz},
    title = {{Characterization of the Teledyne COSMOS camera: a large format CMOS image sensor for astronomy}},
    volume = {11},
    journal = {Journal of Astronomical Telescopes, Instruments, and Systems},
    number = {2},
    publisher = {SPIE},
    pages = {026003},
    year = {2025},
    doi = {10.1117/1.JATIS.11.2.026003},
    URL = {https://doi.org/10.1117/1.JATIS.11.2.026003}
}

@article{jahne2010emva,
  title={Emva 1288 standard for machine vision: Objective specification of vital camera data},
  author={J{\"a}hne, Bernd},
  journal={Optik \& Photonik},
  volume={5},
  number={1},
  pages={53--54},
  year={2010},
  publisher={Wiley Online Library}
}

@ARTICLE{2019AN....340..638K,
       author = {{Karpov}, Sergey and {Bajat}, Armelle and {Christov}, Asen and {Prouza}, Michael},
        title = "{Evaluation of scientific complementary metal{\textendash}oxide{\textendash}semiconductor sensors for sky survey applications}",
      journal = {Astronomische Nachrichten},
         year = 2019,
        month = aug,
       volume = {340},
       number = {7},
        pages = {638-645},
          doi = {10.1002/asna.201913669},
archivePrefix = {arXiv},
       eprint = {1909.00729},
 primaryClass = {astro-ph.IM},
       adsurl = {https://ui.adsabs.harvard.edu/abs/2019AN....340..638K}
}

@inproceedings{wang20154m,
  title={A 4M, 1.4 e-noise, 96dB dynamic range, back-side illuminated CMOS image sensor},
  author={Wang, Xinyang and Ma, Cheng and Liu, Yang and Li, Yang and Zhou, Quan},
  booktitle={Proceedings of the International Image Sensor Workshop},
  year={2015}
}

@INPROCEEDINGS{2023sndd.confE...3A,
       author = {{Alarcon}, Miguel R. and {Licandro}, Javier and {Serra-ricart}, Miquel and {Joven}, Enrique and {Gaitan}, Vicens},
        title = "{Scientific CMOS sensors in Astronomy: QHY600 and QHY411}",
    booktitle = {2nd NEO and Debris Detection Conference},
         year = 2023,
        month = jan,
          eid = {3},
        pages = {3},
       adsurl = {https://ui.adsabs.harvard.edu/abs/2023sndd.confE...3A}
}

@ARTICLE{2021RAA....21..268Q,
       author = {{Qiu}, Peng and {Zhao}, Yong and {Zheng}, Jie and {Wang}, Jian-Feng and {Jiang}, Xiao-Jun},
        title = "{Research on performances of back-illuminated scientific CMOS for astronomical observations}",
      journal = {Research in Astronomy and Astrophysics},
         year = 2021,
        month = nov,
       volume = {21},
       number = {10},
          eid = {268},
        pages = {268},
          doi = {10.1088/1674-4527/21/10/268},
       adsurl = {https://ui.adsabs.harvard.edu/abs/2021RAA....21..268Q}
}

@ARTICLE{338418,
  author={Ming-Horn Tsai and Ma, T.P. and Hook, T.B.},
  journal={IEEE Electron Device Letters}, 
  title={Channel length dependence of random telegraph signal in sub-micron MOSFET's}, 
  year={1994},
  volume={15},
  number={12},
  pages={504-506},
  doi={10.1109/55.338418}}

@INPROCEEDINGS{2007SPIE.6690E..03J,
       author = {{Janesick}, James and {Andrews}, James and {Tower}, John and {Grygon}, Mark and {Elliott}, Tom and {Cheng}, John and {Lesser}, Michael and {Pinter}, Jeff},
        title = "{Fundamental performance differences between CMOS and CCD imagers: Part II}",
    booktitle = {Focal Plane Arrays for Space Telescopes III},
         year = 2007,
       editor = {{Grycewicz}, Thomas J. and {Marshall}, Cheryl J. and {Warren}, Penny G.},
       series = {Society of Photo-Optical Instrumentation Engineers (SPIE) Conference Series},
       volume = {6690},
        month = sep,
          eid = {669003},
        pages = {669003},
          doi = {10.1117/12.740218},
       adsurl = {https://ui.adsabs.harvard.edu/abs/2007SPIE.6690E..03J}
}

@INPROCEEDINGS{1985SPIE..570....7J,
       author = {{Janesick}, J. and {Klaasen}, K. and {Elliott}, T.},
        title = "{CCD charge collection efficiency and the photon transfer technique}",
    booktitle = {Solid state imaging arrays},
         year = 1985,
       series = {Society of Photo-Optical Instrumentation Engineers (SPIE) Conference Series},
       volume = {570},
        month = jan,
        pages = {7-19},
          doi = {10.1117/12.950297},
       adsurl = {https://ui.adsabs.harvard.edu/abs/1985SPIE..570....7J}
}

@INPROCEEDINGS{2020SPIE11525E..2IW,
       author = {{Wang}, Shushu and {Ping}, Yiding and {Men}, Jinrui and {Zhang}, Chen and {Zhao}, Changyin},
        title = "{The test of the 4K sCMOS camera for astronomical application}",
    booktitle = {SPIE Future Sensing Technologies},
         year = 2020,
       editor = {{Kimata}, Masafumi and {Shaw}, Joseph A. and {Valenta}, Christopher R.},
       series = {Society of Photo-Optical Instrumentation Engineers (SPIE) Conference Series},
       volume = {11525},
        month = nov,
          eid = {115252I},
        pages = {115252I},
          doi = {10.1117/12.2583913},
       adsurl = {https://ui.adsabs.harvard.edu/abs/2020SPIE11525E..2IW}
}

@INPROCEEDINGS{4154408,
  author={Wang, Xinyang and Rao, Padmakumar R. and Mierop, Adri and Theuwissen, Albert J.P.},
  booktitle={2006 International Electron Devices Meeting}, 
  title={Random Telegraph Signal in CMOS Image Sensor Pixels}, 
  year={2006},
  volume={},
  number={},
  pages={1-4},
  doi={10.1109/IEDM.2006.346973}}

@article{fowler2011single,
  title={Single photon CMOS imaging through noise minimization},
  author={Fowler, Boyd},
  journal={Single-Photon Imaging},
  pages={159--195},
  year={2011},
  publisher={Springer}
}

@article{ardilanov2021high,
  title={High-speed camera based on a large-format low-noise CMOS imager},
  author={Ardilanov, VI and Murzin, VA and Afanasyeva, IV and Ivaschenko, NG and Pritychenko, MA and Dodonov, SN},
  journal={Astronomy at the Epoch of Multimessenger Studies},
  pages={52},
  year={2021}
}

@article{townsend2023electro,
  title={Electro-optical characterization of a CMOS image sensor optimized for soft x-ray astronomy},
  author={Townsend-Rose, Charles and Buggey, Thomas and Ivory, James and Stefanov, Konstantin D and Jones, Lawrence and Hetherington, Oliver and Holland, Andrew D and Prod’homme, Thibaut},
  journal={Journal of Astronomical Telescopes, Instruments, and Systems},
  volume={9},
  number={4},
  pages={046001--046001},
  year={2023},
  publisher={Society of Photo-Optical Instrumentation Engineers}
}

@inproceedings{greffe2022characterization,
  title={Characterization of low light performance of a CMOS Sensor for ultraviolet astronomical applications},
  author={Greffe, Timothee and Smith, Roger and Sherman, Myles and Harrison, Fiona and Earnshaw, Hannah and Grefenstette, Brian and Hennessy, John and Nikzad, Shouleh},
  booktitle={X-Ray, Optical, and Infrared Detectors for Astronomy X},
  volume={12191},
  pages={72--112},
  year={2022},
  organization={SPIE}
}

@inproceedings{gimeno2016sky,
  title={On-sky commissioning of Hamamatsu CCDs in GMOS-S},
  author={Gimeno, German and Roth, Katherine and Chiboucas, Kristin and Hibon, Pascale and Boucher, Luc and White, John and Rippa, Matthew and Labrie, Kathleen and Turner, James and Hanna, Kevin and others},
  booktitle={Ground-based and Airborne Instrumentation for Astronomy VI},
  volume={9908},
  pages={872--885},
  year={2016},
  organization={SPIE}
}

@ARTICLE{2010Sci...327..977B,
       author = {{Borucki}, William J. and {Koch}, David and {Basri}, Gibor and {Batalha}, Natalie and {Brown}, Timothy and {Caldwell}, Douglas and {Caldwell}, John and {Christensen-Dalsgaard}, J{\o}rgen and {Cochran}, William D. and {DeVore}, Edna and {Dunham}, Edward W. and {Dupree}, Andrea K. and {Gautier}, Thomas N. and {Geary}, John C. and {Gilliland}, Ronald and {Gould}, Alan and {Howell}, Steve B. and {Jenkins}, Jon M. and {Kondo}, Yoji and {Latham}, David W. and {Marcy}, Geoffrey W. and {Meibom}, S{\o}ren and {Kjeldsen}, Hans and {Lissauer}, Jack J. and {Monet}, David G. and {Morrison}, David and {Sasselov}, Dimitar and {Tarter}, Jill and {Boss}, Alan and {Brownlee}, Don and {Owen}, Toby and {Buzasi}, Derek and {Charbonneau}, David and {Doyle}, Laurance and {Fortney}, Jonathan and {Ford}, Eric B. and {Holman}, Matthew J. and {Seager}, Sara and {Steffen}, Jason H. and {Welsh}, William F. and {Rowe}, Jason and {Anderson}, Howard and {Buchhave}, Lars and {Ciardi}, David and {Walkowicz}, Lucianne and {Sherry}, William and {Horch}, Elliott and {Isaacson}, Howard and {Everett}, Mark E. and {Fischer}, Debra and {Torres}, Guillermo and {Johnson}, John Asher and {Endl}, Michael and {MacQueen}, Phillip and {Bryson}, Stephen T. and {Dotson}, Jessie and {Haas}, Michael and {Kolodziejczak}, Jeffrey and {Van Cleve}, Jeffrey and {Chandrasekaran}, Hema and {Twicken}, Joseph D. and {Quintana}, Elisa V. and {Clarke}, Bruce D. and {Allen}, Christopher and {Li}, Jie and {Wu}, Haley and {Tenenbaum}, Peter and {Verner}, Ekaterina and {Bruhweiler}, Frederick and {Barnes}, Jason and {Prsa}, Andrej},
        title = "{Kepler Planet-Detection Mission: Introduction and First Results}",
      journal = {Science},
         year = 2010,
        month = feb,
       volume = {327},
       number = {5968},
        pages = {977},
          doi = {10.1126/science.1185402},
       adsurl = {https://ui.adsabs.harvard.edu/abs/2010Sci...327..977B}
}

@inproceedings{jorden2014e2v,
  title={e2v new CCD and CMOS technology developments for astronomical sensors},
  author={Jorden, Paul R and Jordan, Doug and Jerram, Paul A and Pratlong, J{\'e}r{\^o}me and Swindells, Ian},
  booktitle={High energy, optical, and infrared detectors for astronomy VI},
  volume={9154},
  pages={201--215},
  year={2014},
  organization={SPIE}
}

@BOOK{2000hccd.book.....H,
       author = {{Howell}, Steve B.},
        title = "{Handbook of CCD Astronomy}",
         year = 2000,
    publisher =  "{Cambridge University Press}",
       adsurl = {https://ui.adsabs.harvard.edu/abs/2000hccd.book.....H}
}

@inproceedings{janesick2006fundamental,
  title={Fundamental performance differences between CMOS and CCD imagers: Part 1},
  author={Janesick, James and Andrews, James T and Elliott, Tom},
  booktitle={High Energy, Optical, and Infrared Detectors for Astronomy II},
  volume={6276},
  pages={208--226},
  year={2006},
  organization={SPIE}
}

@ARTICLE{1985ITED...32.1525T,
       author = {{Tsoi}, H. -Y. and {Ellul}, J.~P. and {King}, M.~I. and {White}, J.~J. and {Bradley}, W.~C.},
        title = "{A deep-depletion CCD imager for soft X-ray, visible, and near-infrared sensing}",
      journal = {IEEE Transactions on Electron Devices},
         year = 1985,
        month = aug,
       volume = {32},
        pages = {1525-1530},
       adsurl = {https://ui.adsabs.harvard.edu/abs/1985ITED...32.1525T}
}

@INPROCEEDINGS{1480428,
  author={Peckerar, M.C. and McCann, D. and Blaha, F. and Mend, W. and Fulton, R.},
  booktitle={1979 International Electron Devices Meeting}, 
  title={Deep depletion charge-coupled devices for X-ray and IR sensing applications}, 
  year={1979},
  volume={},
  number={},
  pages={144-146},
  doi={10.1109/IEDM.1979.189563}}

@article{peckerar1981x,
  title={X-ray imaging with a charge-coupled device fabricated on a high-resistivity silicon substrate},
  author={Peckerar, MC and McCann, DH and Yu, Leepo},
  journal={Applied Physics Letters},
  volume={39},
  number={1},
  pages={55--57},
  year={1981},
  publisher={American Institute of Physics}
}

@ARTICLE{2025arXiv250312449X,
       author = {{Xiao}, Kai and {Huang}, Yang and {Yuan}, Haibo and {Li}, Zhirui and {Sun}, Yongkang and {Beers}, Timothy C. and {He}, Min and {Liu}, Jifeng and {Wu}, Hong and {Mao}, Yongna and {Huang}, Bowen and {Ma}, Mingyang and {Zheng}, Chuanjie and {Gu}, Hongrui and {Wang}, Beichuan and {Yang}, Lin and {Xu}, Shuai},
        title = "{Calibration of Complementary Metal-oxide-semiconductor Sensor-based Photometry to a Few-millimagnitude Precision: The Case of the Mini-SiTian Array}",
      journal = {arXiv e-prints},
         year = 2025,
        month = mar,
          eid = {arXiv:2503.12449},
        pages = {arXiv:2503.12449},
          doi = {10.48550/arXiv.2503.12449},
archivePrefix = {arXiv},
       eprint = {2503.12449},
 primaryClass = {astro-ph.IM},
       adsurl = {https://ui.adsabs.harvard.edu/abs/2025arXiv250312449X}
}

@article{Shao_2024,
doi = {10.1088/1674-4527/ad1793},
url = {https://dx.doi.org/10.1088/1674-4527/ad1793},
year = {2024},
month = {feb},
publisher = {National Astromonical Observatories, CAS and IOP Publishing},
volume = {24},
number = {2},
pages = {025021},
author = {Shao, Li and Zhan, Hu and Liu, Chao and Chi, Haonan and Luo, Qiuyan and Mu, Huaipu and Shi, Wenzhong},
title = {The Impact of Bias Row Noise to Photometric Accuracy: Case Study Based on a Scientific CMOS Detector},
journal = {Research in Astronomy and Astrophysics}
}

@INPROCEEDINGS{2014SPIE.9154E..2IW,
       author = {{Wang}, Shiang-Yu and {Ling}, Hung-Hsu and {Hu}, Yen-Shan and {Geary}, John C. and {Amato}, Stephen M. and {Pratlong}, Jerome and {Pike}, Andrew and {Jordan}, Paul and {Lehner}, Matthew J.},
        title = "{Characteristic of e2v CMOS sensors for astronomical applications}",
    booktitle = {High Energy, Optical, and Infrared Detectors for Astronomy VI},
         year = 2014,
       editor = {{Holland}, Andrew D. and {Beletic}, James},
       series = {Society of Photo-Optical Instrumentation Engineers (SPIE) Conference Series},
       volume = {9154},
        month = jul,
          eid = {91542I},
        pages = {91542I},
          doi = {10.1117/12.2057361},
       adsurl = {https://ui.adsabs.harvard.edu/abs/2014SPIE.9154E..2IW}
}

@article{geurts201798db,
  title={A 98dB Linear Dynamic Range, High Speed CMOS Image Sensor},
  author={Geurts, Tomas and Cremers, Bart and Innocent, Manuel and Vroom, Wiet and Esquenet, Cedric and Cools, Thomas and Compiet, John and Okcan, Burak and Chapinal, Genis and Luypaert, Carl and others},
  journal={Proceedings of the International Image Sensor Worshop, Hiroshima, Japan},
  volume={30},
  pages={43},
  year={2017}
}

@ARTICLE{2022arXiv220606693G,
       author = {{Ge}, Jian and {Zhang}, Hui and {Zang}, Weicheng and {Deng}, Hongping and {Mao}, Shude and {Xie}, Ji-Wei and {Liu}, Hui-Gen and {Zhou}, Ji-Lin and {Willis}, Kevin and {Huang}, Chelsea and {Howell}, Steve B. and {Feng}, Fabo and {Zhu}, Jiapeng and {Yao}, Xinyu and {Liu}, Beibei and {Aizawa}, Masataka and {Zhu}, Wei and {Li}, Ya-Ping and {Ma}, Bo and {Ye}, Quanzhi and {Yu}, Jie and {Xiang}, Maosheng and {Yu}, Cong and {Liu}, Shangfei and {Yang}, Ming and {Wang}, Mu-Tian and {Shi}, Xian and {Fang}, Tong and {Zong}, Weikai and {Liu}, Jinzhong and {Zhang}, Yu and {Zhang}, Liyun and {El-Badry}, Kareem and {Shen}, Rongfeng and {Tam}, Pak-Hin Thomas and {Hu}, Zhecheng and {Yang}, Yanlv and {Zou}, Yuan-Chuan and {Wu}, Jia-Li and {Lei}, Wei-Hua and {Wei}, Jun-Jie and {Wu}, Xue-Feng and {Sun}, Tian-Rui and {Wang}, Fa-Yin and {Zhang}, Bin-Bin and {Xu}, Dong and {Yang}, Yuan-Pei and {Li}, Wen-Xiong and {Xiang}, Dan-Feng and {Wang}, Xiaofeng and {Wang}, Tinggui and {Zhang}, Bing and {Jia}, Peng and {Yuan}, Haibo and {Zhang}, Jinghua and {Xuesong Wang}, Sharon and {Gan}, Tianjun and {Wang}, Wei and {Zhao}, Yinan and {Liu}, Yujuan and {Wei}, Chuanxin and {Kang}, Yanwu and {Yang}, Baoyu and {Qi}, Chao and {Liu}, Xiaohua and {Zhang}, Quan and {Zhu}, Yuji and {Zhou}, Dan and {Zhang}, Congcong and {Yu}, Yong and {Zhang}, Yongshuai and {Li}, Yan and {Tang}, Zhenghong and {Wang}, Chaoyan and {Wang}, Fengtao and {Li}, Wei and {Cheng}, Pengfei and {Shen}, Chao and {Li}, Baopeng and {Pan}, Yue and {Yang}, Sen and {Gao}, Wei and {Song}, Zongxi and {Wang}, Jian and {Zhang}, Hongfei and {Chen}, Cheng and {Wang}, Hui and {Zhang}, Jun and {Wang}, Zhiyue and {Zeng}, Feng and {Zheng}, Zhenhao and {Zhu}, Jie and {Guo}, Yingfan and {Zhang}, Yihao and {Li}, Yudong and {Wen}, Lin and {Feng}, Jie and {Chen}, Wen and {Chen}, Kun and {Han}, Xingbo and {Yang}, Yingquan and {Wang}, Haoyu and {Duan}, Xuliang and {Huang}, Jiangjiang and {Liang}, Hong and {Bi}, Shaolan and {Gai}, Ning and {Ge}, Zhishuai and {Guo}, Zhao and {Huang}, Yang and {Li}, Gang and {Li}, Haining and {Li}, Tanda and {Yuxi} and {Lu} and {Rix}, Hans-Walter and {Shi}, Jianrong and {Song}, Fen and {Tang}, Yanke and {Ting}, Yuan-Sen and {Wu}, Tao and {Wu}, Yaqian and {Yang}, Taozhi and {Yin}, Qing-Zhu and {Gould}, Andrew and {Lee}, Chung-Uk and {Dong}, Subo and {Yee}, Jennifer C. and {Shvartzvald}, Yossi and {Yang}, Hongjing and {Kuang}, Renkun and {Zhang}, Jiyuan and {Liao}, Shilong and {Qi}, Zhaoxiang and {Yang}, Jun and {Zhang}, Ruisheng and {Jiang}, Chen and {Ou}, Jian-Wen and {Li}, Yaguang and {Beck}, Paul and {Bedding}, Timothy R. and {Campante}, Tiago L. and {Chaplin}, William J. and {Christensen-Dalsgaard}, J{\o}rgen and {Garc{\'\i}a}, Rafael A. and {Gaulme}, Patrick and {Gizon}, Laurent and {Hekker}, Saskia and {Huber}, Daniel and {Khanna}, Shourya and {Li}, Yan and {Mathur}, Savita and {Miglio}, Andrea and {Mosser}, Beno{\^\i}t and {Ong}, J.~M. Joel and {Santos}, {\^A}ngela R.~G. and {Stello}, Dennis and {Bowman}, Dominic M. and {Lares-Martiz}, Mariel and {Murphy}, Simon and {Niu}, Jia-Shu and {Ma}, Xiao-Yu and {Moln{\'a}r}, L{\'a}szl{\'o} and {Fu}, Jian-Ning and {De Cat}, Peter and {Su}, Jie and {consortium}, the ET},
        title = "{ET White Paper: To Find the First Earth 2.0}",
      journal = {arXiv e-prints},
         year = 2022,
        month = jun,
          eid = {arXiv:2206.06693},
        pages = {arXiv:2206.06693},
          doi = {10.48550/arXiv.2206.06693},
archivePrefix = {arXiv},
       eprint = {2206.06693},
 primaryClass = {astro-ph.IM},
       adsurl = {https://ui.adsabs.harvard.edu/abs/2022arXiv220606693G}
}

@INPROCEEDINGS{2024SPIE13103E..0RK,
       author = {{Khandelwal}, Aditya and {Jeram}, Sarik and {Dungee}, Ryan and {Lau}, Albert W.~K. and {Lau}, Allison and {Sun}, Ethen and {Van-Lane}, Phil and {Chen}, Shaojie and {Tohuvavohu}, Aaron and {Li}, Ting S.},
        title = "{Beyond CCDs: characterization of sCMOS detectors for optical astronomy}",
    booktitle = {X-Ray, Optical, and Infrared Detectors for Astronomy XI},
         year = 2024,
       editor = {{Holland}, Andrew D. and {Minoglou}, Kyriaki},
       series = {Society of Photo-Optical Instrumentation Engineers (SPIE) Conference Series},
       volume = {13103},
        month = aug,
          eid = {131030R},
        pages = {131030R},
          doi = {10.1117/12.3018522},
archivePrefix = {arXiv},
       eprint = {2409.16449},
 primaryClass = {astro-ph.IM},
       adsurl = {https://ui.adsabs.harvard.edu/abs/2024SPIE13103E..0RK}
}

@inproceedings{Gill_2022,
author = {Ajay S. Gill and Mohamed M. Shaaban and Aaron Tohuvavohu and Suresh Sivanandam and Roberto G. Abraham and Seery Chen and Maria R. Drout and Deborah Lokhorst and Christopher D. Matzner and Stefan W. Mochnacki and Calvin B. Netterfield},
title = {{A low-cost ultraviolet-to-infrared absolute quantum efficiency characterization system of detectors}},
volume = {12191},
booktitle = {X-Ray, Optical, and Infrared Detectors for Astronomy X},
editor = {Andrew D. Holland and James Beletic},
organization = {International Society for Optics and Photonics},
publisher = {SPIE},
pages = {1219114},
year = {2022},
doi = {10.1117/12.2627564},
URL = {https://doi.org/10.1117/12.2627564},
series = {Proceedings of SPIE}
}

@ARTICLE{Ioannis24,
       author = {{Apergis}, Ioannis and {Bayliss}, Daniel and {Asimakoulas}, Leonidas and {Chote}, Paul and {McCormac}, James and {Mitchell}, Morgan A. and {Gill}, Sam and {Steen}, Philip G. and {Wheatley}, Peter J.},
        title = "{High-precision photometry with a scientific CMOS camera: I lab testing of the Marana camera}",
      journal = {RAS Techniques and Instruments},
         year = 2025,
        month = jan,
       volume = {4},
          eid = {rzaf049},
        pages = {rzaf049},
          doi = {10.1093/rasti/rzaf049},
archivePrefix = {arXiv},
       eprint = {2510.14484},
 primaryClass = {astro-ph.IM},
       adsurl = {https://ui.adsabs.harvard.edu/abs/2025RASTI...4af049A}
}

@ARTICLE{2015MNRASOsborn,
       author = {{Osborn}, J. and {F{\"o}hring}, D. and {Dhillon}, V.~S. and {Wilson}, R.~W.},
        title = "{Atmospheric scintillation in astronomical photometry}",
      journal = {\mnras},
         year = 2015,
        month = sep,
       volume = {452},
       number = {2},
        pages = {1707-1716},
          doi = {10.1093/mnras/stv1400},
archivePrefix = {arXiv},
       eprint = {1506.06921},
 primaryClass = {astro-ph.IM},
       adsurl = {https://ui.adsabs.harvard.edu/abs/2015MNRAS.452.1707O}
}

@ARTICLE{2018AJ....156..102S,
       author = {{Stassun}, Keivan G. and {Oelkers}, Ryan J. and {Pepper}, Joshua and {Paegert}, Martin and {De Lee}, Nathan and {Torres}, Guillermo and {Latham}, David W. and {Charpinet}, St{\'e}phane and {Dressing}, Courtney D. and {Huber}, Daniel and {Kane}, Stephen R. and {L{\'e}pine}, S{\'e}bastien and {Mann}, Andrew and {Muirhead}, Philip S. and {Rojas-Ayala}, B{\'a}rbara and {Silvotti}, Roberto and {Fleming}, Scott W. and {Levine}, Al and {Plavchan}, Peter},
        title = "{The TESS Input Catalog and Candidate Target List}",
      journal = {\aj},
         year = 2018,
        month = sep,
       volume = {156},
       number = {3},
          eid = {102},
        pages = {102},
          doi = {10.3847/1538-3881/aad050},
archivePrefix = {arXiv},
       eprint = {1706.00495},
 primaryClass = {astro-ph.EP},
       adsurl = {https://ui.adsabs.harvard.edu/abs/2018AJ....156..102S}
}

@ARTICLE{PLATO_MISSION_2025,
       author = {{Rauer}, Heike and {Aerts}, Conny and {Cabrera}, Juan and {Deleuil}, Magali and {Erikson}, Anders and {Gizon}, Laurent and {Goupil}, Mariejo and {Heras}, Ana and {Walloschek}, Thomas and {Lorenzo-Alvarez}, Jose and {Marliani}, Filippo and {Martin-Garcia}, C{\'e}sar and {Mas-Hesse}, J. Miguel and {O'Rourke}, Laurence and {Osborn}, Hugh and {Pagano}, Isabella and {Piotto}, Giampaolo and {Pollacco}, Don and {Ragazzoni}, Roberto and {Ramsay}, Gavin and {Udry}, St{\'e}phane and {Appourchaux}, Thierry and {Benz}, Willy and {Brandeker}, Alexis and {G{\"u}del}, Manuel and {Janot-Pacheco}, Eduardo and {Kabath}, Petr and {Kjeldsen}, Hans and {Min}, Michiel and {Santos}, Nuno and {Smith}, Alan and {Suarez}, Juan-Carlos and {Werner}, Stephanie C. and {Aboudan}, Alessio and {Abreu}, Manuel and {Acu{\~n}a}, Lorena and {Adams}, Moritz and {Adibekyan}, Vardan and {Affer}, Laura and {Agneray}, Fran{\c{c}}ois and {Agnor}, Craig and {Aguirre B{\o}rsen-Koch}, Victor and {Ahmed}, Saad and {Aigrain}, Suzanne and {Al-Bahlawan}, Ashraf and {Alcacera Gil}, Ma de los Angeles and {Alei}, Eleonora and {Alencar}, Silvia and {Alexander}, Richard and {Alfonso-Garz{\'o}n}, Julia and {Alibert}, Yann and {Allende Prieto}, Carlos and {Almeida}, Leonardo and {Alonso Sobrino}, Roi and {Altavilla}, Giuseppe and {Althaus}, Christian and {Alvarez Trujillo}, Luis Alonso and {Amarsi}, Anish and {Ammler-von Eiff}, Matthias and {Am{\^o}res}, Eduardo and {Andrade}, Laerte and {Antoniadis-Karnavas}, Alexandros and {Ant{\'o}nio}, Carlos and {Aparicio del Moral}, Beatriz and {Appolloni}, Matteo and {Arena}, Claudio and {Armstrong}, David and {Aroca Aliaga}, Jose and {Asplund}, Martin and {Audenaert}, Jeroen and {Auricchio}, Natalia and {Avelino}, Pedro and {Baeke}, Ann and {Bailli{\'e}}, Kevin and {Balado}, Ana and {Ballber Balaguer{\'o}}, Pau and {Balestra}, Andrea and {Ball}, Warrick and {Ballans}, Herve and {Ballot}, Jerome and {Barban}, Caroline and {Barbary}, Ga{\"e}le and {Barbieri}, Mauro and {Barcel{\'o} Forteza}, Sebasti{\`a} and {Barker}, Adrian and {Barklem}, Paul and {Barnes}, Sydney and {Barrado Navascues}, David and {Barragan}, Oscar and {Baruteau}, Cl{\'e}ment and {Basu}, Sarbani and {Baudin}, Frederic and {Baumeister}, Philipp and {Bayliss}, Daniel and {Bazot}, Michael and {Beck}, Paul G. and {Belkacem}, Kevin and {Bellinger}, Earl and {Benatti}, Serena and {Benomar}, Othman and {B{\'e}rard}, Diane and {Bergemann}, Maria and {Bergomi}, Maria and {Bernardo}, Pierre and {Biazzo}, Katia and {Bignamini}, Andrea and {Bigot}, Lionel and {Billot}, Nicolas and {Binet}, Martin and {Biondi}, David and {Biondi}, Federico and {Birch}, Aaron C. and {Bitsch}, Bertram and {Bluhm Ceballos}, Paz Victoria and {B{\'o}di}, Attila and {Bogn{\'a}r}, Zs{\'o}fia and {Boisse}, Isabelle and {Bolmont}, Emeline and {Bonanno}, Alfio and {Bonavita}, Mariangela and {Bonfanti}, Andrea and {Bonfils}, Xavier and {Bonito}, Rosaria and {Bonomo}, Aldo Stefano and {B{\"o}rner}, Anko and {Boro Saikia}, Sudeshna and {Borreguero Mart{\'\i}n}, Elisa and {Borsa}, Francesco and {Borsato}, Luca and {Bossini}, Diego and {Bouchy}, Francois and {Bou{\'e}}, Gwena{\"e}l and {Boufleur}, Rodrigo and {Boumier}, Patrick and {Bourrier}, Vincent and {Bowman}, Dominic M. and {Bozzo}, Enrico and {Bradley}, Louisa and {Bray}, John and {Bressan}, Alessandro and {Breton}, Sylvain and {Brienza}, Daniele and {Brito}, Ana and {Brogi}, Matteo and {Brown}, Beverly and {Brown}, David J.~A. and {Brun}, Allan Sacha and {Bruno}, Giovanni and {Bruns}, Michael and {Buchhave}, Lars A. and {Bugnet}, Lisa and {Buldgen}, Ga{\"e}l and {Burgess}, Patrick and {Busatta}, Andrea and {Busso}, Giorgia and {Buzasi}, Derek and {Caballero}, Jos{\'e} A. and {Cabral}, Alexandre and {Cabrero Gomez}, Juan-Francisco and {Calderone}, Flavia and {Cameron}, Robert and {Cameron}, Andrew and {Campante}, Tiago and {Campos Gestal}, N{\'e}stor and {Canto Martins}, Bruno Leonardo and {Cara}, Christophe and {Carone}, Ludmila and {Carrasco}, Josep Manel and {Casagrande}, Luca and {Casewell}, Sarah L. and {Cassisi}, Santi and {Castellani}, Marco and {Castro}, Matthieu and {Catala}, Claude and {Catal{\'a}n Fern{\'a}ndez}, Irene and {Catelan}, M{\'a}rcio and {Cegla}, Heather and {Cerruti}, Chiara and {Cessa}, Virginie and {Chadid}, Merieme and {Chaplin}, William and {Charpinet}, Stephane and {Chiappini}, Cristina and {Chiarucci}, Simone and {Chiavassa}, Andrea and {Chinellato}, Simonetta and {Chirulli}, Giovanni and {Christensen-Dalsgaard}, J{\o}rgen and {Church}, Ross and {Claret}, Antonio and {Clarke}, Cathie and {Claudi}, Riccardo and {Clermont}, Lionel and {Coelho}, Hugo and {Coelho}, Joao and {Cogato}, Fabrizio and {Colom{\'e}}, Josep and {Condamin}, Mathieu and {Conde Garc{\'\i}a}, Fernando and {Conseil}, Simon},
        title = "{The PLATO mission}",
      journal = {Experimental Astronomy},
         year = 2025,
        month = jun,
       volume = {59},
       number = {3},
          eid = {26},
        pages = {26},
          doi = {10.1007/s10686-025-09985-9},
archivePrefix = {arXiv},
       eprint = {2406.05447},
 primaryClass = {astro-ph.IM},
       adsurl = {https://ui.adsabs.harvard.edu/abs/2025ExA....59...26R}
}

@INCOLLECTION{KELT2018,
       author = {{Pepper}, Joshua and {Stassun}, Keivan G. and {Gaudi}, B.~S.},
        title = "{KELT: The Kilodegree Extremely Little Telescope, a Survey for Exoplanets Transiting Bright, Hot Stars}",
    booktitle = {Handbook of Exoplanets},
      publisher = {arXiv preprint arXiv:1802.10158},
         year = 2018,
       editor = {{Deeg}, Hans J. and {Belmonte}, Juan Antonio},
          eid = {128},
        pages = {128},
          doi = {10.1007/978-3-319-55333-7_128},
       adsurl = {https://ui.adsabs.harvard.edu/abs/2018haex.bookE.128P}
}

@article{mitchell_search_2025,
	title = {A search for photometric variability towards the globular cluster {M3} with the {TWenty} {Inch} {Survey} {Telescope}},
	issn = {0035-8711},
	url = {https://doi.org/10.1093/mnras/staf2100},
	doi = {10.1093/mnras/staf2100},
	urldate = {2025-11-26},
	journal = {Monthly Notices of the Royal Astronomical Society},
	author = {Mitchell, Morgan A and Chote, Paul and McCormac, James and Pollacco, Don and Apergis, Ioannis and Lyman, J D and Lockley, Isobel S and Gill, Samuel and Blake, James A and Claringbold, Alastair B and Steeghs, D T H and Casares, J},
	month = nov,
	year = {2025},
	pages = {staf2100},
}

@ARTICLE{2023A&A...670A.119B,
       author = {{Betoule}, Marc and {Antier}, Sarah and {Bertin}, Emmanuel and {Blanc}, Pierre {\'E}ric and {Bongard}, S{\'e}bastien and {Cohen Tanugi}, Johann and {Dagoret-Campagne}, Sylvie and {Feinstein}, Fabrice and {Hardin}, Delphine and {Juramy}, Claire and {Le Guillou}, Laurent and {Le Van Suu}, Auguste and {Moniez}, Marc and {Neveu}, J{\'e}r{\'e}my and {Nuss}, {\'E}ric and {Plez}, Bertrand and {Regnault}, Nicolas and {Sepulveda}, Eduardo and {Sommer}, K{\'e}lian and {Souverin}, Thierry and {Wang}, Xiao Feng},
        title = "{StarDICE. I. Sensor calibration bench and absolute photometric calibration of a Sony IMX411 sensor}",
      journal = {\aap},
         year = 2023,
        month = feb,
       volume = {670},
          eid = {A119},
        pages = {A119},
          doi = {10.1051/0004-6361/202244973},
archivePrefix = {arXiv},
       eprint = {2211.04913},
 primaryClass = {astro-ph.IM},
       adsurl = {https://ui.adsabs.harvard.edu/abs/2023A&A...670A.119B}
}

@ARTICLE{maxted2016ellc,
       author = {{Maxted}, P.~F.~L.},
        title = "{ellc: A fast, flexible light curve model for detached eclipsing binary stars and transiting exoplanets}",
      journal = {\aap},
         year = 2016,
        month = jun,
       volume = {591},
          eid = {A111},
        pages = {A111},
          doi = {10.1051/0004-6361/201628579},
archivePrefix = {arXiv},
       eprint = {1603.08484},
 primaryClass = {astro-ph.IM},
       adsurl = {https://ui.adsabs.harvard.edu/abs/2016A&A...591A.111M}
}

@ARTICLE{speagle2020dynesty,
       author = {{Speagle}, Joshua S.},
        title = "{DYNESTY: a dynamic nested sampling package for estimating Bayesian posteriors and evidences}",
      journal = {\mnras},
         year = 2020,
        month = apr,
       volume = {493},
       number = {3},
        pages = {3132-3158},
          doi = {10.1093/mnras/staa278},
archivePrefix = {arXiv},
       eprint = {1904.02180},
 primaryClass = {astro-ph.IM},
       adsurl = {https://ui.adsabs.harvard.edu/abs/2020MNRAS.493.3132S}
}

@INPROCEEDINGS{Skilling2004nestedsampling,
       author = {{Skilling}, John},
        title = "{Nested Sampling}",
    booktitle = {American Institute of Physics Conference Series},
         year = "2004",
       editor = {{Fischer}, Rainer and {Preuss}, Roland and {Toussaint}, Udo Von},
       series = {American Institute of Physics Conference Series},
       volume = {735},
        month = "Nov",
        pages = {395-405},
          doi = {10.1063/1.1835238},
       adsurl = {https://ui.adsabs.harvard.edu/abs/2004AIPC..735..395S}
}

@article{Skilling2006nestedsampling,
       author = "Skilling, John",
          doi = "10.1214/06-BA127",
     fjournal = "Bayesian Analysis",
      journal = "Bayesian Anal.",
        month = "12",
       number = "4",
        pages = "833--859",
    publisher = "International Society for Bayesian Analysis",
        title = "Nested sampling for general Bayesian computation",
          url = "https://doi.org/10.1214/06-BA127",
       volume = "1",
         year = "2006"
}

@ARTICLE{1989PASP..101..616H,
       author = {{Howell}, Steve B.},
        title = "{Two-Dimensional Aperture Photometry: Signal-to-Noise Ratio of Point-Source Observations and Optimal Data-Extraction Techniques}",
      journal = {\pasp},
         year = 1989,
        month = jun,
       volume = {101},
        pages = {616},
          doi = {10.1086/132477},
       adsurl = {https://ui.adsabs.harvard.edu/abs/1989PASP..101..616H}
}

@ARTICLE{1967AJ.....72..747Y,
       author = {{Young}, Andrew T.},
        title = "{Photometric error analysis. VI. Confirmation of Reiger's theory of scintillation}",
      journal = {\aj},
         year = 1967,
        month = aug,
       volume = {72},
        pages = {747},
          doi = {10.1086/110303},
       adsurl = {https://ui.adsabs.harvard.edu/abs/1967AJ.....72..747Y}
}

@ARTICLE{kepler670,
       author = {{Morton}, Timothy D. and {Bryson}, Stephen T. and {Coughlin}, Jeffrey L. and {Rowe}, Jason F. and {Ravichandran}, Ganesh and {Petigura}, Erik A. and {Haas}, Michael R. and {Batalha}, Natalie M.},
        title = "{False Positive Probabilities for all Kepler Objects of Interest: 1284 Newly Validated Planets and 428 Likely False Positives}",
      journal = {\apj},
         year = 2016,
        month = may,
       volume = {822},
       number = {2},
          eid = {86},
        pages = {86},
          doi = {10.3847/0004-637X/822/2/86},
archivePrefix = {arXiv},
       eprint = {1605.02825},
 primaryClass = {astro-ph.EP},
       adsurl = {https://ui.adsabs.harvard.edu/abs/2016ApJ...822...86M}
}

@ARTICLE{kelt1,
       author = {{Siverd}, Robert J. and {Beatty}, Thomas G. and {Pepper}, Joshua and {Eastman}, Jason D. and {Collins}, Karen and {Bieryla}, Allyson and {Latham}, David W. and {Buchhave}, Lars A. and {Jensen}, Eric L.~N. and {Crepp}, Justin R. and {Street}, Rachel and {Stassun}, Keivan G. and {Gaudi}, B. Scott and {Berlind}, Perry and {Calkins}, Michael L. and {DePoy}, D.~L. and {Esquerdo}, Gilbert A. and {Fulton}, Benjamin J. and {F{\H{u}}r{\'e}sz}, G{\'a}bor and {Geary}, John C. and {Gould}, Andrew and {Hebb}, Leslie and {Kielkopf}, John F. and {Marshall}, Jennifer L. and {Pogge}, Richard and {Stanek}, K.~Z. and {Stefanik}, Robert P. and {Szentgyorgyi}, Andrew H. and {Trueblood}, Mark and {Trueblood}, Patricia and {Stutz}, Amelia M. and {van Saders}, Jennifer L.},
        title = "{KELT-1b: A Strongly Irradiated, Highly Inflated, Short Period, 27 Jupiter-mass Companion Transiting a Mid-F Star}",
      journal = {\apj},
         year = 2012,
        month = dec,
       volume = {761},
       number = {2},
          eid = {123},
        pages = {123},
          doi = {10.1088/0004-637X/761/2/123},
archivePrefix = {arXiv},
       eprint = {1206.1635},
 primaryClass = {astro-ph.EP},
       adsurl = {https://ui.adsabs.harvard.edu/abs/2012ApJ...761..123S}
}

@ARTICLE{kepler76,
       author = {{Faigler}, S. and {Tal-Or}, L. and {Mazeh}, T. and {Latham}, D.~W. and {Buchhave}, L.~A.},
        title = "{BEER Analysis of Kepler and CoRoT Light Curves. I. Discovery of Kepler-76b: A Hot Jupiter with Evidence for Superrotation}",
      journal = {\apj},
         year = 2013,
        month = jul,
       volume = {771},
       number = {1},
          eid = {26},
        pages = {26},
          doi = {10.1088/0004-637X/771/1/26},
archivePrefix = {arXiv},
       eprint = {1304.6841},
 primaryClass = {astro-ph.EP},
       adsurl = {https://ui.adsabs.harvard.edu/abs/2013ApJ...771...26F}
}

@ARTICLE{kepler41,
       author = {{Santerne}, A. and {Bonomo}, A.~S. and {H{\'e}brard}, G. and {Deleuil}, M. and {Moutou}, C. and {Almenara}, J. -M. and {Bouchy}, F. and {D{\'\i}az}, R.~F.},
        title = "{SOPHIE velocimetry of Kepler transit candidates. IV. KOI-196b: a non-inflated hot Jupiter with a high albedo}",
      journal = {\aap},
         year = 2011,
        month = dec,
       volume = {536},
          eid = {A70},
        pages = {A70},
          doi = {10.1051/0004-6361/201117807},
archivePrefix = {arXiv},
       eprint = {1108.0550},
 primaryClass = {astro-ph.EP},
       adsurl = {https://ui.adsabs.harvard.edu/abs/2011A&A...536A..70S}
}

@ARTICLE{2025AdSpR..76..764C,
       author = {{Chote}, Paul and {Airey}, Robert and {McCormac}, James and {Pollacco}, Don and {West}, Richard and {Ulaczyk}, Krzysztof and {Dyer}, Martin J. and {Agathanggelou}, Alexander and {Feline}, William and {George}, Simon and {Meredith}, Calum and {Privett}, Grant},
        title = "{High-precision light curves of geostationary objects: The PHANTOM ECHOES 2 RPO campaign}",
      journal = {Advances in Space Research},
         year = 2025,
        month = jul,
       volume = {76},
       number = {2},
        pages = {764-783},
          doi = {10.1016/j.asr.2025.05.026},
archivePrefix = {arXiv},
       eprint = {2506.01549},
 primaryClass = {astro-ph.IM},
       adsurl = {https://ui.adsabs.harvard.edu/abs/2025AdSpR..76..764C}
}

@ARTICLE{KELT-16_2017,
       author = {{Oberst}, Thomas E. and {Rodriguez}, Joseph E. and {Col{\'o}n}, Knicole D. and {Angerhausen}, Daniel and {Bieryla}, Allyson and {Ngo}, Henry and {Stevens}, Daniel J. and {Stassun}, Keivan G. and {Gaudi}, B. Scott and {Pepper}, Joshua and {Penev}, Kaloyan and {Mawet}, Dimitri and {Latham}, David W. and {Heintz}, Tyler M. and {Osei}, Baffour W. and {Collins}, Karen A. and {Kielkopf}, John F. and {Visgaitis}, Tiffany and {Reed}, Phillip A. and {Escamilla}, Alejandra and {Yazdi}, Sormeh and {McLeod}, Kim K. and {Lunsford}, Leanne T. and {Spencer}, Michelle and {Joner}, Michael D. and {Gregorio}, Joao and {Gaillard}, Clement and {Matt}, Kyle and {Dumont}, Mary Thea and {Stephens}, Denise C. and {Cohen}, David H. and {Jensen}, Eric L.~N. and {Calchi Novati}, Sebastiano and {Bozza}, Valerio and {Labadie-Bartz}, Jonathan and {Siverd}, Robert J. and {Lund}, Michael B. and {Beatty}, Thomas G. and {Eastman}, Jason D. and {Penny}, Matthew T. and {Manner}, Mark and {Zambelli}, Roberto and {Fulton}, Benjamin J. and {Stockdale}, Christopher and {DePoy}, D.~L. and {Marshall}, Jennifer L. and {Pogge}, Richard W. and {Gould}, Andrew and {Trueblood}, Mark and {Trueblood}, Patricia},
        title = "{KELT-16b: A Highly Irradiated, Ultra-short Period Hot Jupiter Nearing Tidal Disruption}",
      journal = {\aj},
         year = 2017,
        month = mar,
       volume = {153},
       number = {3},
          eid = {97},
        pages = {97},
          doi = {10.3847/1538-3881/153/3/97},
archivePrefix = {arXiv},
       eprint = {1608.00618},
 primaryClass = {astro-ph.EP},
       adsurl = {https://ui.adsabs.harvard.edu/abs/2017AJ....153...97O}
}

@INPROCEEDINGS{Guyon_2010,
       author = {{Guyon}, Olivier and {Martinache}, Frantz and {Garrel}, Vincent and {Vogt}, Frederic and {Yokochi}, Kaito and {Yoshikawa}, Takashi},
        title = "{The Subaru coronagraphic extreme AO (SCExAO) system: wavefront control and detection of exoplanets with coherent light modulation in the focal plane}",
    booktitle = {Adaptive Optics Systems II},
         year = 2010,
       editor = {{Ellerbroek}, Brent L. and {Hart}, Michael and {Hubin}, Norbert and {Wizinowich}, Peter L.},
       series = {Society of Photo-Optical Instrumentation Engineers (SPIE) Conference Series},
       volume = {7736},
        month = jul,
          eid = {773624},
        pages = {773624},
          doi = {10.1117/12.857878},
       adsurl = {https://ui.adsabs.harvard.edu/abs/2010SPIE.7736E..24G}
}

@INPROCEEDINGS{Downing_2016,
       author = {{Downing}, Mark and {Kolb}, Johann and {Dierickx}, Bart and {Defernez}, Arnaud and {Feautrier}, Philippe and {Fryer}, Martin and {Gach}, Jean-Luc and {Jerram}, Paul and {Jorden}, Paul and {Meyer}, Manfred and {Pike}, Andrew and {Reyes}, Javier and {Stadler}, Eric and {Swift}, Nick},
        title = "{LGSD/NGSD: high speed visible CMOS imagers for E-ELT adaptive optics}",
    booktitle = {High Energy, Optical, and Infrared Detectors for Astronomy VII},
         year = 2016,
       editor = {{Holland}, Andrew D. and {Beletic}, James},
       series = {Society of Photo-Optical Instrumentation Engineers (SPIE) Conference Series},
       volume = {9915},
        month = aug,
          eid = {991512},
        pages = {991512},
          doi = {10.1117/12.2231960},
       adsurl = {https://ui.adsabs.harvard.edu/abs/2016SPIE.9915E..12D}
}

@article{alarcon2023scientific,
  title={Scientific CMOS Sensors in Astronomy: IMX455 and IMX411},
  author={Alarcon, Miguel R and Licandro, Javier and Serra-Ricart, Miquel and Joven, Enrique and Gaitan, Vicens and de Sousa, Rebeca},
  journal={Publications of the Astronomical Society of the Pacific},
  volume={135},
  number={1047},
  pages={055001},
  year={2023},
  publisher={IOP Publishing}
}

@ARTICLE{2006MNRAS.373..231P,
       author = {{Pont}, Fr{\'e}d{\'e}ric and {Zucker}, Shay and {Queloz}, Didier},
        title = "{The effect of red noise on planetary transit detection}",
      journal = {\mnras},
         year = 2006,
        month = nov,
       volume = {373},
       number = {1},
        pages = {231-242},
          doi = {10.1111/j.1365-2966.2006.11012.x},
archivePrefix = {arXiv},
       eprint = {astro-ph/0608597},
 primaryClass = {astro-ph},
       adsurl = {https://ui.adsabs.harvard.edu/abs/2006MNRAS.373..231P}
}

@ARTICLE{allesfitter-paper,
       author = {{G{\"u}nther}, Maximilian N. and {Daylan}, Tansu},
        title = "{Allesfitter: Flexible Star and Exoplanet Inference from Photometry and Radial Velocity}",
      journal = {\apjs},
         year = 2021,
        month = may,
       volume = {254},
       number = {1},
          eid = {13},
        pages = {13},
          doi = {10.3847/1538-4365/abe70e},
archivePrefix = {arXiv},
       eprint = {2003.14371},
 primaryClass = {astro-ph.EP},
       adsurl = {https://ui.adsabs.harvard.edu/abs/2021ApJS..254...13G}
}

@MISC{allesfitter-code,
 author = {{G{\"u}nther}, Maximilian~N. and {Daylan}, Tansu},
 title = "{Allesfitter: Flexible Star and Exoplanet Inference From Photometry and Radial Velocity}",
 howpublished = {Astrophysics Source Code Library},
 year = 2019,
 month = mar,
 archivePrefix = "ascl",
 eprint = {1903.003},
 adsurl = {http://adsabs.harvard.edu/abs/2019ascl.soft03003G}

}

@inproceedings{karpov2020,
  title={Evaluation of scientific CMOS sensors for sky survey applications},
  author={Karpov, Sergey and Bajat, Armelle and Christov, Asen and Prouza, Michael and Beskin, Grigory},
  booktitle={X-Ray, Optical, and Infrared Detectors for Astronomy IX},
  volume={11454},
  pages={86--98},
  year={2020},
  organization={SPIE}
}

@article{Ioannis26,
       author = {{Apergis}, Ioannis and {Bayliss}, Daniel and {Chote}, Paul and {McCormac}, James and {Wheatley}, Peter J. and {Mitchell}, Morgan A. and {Fern{\'a}ndez Fern{\'a}ndez}, Jorge and {Gill}, Sam and {Bryant}, Edward M. and {Rodel}, Toby and {Asimakoulas}, Leonidas and {Anderson}, David R. and {Blake}, James A. and {Casewell}, Sarah L. and {Eeles-Nolle}, Fintan and {Hawthorn}, Faith and {Jenkins}, James S. and {Lendl}, Monika and {Lockley}, Isobel S. and {Moyano}, Maximiliano and {O'Brien}, Sean M. and {Saha}, Suman and {Smith}, Alexis M.~S. and {Steen}, Philip G. and {Vines}, Jose I. and {West}, Richard G. and {Zivave}, Tafadzwa},
        title = "{High-precision photometry with a scientific CMOS camera: II on-sky testing of the Marana camera at the NGTS facility}",
      journal = {RAS Techniques and Instruments},
         year = 2026,
        month = jan,
       volume = {5},
          eid = {rzag022},
        pages = {rzag022},
          doi = {10.1093/rasti/rzag022},
archivePrefix = {arXiv},
       eprint = {2603.16361},
 primaryClass = {astro-ph.IM},
       adsurl = {https://ui.adsabs.harvard.edu/abs/2026RASTI...5ag022A}
}

@article{Ricker2015TESS,
	title        = {{Transiting Exoplanet Survey Satellite (TESS)}},
	author       = {{Ricker}, George R. and {Winn}, Joshua N. and {Vanderspek}, Roland and {Latham}, David W. and {Bakos}, G{\'a}sp{\'a}r {\'A}. and {Bean}, Jacob L. and {Berta-Thompson}, Zachory K. and {Brown}, Timothy M. and {Buchhave}, Lars and {Butler}, Nathaniel R. and {Butler}, R. Paul and {Chaplin}, William J. and {Charbonneau}, David and {Christensen-Dalsgaard}, J{\o}rgen and {Clampin}, Mark and {Deming}, Drake and {Doty}, John and {De Lee}, Nathan and {Dressing}, Courtney and {Dunham}, Edward W. and {Endl}, Michael and {Fressin}, Francois and {Ge}, Jian and {Henning}, Thomas and {Holman}, Matthew J. and {Howard}, Andrew W. and {Ida}, Shigeru and {Jenkins}, Jon M. and {Jernigan}, Garrett and {Johnson}, John Asher and {Kaltenegger}, Lisa and {Kawai}, Nobuyuki and {Kjeldsen}, Hans and {Laughlin}, Gregory and {Levine}, Alan M. and {Lin}, Douglas and {Lissauer}, Jack J. and {MacQueen}, Phillip and {Marcy}, Geoffrey and {McCullough}, Peter R. and {Morton}, Timothy D. and {Narita}, Norio and {Paegert}, Martin and {Palle}, Enric and {Pepe}, Francesco and {Pepper}, Joshua and {Quirrenbach}, Andreas and {Rinehart}, Stephen A. and {Sasselov}, Dimitar and {Sato}, Bun'ei and {Seager}, Sara and {Sozzetti}, Alessandro and {Stassun}, Keivan G. and {Sullivan}, Peter and {Szentgyorgyi}, Andrew and {Torres}, Guillermo and {Udry}, Stephane and {Villasenor}, Joel},
	year         = 2015,
	month        = jan,
	journal      = {Journal of Astronomical Telescopes, Instruments, and Systems},
	volume       = 1,
	pages        = {014003},
	doi          = {10.1117/1.JATIS.1.1.014003},
	eid          = {014003},
	adsurl       = {https://ui.adsabs.harvard.edu/abs/2015JATIS...1a4003R}
}

@ARTICLE{DONUTS,
       author = {{McCormac}, J. and {Pollacco}, D. and {Skillen}, I. and {Faedi}, F. and {Todd}, I. and {Watson}, C.~A.},
        title = "{DONUTS: A Science Frame Autoguiding Algorithm with Sub-Pixel Precision, Capable of Guiding on Defocused Stars}",
      journal = {\pasp},
         year = 2013,
        month = may,
       volume = {125},
       number = {927},
        pages = {548},
          doi = {10.1086/670940},
archivePrefix = {arXiv},
       eprint = {1304.2405},
 primaryClass = {astro-ph.IM},
       adsurl = {https://ui.adsabs.harvard.edu/abs/2013PASP..125..548M}
}

@ARTICLE{sex,
       author = {{Bertin}, E. and {Arnouts}, S.},
        title = "{SExtractor: Software for source extraction.}",
      journal = {\aaps},
         year = 1996,
        month = jun,
       volume = {117},
        pages = {393-404},
          doi = {10.1051/aas:1996164},
       adsurl = {https://ui.adsabs.harvard.edu/abs/1996A&AS..117..393B}
}

@ARTICLE{sex2016,
       author = {{Barbary}, Kyle},
        title = "{SEP: Source Extractor as a library}",
      journal = {The Journal of Open Source Software},
         year = 2016,
        month = oct,
       volume = {1},
       number = {6},
          eid = {58},
        pages = {58},
          doi = {10.21105/joss.00058},
       adsurl = {https://ui.adsabs.harvard.edu/abs/2016JOSS....1...58B}
}

@ARTICLE{astrometry,
       author = {{Lang}, Dustin and {Hogg}, David W. and {Mierle}, Keir and {Blanton}, Michael and {Roweis}, Sam},
        title = "{Astrometry.net: Blind Astrometric Calibration of Arbitrary Astronomical Images}",
      journal = {\aj},
         year = 2010,
        month = may,
       volume = {139},
       number = {5},
        pages = {1782-1800},
          doi = {10.1088/0004-6256/139/5/1782},
archivePrefix = {arXiv},
       eprint = {0910.2233},
 primaryClass = {astro-ph.IM},
       adsurl = {https://ui.adsabs.harvard.edu/abs/2010AJ....139.1782L}
}

@ARTICLE{Castro-Ginard_2024,
       author = {{Castro-Ginard}, Alfred and {Penoyre}, Zephyr and {Casey}, Andrew R. and {Brown}, Anthony G.~A. and {Belokurov}, Vasily and {Cantat-Gaudin}, Tristan and {Drimmel}, Ronald and {Fouesneau}, Morgan and {Khanna}, Shourya and {Kurbatov}, Evgeny P. and {Price-Whelan}, Adrian M. and {Rix}, Hans-Walter and {Smart}, Richard L.},
        title = "{Gaia DR3 detectability of unresolved binary systems}",
      journal = {\aap},
         year = 2024,
        month = aug,
       volume = {688},
          eid = {A1},
        pages = {A1},
          doi = {10.1051/0004-6361/202450172},
archivePrefix = {arXiv},
       eprint = {2404.14127},
 primaryClass = {astro-ph.GA},
       adsurl = {https://ui.adsabs.harvard.edu/abs/2024A&A...688A...1C}
}

%\begin{thebibliography}{99}
%\end{thebibliography}

%%%%%%%%%%%%%%%%%%%%%%%%%%%%%%%%%%%%%%%%%%%%%%%%%%

%%%%%%%%%%%%%%%%% APPENDICES %%%%%%%%%%%%%%%%%%%%%

\appendix
\section{Extra Material}
\label{sec:extra_material}

\begin{figure*}
\includegraphics[width=\textwidth]{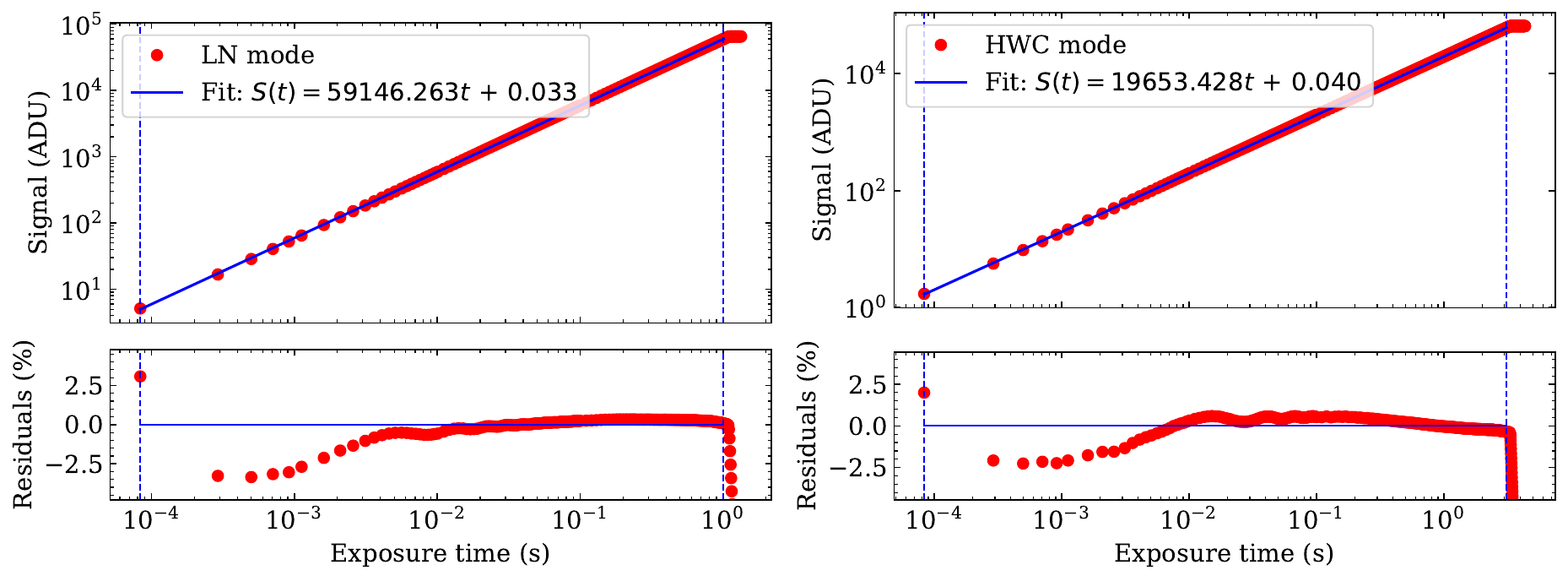}
\caption[Linearity for the Toran 26B 0-95]{Linearity measurements for the Toran 26B camera at 0\,$^{\circ}$C. \textbf{Left:} The LN image mode. \textbf{Right:} The HWC image mode. The best fit linear model between 0 and 95 \% of saturation is plotted with a blue solid line. Residuals to the linear model are shown in the bottom panels. The dashed horizontal blue lines indicate the 0 to 95\% boundaries we set to measure fit the linearity.}
\label{fig:linearity_toran_0-95}
\end{figure*}

\begin{figure}\centering
\includegraphics[width=\columnwidth]{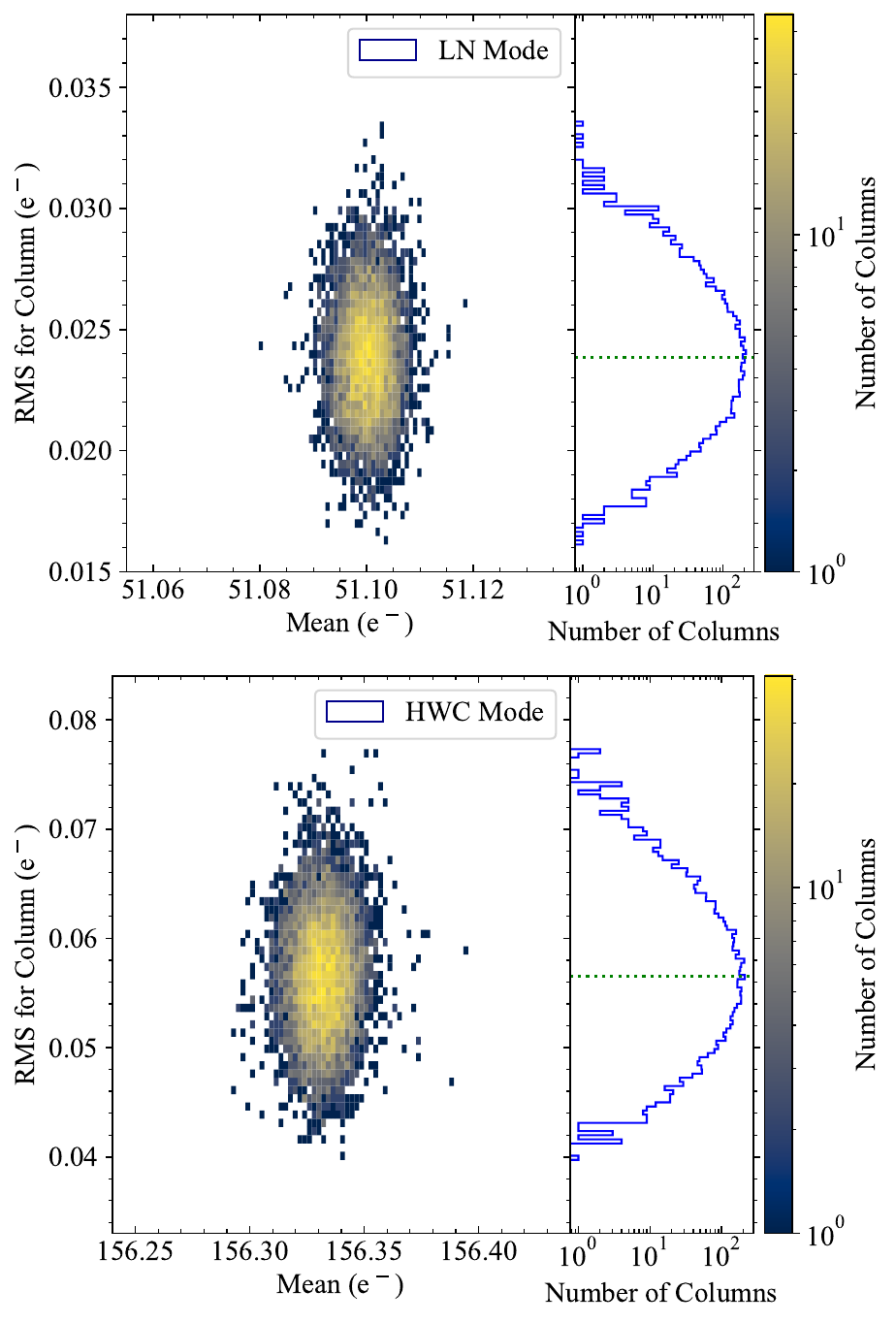}
\caption[Column noise for Toran 26B]{Column noise distribution using 75 consecutive bias frames. The 2-D histogram corresponds to row temporal noise and the mean of each row from these frames while the colour bar indicates the number of rows. The logarithmic row noise distribution is show in the blue histograms, with the median value as a green dotted line. \textbf{Top:} Mean row noise distribution for the FFR mode. \textbf{Bottom:} Mean row noise distribution for the HDR mode.}
\label{fig:col_col}
\end{figure}

\begin{figure}\centering
\includegraphics[width=\columnwidth]{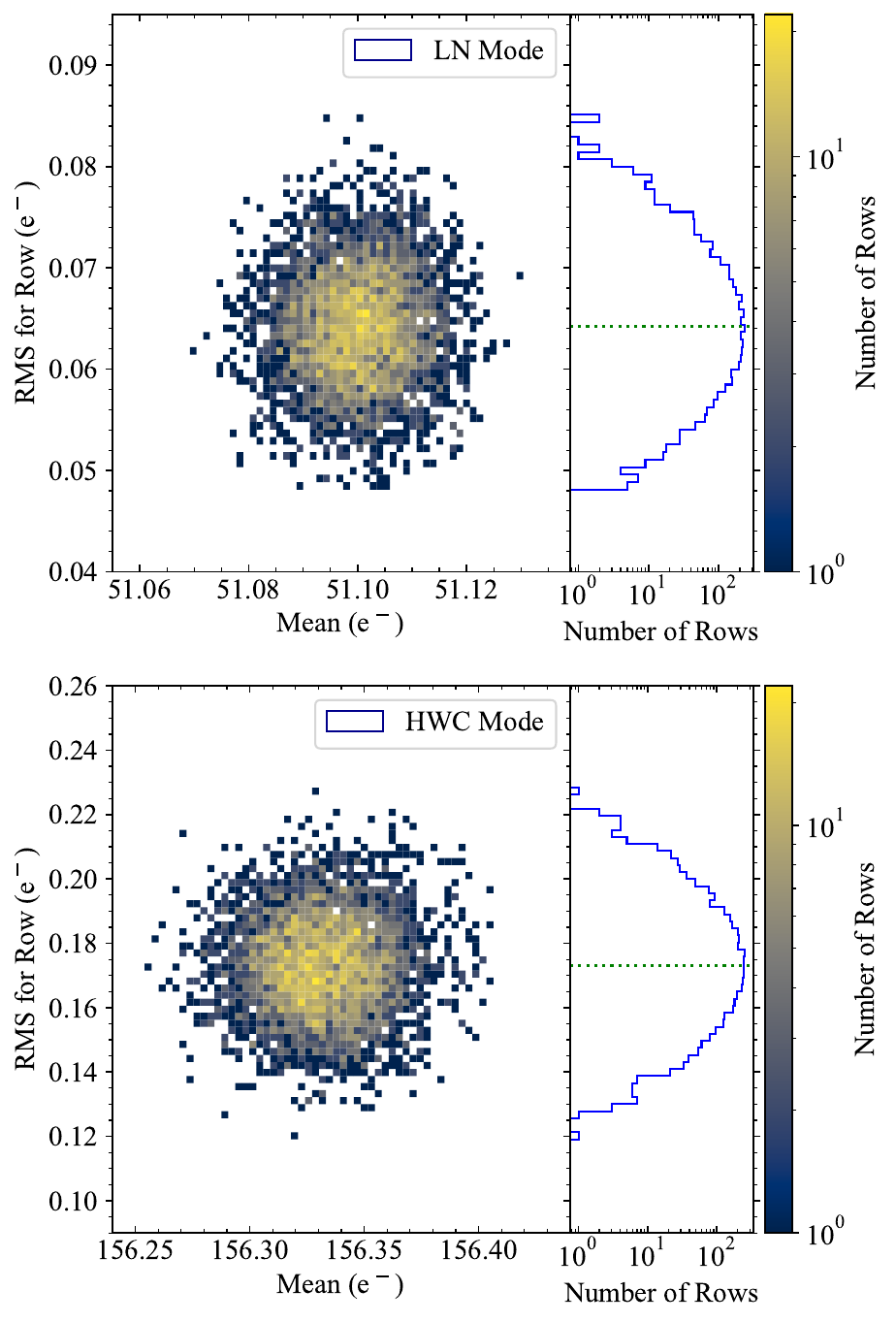}
\caption[Row noise for Toran 26B]{Row noise distribution using 75 consecutive bias frames. The 2-D histogram corresponds to row temporal noise and the mean of each row from these frames while the colour bar indicates the number of rows. The logarithmic row noise distribution is show in the blue histograms, with the median value as a green dotted line. \textbf{Top:} Mean row noise distribution for the LN mode. \textbf{Bottom:} Mean row noise distribution for the HWC mode.}
\label{fig:row_row_rn}
\end{figure}

\begin{figure}\centering
\includegraphics[width=\columnwidth]{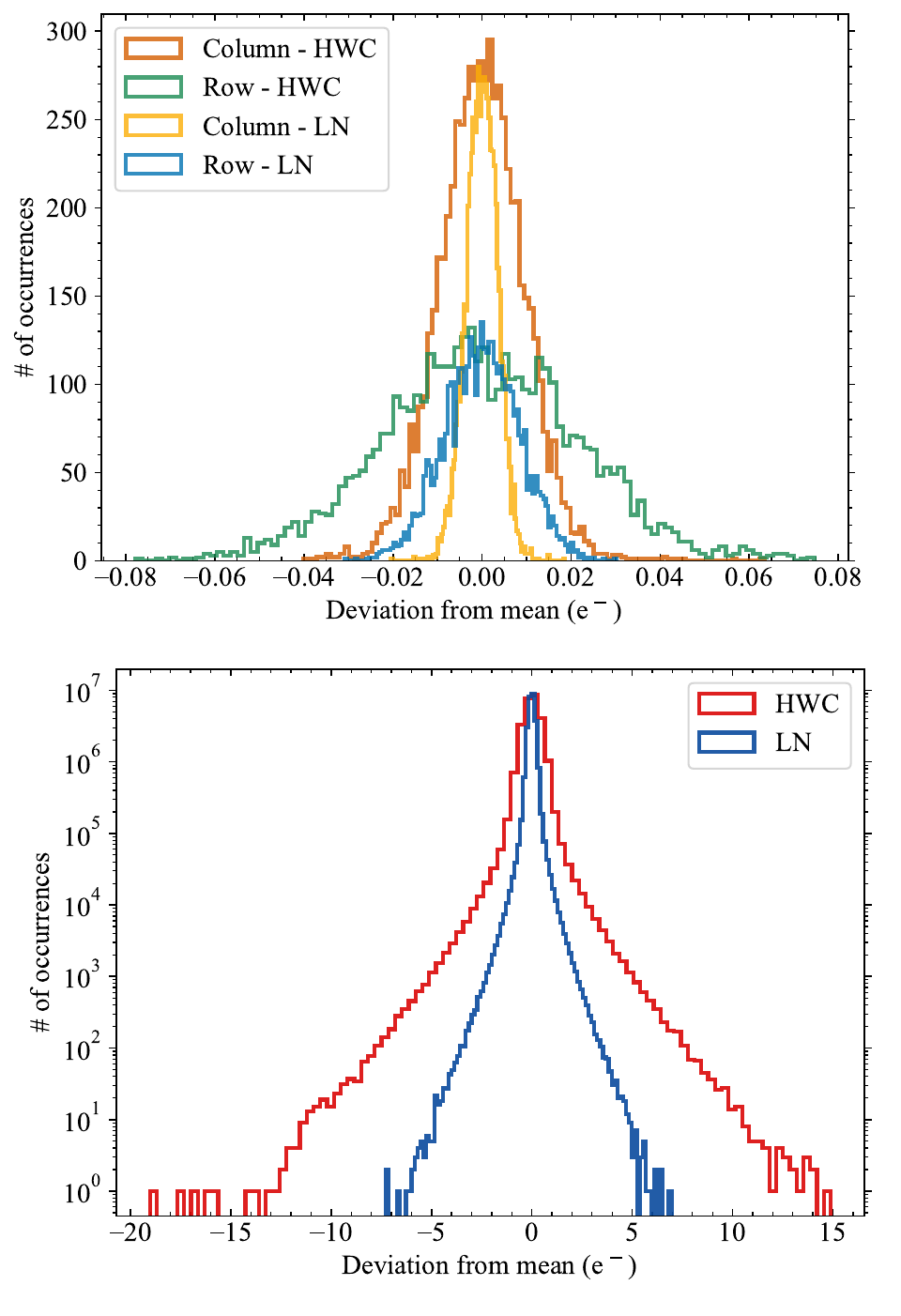}
\caption[DSNU for Toran 26B]{Histogram distribution of the DSNU for both modes modes. \textbf{The top} panel displays the distribution of row and column DSNU after subtracting the spatial mean level. \textbf{The bottom} panel shows the distribution of image DSNU, also after subtracting the spatial mean level.}
\label{fig:dsnu_rn}
\end{figure}

\begin{figure*}
\includegraphics[width=\textwidth]{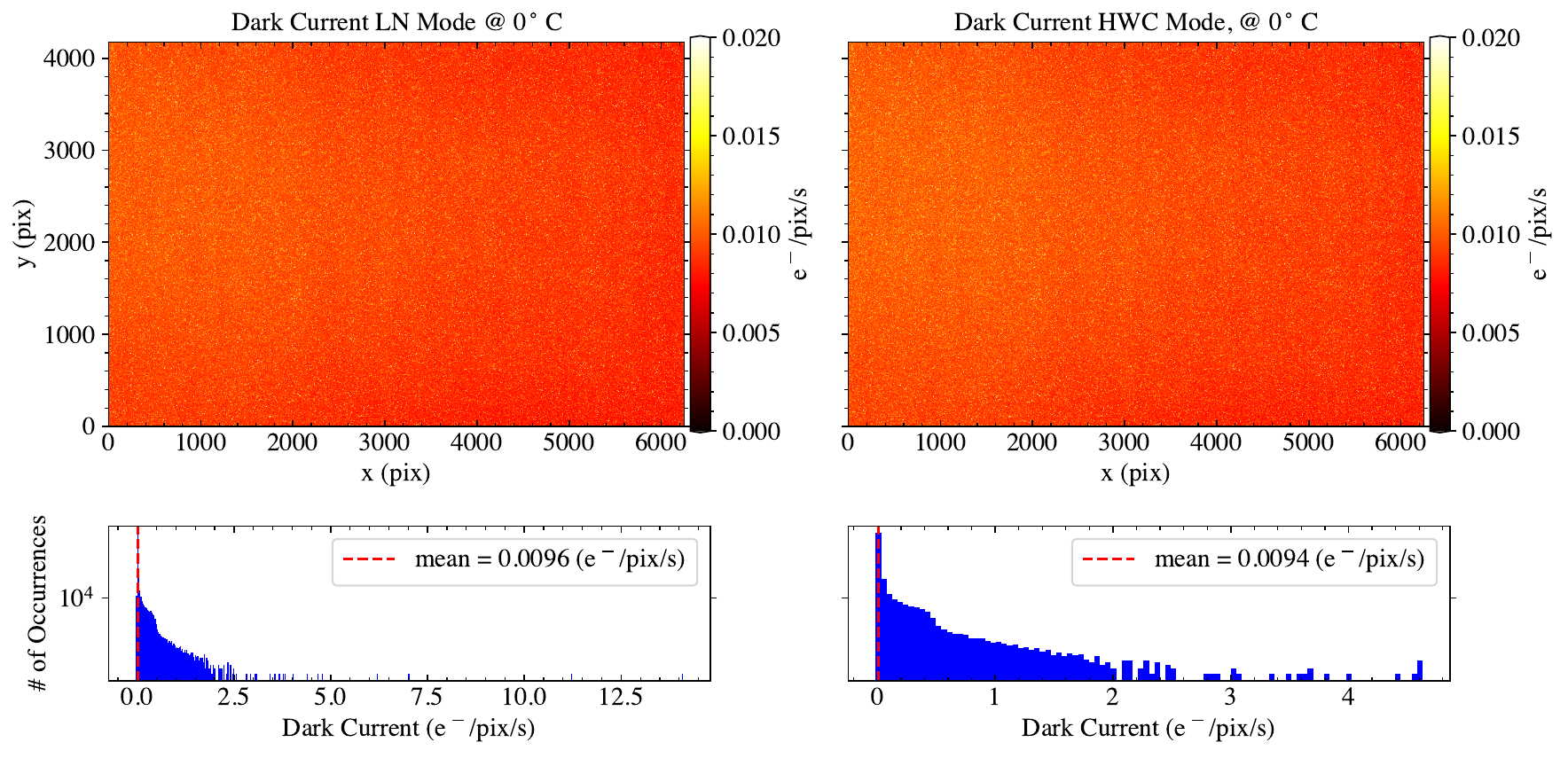}
\caption{Dark current maps taken for new Andor camera without the LDC at temperature of 0\,$^\circ$\,C. \textbf{Left} is the LN mode. \textbf{Right} is the HWC mode. Below each are the histograms of pixel value distributions from each frame, along with the mean (red dotted line).}
\label{fig:Dark_LN_HWC}
\end{figure*}

% \begin{figure*}
% \includegraphics[width=\columnwidth]{Data/donuts_KELT-16-20250806210525955.fits_20251019.pdf}
% \caption{Dark images taken for new Andor camera without the LDC at temperature of 0\,$^\circ$\,C. \textbf{Left} is the LN mode. \textbf{Right} is the HWC mode. Below each are the histograms of pixel value distributions from each frame, along with the mean (red dotted line).}
% \label{fig:donut}
% \end{figure*}

\begin{figure}
\includegraphics[width=\columnwidth]{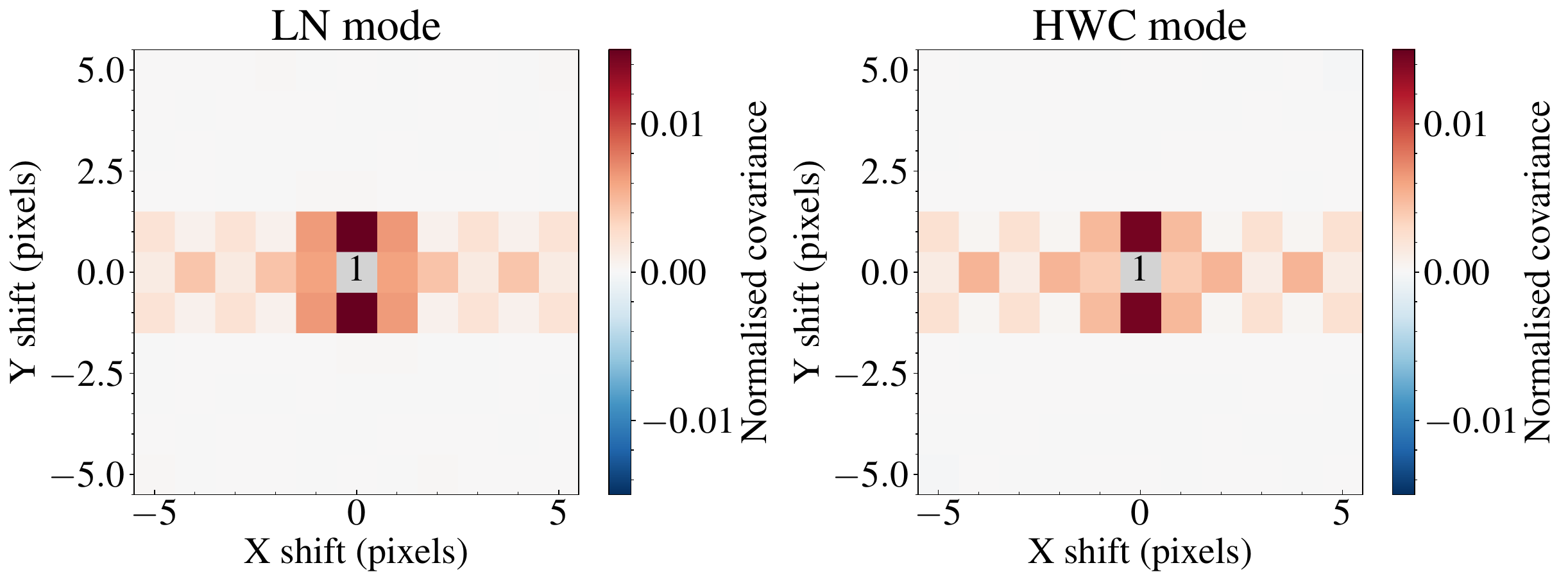}
\caption{Spatial autocorrelation of the bias data residuals after subtracting
the per-pixel temporal mean, averaged over the full frame and normalised
by the covariance. \textbf{Left:} LN mode. \textbf{Right:} HWC mode.
The axes indicate spatial shifts in pixels. The central cell corresponds
to zero separation and has an autocorrelation of unity by definition and
it is excluded from the colour scale to highlight correlations at
nonzero separations.}
\label{fig:cross_corr_dark}
\end{figure}

% \begin{figure}
% \includegraphics[width=\columnwidth]{Data/zp_vs_time.pdf}
% \caption{Diagnostic output plots of an observing night. Zeropoint as a function of time for the night 6 August 2025 observing the KELT-16 field.}
% \label{fig:zp}
% \end{figure}

% \begin{figure}
% \includegraphics[width=\columnwidth]{Data/hfd_vs_time.pdf}
% \caption{Diagnostic output plots of an observing night. Half flux diameter as a function of time for the night 6 August 2025 observing the KELT-16 field.}
% \label{fig:hfd}
% \end{figure}

\begin{figure}
\includegraphics[width=\columnwidth]{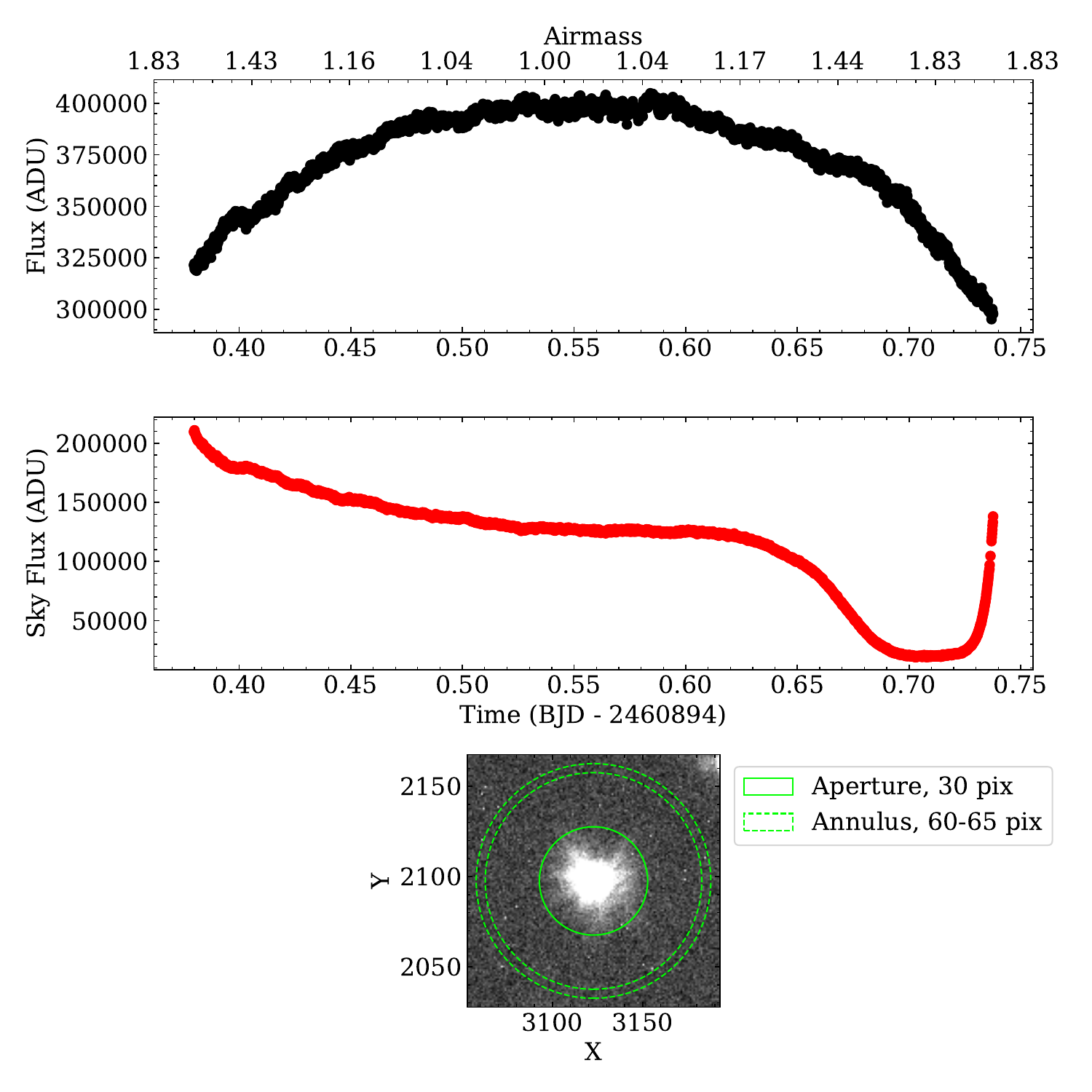}
\caption{Diagnostic output from the photometry. \textbf{Upper panel} shows the raw lightcurve for the star KELT-16 in the night of 6 August 2025. \textbf{Middle panel} shows the sky background flux. \textbf{Lower panel} shows the x-y pixel position of the star with the aperture and annulus radii.}
\label{fig:lc_kelt-16}
\end{figure}

\begin{figure}
\includegraphics[width=\columnwidth]{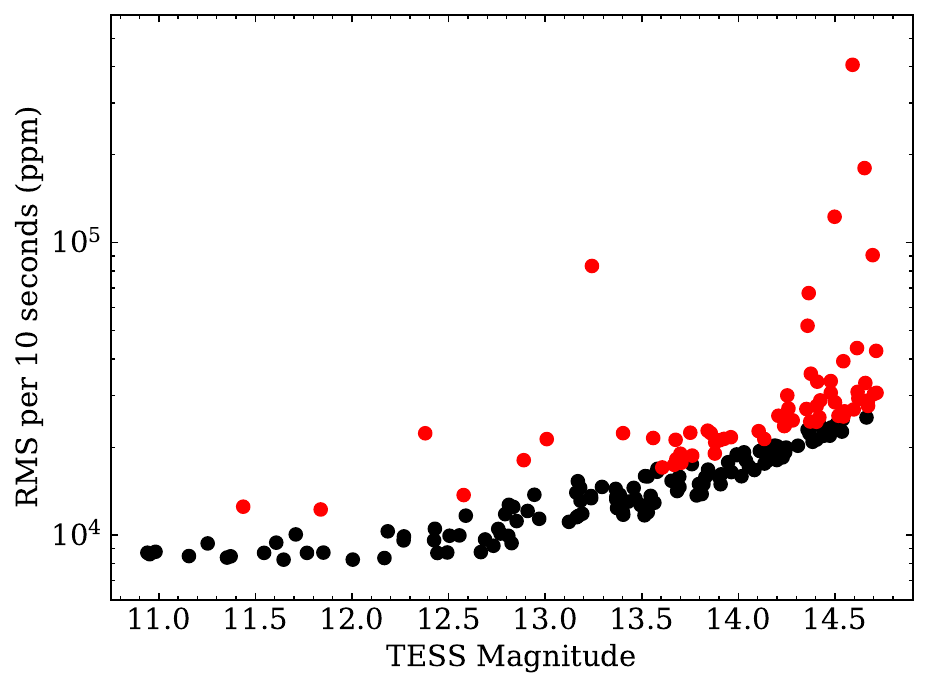}
\caption{Noise model for performing relative photometry on a target star. The RMS noise of the comparison stars is plotted against their magnitude, with the filtering procedure highlighting noisy stars to be removed (red solid points) and stable stars retained for relative photometry (black solid points).}
\label{fig:rms_vs_mags_material}
\end{figure}

\begin{figure}
\centering
\includegraphics[width=\columnwidth]{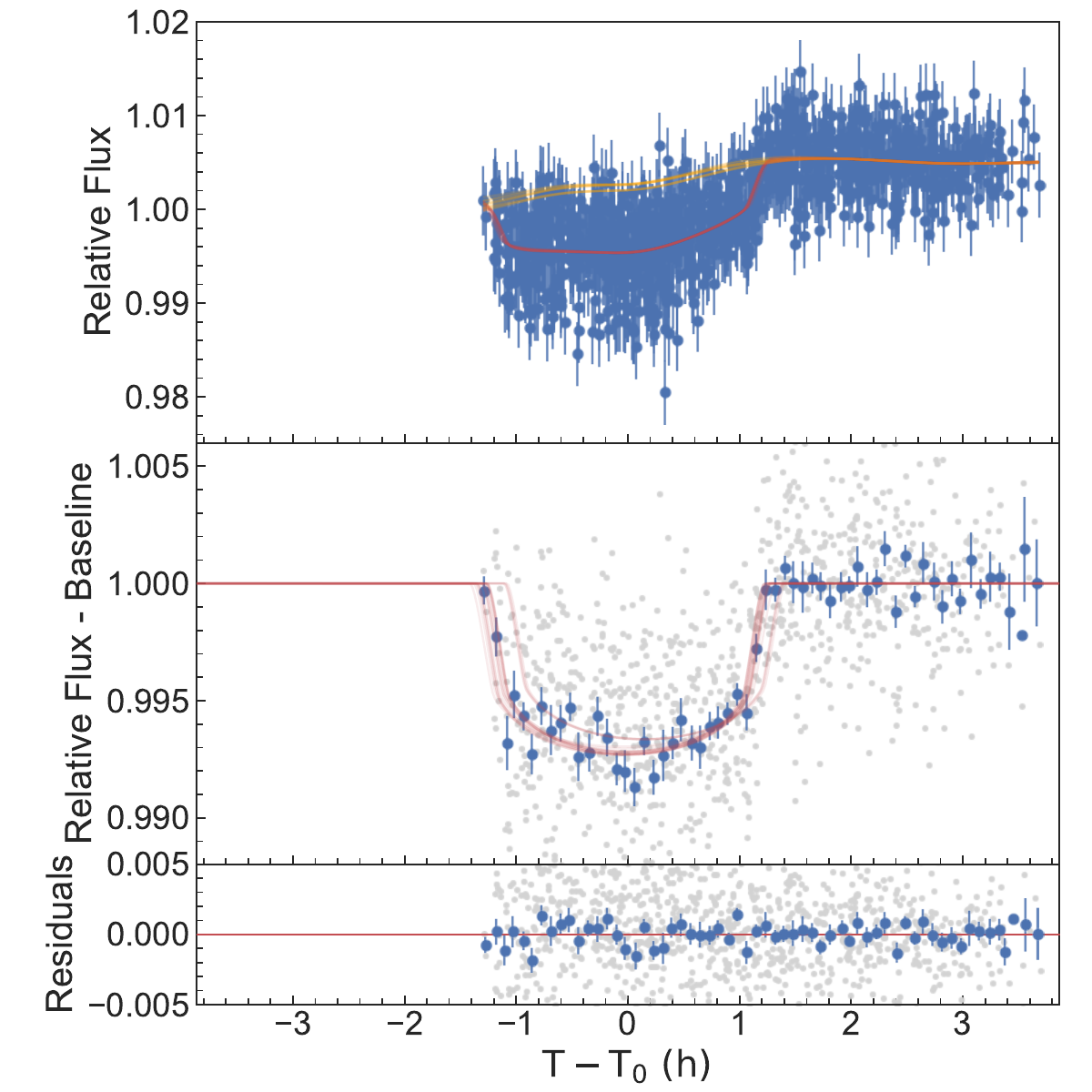}
\caption{Same as in \autoref{fig:kelt-16}. Normalized lightcurves for the transiting system of KELT-1 (TIC 432549364) \citep{kelt1} from TWIST on 2025 August 10.}
\label{fig:kelt-1}
\end{figure}

\begin{table*}
    \centering
    \caption{\texttt{allesfitter} model priors. Planetary system properties
    for the transiting system KELT-1 (TIC 432549364), as illustrated in
    \autoref{fig:kelt-1}, observed with TWIST on 2025 August 10.}

    {\renewcommand{\arraystretch}{1.25}
    \begin{tabular}{lccc}
        \hline
        \textbf{Parameter} & \textbf{Initial Guess} & \textbf{Prior}
        & \textbf{Fitted value Toran-26B} \\
        \hline
        $R_p/R_{\star}$        & \FITNBrb     & $\mathcal{U}(0.07,0.08)$              & \CMOSNBrb \\
        $(R_\star+R_b)/a_b$    & \FITNBrsuma  & $\mathcal{U}(0.27,0.30)$              & \CMOSNBrsuma \\
        $\cos{i_b}$            & \FITNBinc    & $\mathcal{U}(0.04,0.07)$              & \CMOSNBinc \\
        $T_{\mathrm{c}}$ (BJD) & \FITNBtc     & $\mathcal{U}(2457727.010,2457727.012)$ & \CMOSNBtc \\
        Orbital period (days)  & \FITNBperiod & $\mathcal{U}(1.215,1.219)$            & \CMOSNBperiod \\
        $q_1$                  & \FITNBqone   & $\mathcal{N}(0.318,0.010)$            & \CMOSNBqone \\
        $q_2$                  & \FITNBqtwo   & $\mathcal{N}(0.369,0.080)$            & \CMOSNBqtwo \\
        \hline
    \end{tabular}}
    \label{tab:planetary-kelt-1_0810}
\end{table*}

\begin{figure}
\centering
\includegraphics[width=\columnwidth]{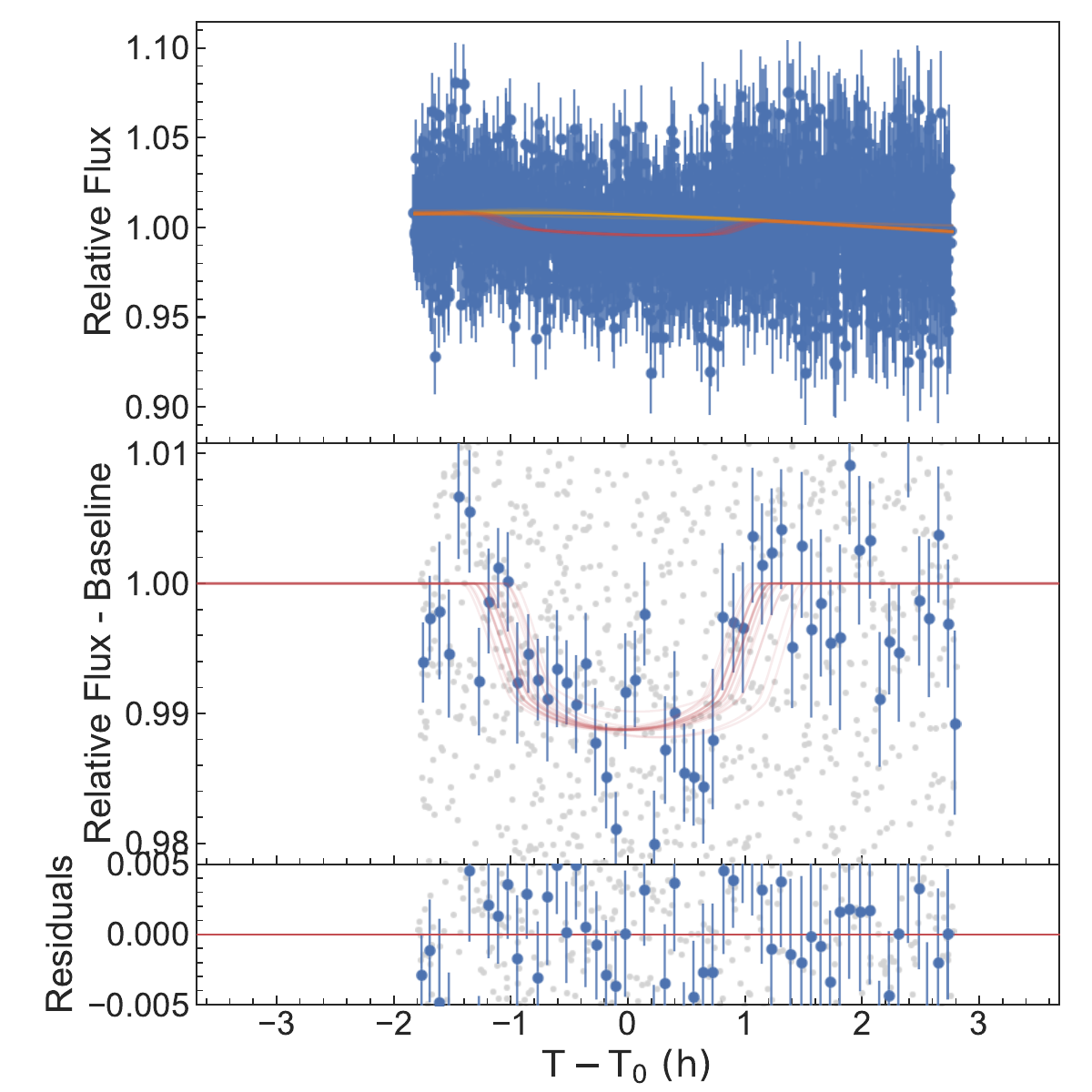}
\caption{Same as in \autoref{fig:kelt-16}. Normalized lightcurves for the transiting system of Kepler-41 (TIC 271042217) \citep{kepler41} from TWIST on 2025 August 12.}
\label{fig:kepler-41}
\end{figure}

\begin{table*}
	\centering
	\caption{\texttt{allesfitter} model priors. Planetary System properties for the transiting system Kepler-41 (TIC 271042217), as illustrated in \autoref{fig:kepler-41}, at TWIST on 2025 August 12.}
    {\renewcommand{\arraystretch}{1.25}
    \begin{tabular}{lccc}
    \hline
	\textbf{Parameter}&\textbf{Initial Guess}&\textbf{Prior}&\textbf{Fitted value Toran-26B} \\
    \hline
    $R_p / R_{\star}$            & \FITNCrb          & $\mathcal{U}\left(0.097,0.107\right)$         & \CMOSNCrb	   \\
	$(R_\star + R_b) / a_b$      & \FITNCrsuma       & $\mathcal{U}\left(0.213,0.223\right)$     & \CMOSNCrsuma  \\
    $\cos{i_b}$                  & \FITNCinc         & $\mathcal{U}\left(0.13,0.14\right)$     & \CMOSNCinc	   \\
    Tc (BJD)		             & \FITNCtc	         & $\mathcal{U}\left(2455664.156,2455664.16\right)$ & \CMOSNCtc \\
    Orbital Period (days)	  	 & \FITNCperiod      & $\mathcal{U}\left(1.85,1.86\right)$   & \CMOSNCperiod \\
	$q_1$                        & \FITNCqone        & $\mathcal{N}\left(0.369,0.008\right)$           & \CMOSNCqone   \\
	$q_2$                        & \FITNCqtwo        & $\mathcal{N}\left(0.396,0.074\right)$           & \CMOSNCqtwo   \\
    \hline
    \end{tabular}}
    \label{tab:planetary-Kepler-41_0812}
\end{table*}

% \begin{figure}
% \centering
%     \includegraphics[width=\columnwidth]{Data/TOI-5688_0815.pdf}
% \caption{Same as in \autoref{fig:kelt-16}. Normalized lightcurves for the transiting system of TOI-5688 A (TIC 193634953) \citep{toi5688} from TWIST on 2025 August 15.}
% \label{fig:toi-5688}
% \end{figure}

\begin{figure}
\centering
\includegraphics[width=\columnwidth]{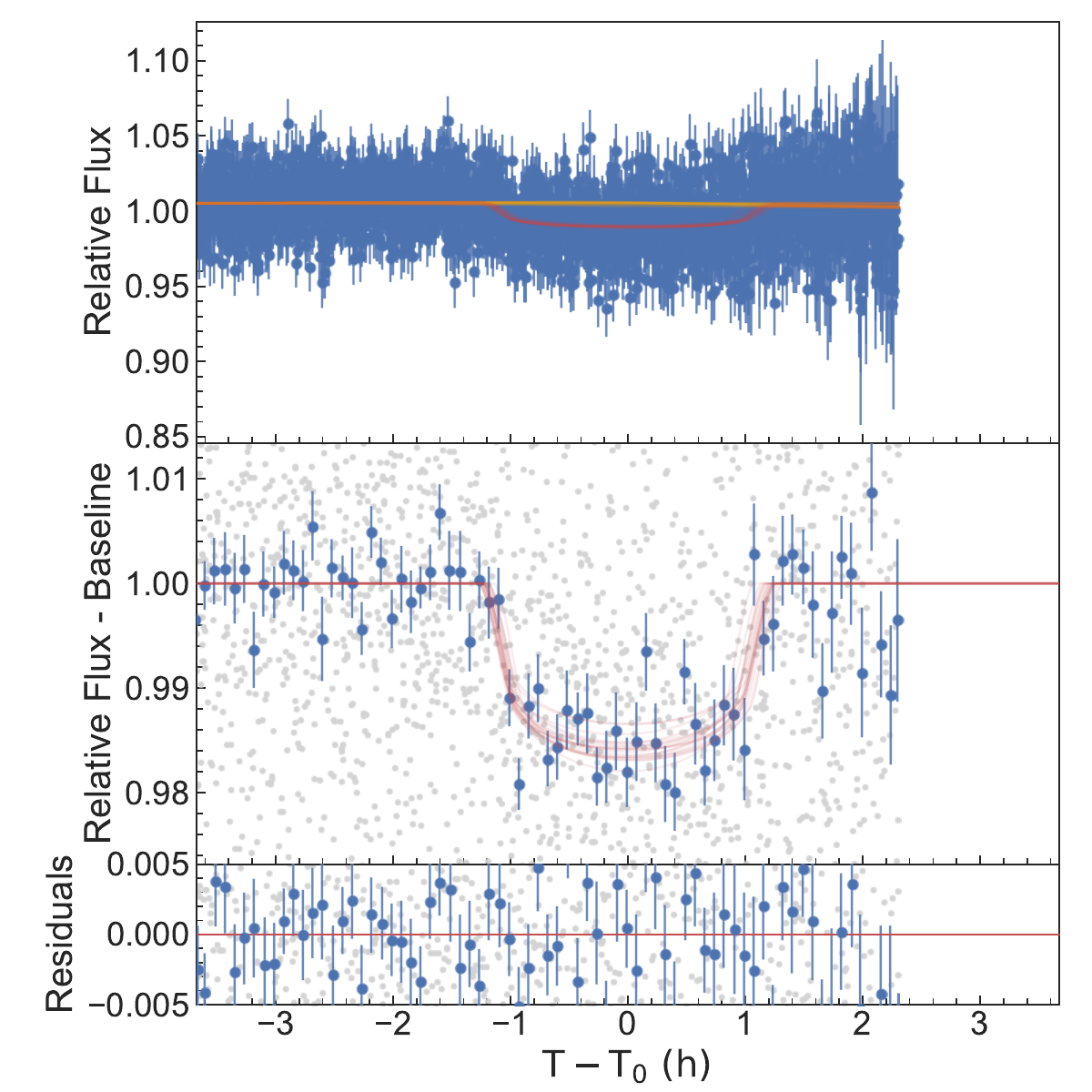}
\caption{Same as in \autoref{fig:kelt-16}. Normalized lightcurves for the transiting system of Kepler-670 (TIC 27990610) \citep{kepler670} from TWIST on 2025 August 16.}
\label{fig:kepler-670}
\end{figure}

\begin{table*}
	\centering
	\caption{\texttt{allesfitter} model priors. Planetary System properties for the transiting system Kepler-670 (TIC 27990610), as illustrated in \autoref{fig:kepler-670}, at TWIST on 2025 August 16.}
    {\renewcommand{\arraystretch}{1.25}
    \begin{tabular}{lccc}
    \hline
	\textbf{Parameter}&\textbf{Initial Guess}&\textbf{Prior}&\textbf{Fitted value Toran-26B} \\
    \hline
    $R_p / R_{\star}$            & \FITNErb          & $\mathcal{U}\left(0.1,0.13\right)$         & \CMOSNErb	   \\
	$(R_\star + R_b) / a_b$      & \FITNErsuma       & $\mathcal{U}\left(0.1,0.12\right)$     & \CMOSNErsuma  \\
    $\cos{i_b}$                  & \FITNEinc         & $\mathcal{U}\left(0.005,0.02\right)$     & \CMOSNEinc	   \\
    Tc (BJD)		             & \FITNEtc	         & $\mathcal{U}\left(2455003.96,2455003.97\right)$ & \CMOSNEtc \\
    Orbital Period (days)	  	 & \FITNEperiod      & $\mathcal{U}\left(2.811,2.818\right)$   & \CMOSNEperiod \\
	$q_1$                        & \FITNEqone        & $\mathcal{N}\left(0.382,0.013\right)$           & \CMOSNEqone   \\
	$q_2$                        & \FITNEqtwo        & $\mathcal{N}\left(0.393,0.09\right)$           & \CMOSNEqtwo   \\
    \hline
    \end{tabular}}
    \label{tab:planetary-Kepler-670_0816}
\end{table*}

\begin{figure}
\centering
\includegraphics[width=\columnwidth]{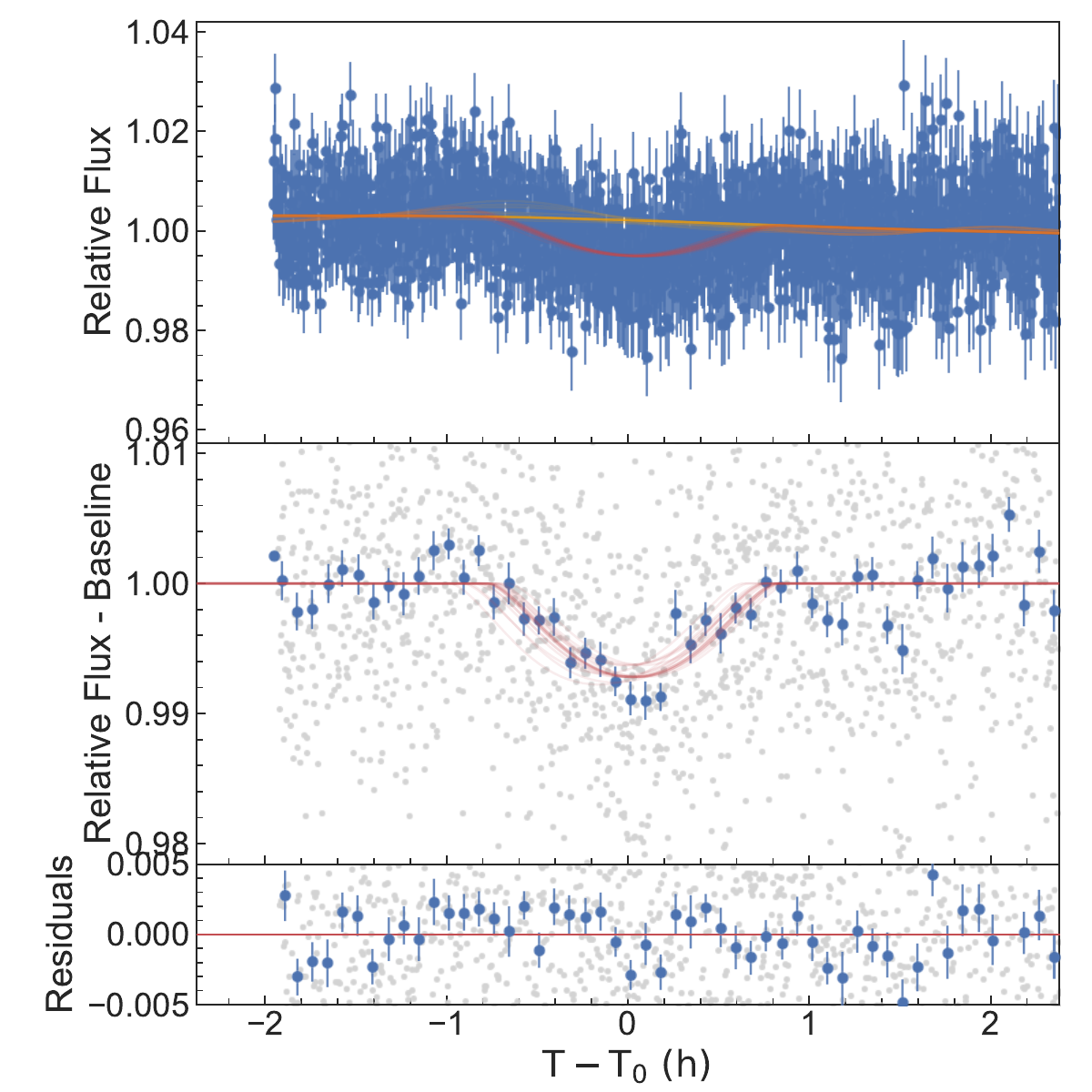}
\caption{Same as in \autoref{fig:kelt-16}. Normalized lightcurves for the transiting system of Kepler-76 (TIC 138644215) \citep{kepler76} from TWIST on 2025 September 9.}
\label{fig:kepler-76}
\end{figure}

\begin{table*}
	\centering
	\caption{\texttt{allesfitter} model priors. Planetary System properties for the transiting system Kepler-76 (TIC 138644215), as illustrated in \autoref{fig:kepler-76}, at TWIST on 2025 September 9.}
    {\renewcommand{\arraystretch}{1.25}
    \begin{tabular}{lccc}
    \hline
	\textbf{Parameter}&\textbf{Initial Guess}&\textbf{Prior}&\textbf{Fitted value Toran-26B} \\
    \hline
    $R_p / R_{\star}$            & \FITNFrb          & $\mathcal{U}\left(0.098,0.108\right)$         & \CMOSNFrb	   \\
	$(R_\star + R_b) / a_b$      & \FITNFrsuma       & $\mathcal{U}\left(0.24,0.25\right)$     & \CMOSNFrsuma  \\
    $\cos{i_b}$                  & \FITNFinc         & $\mathcal{U}\left(0.21,0.22\right)$     & \CMOSNFinc	   \\
    Tc (BJD)		             & \FITNFtc	         & $\mathcal{U}\left(2455654.041,2455654.045\right)$ & \CMOSNFtc \\
    Orbital Period (days)	  	 & \FITNFperiod      & $\mathcal{U}\left(1.54,1.55\right)$   & \CMOSNFperiod \\
	$q_1$                        & \FITNFqone        & $\mathcal{N}\left(0.345,0.009\right)$           & \CMOSNFqone   \\
	$q_2$                        & \FITNFqtwo        & $\mathcal{N}\left(0.384,0.08\right)$           & \CMOSNFqtwo   \\
    \hline
    \end{tabular}}
    \label{tab:planetary-kepler-76_0909}
\end{table*}
%%%%%%%%%%%%%%%%%%%%%%%%%%%%%%%%%%%%%%%%%%%%%%%%%%

% Don't change these lines
\bsp	% typesetting comment
\label{lastpage}
\end{document}